\documentclass[fleqn,usenatbib]{mnras}

\usepackage{newtxtext,newtxmath}
\usepackage[dvipsnames]{xcolor}

\usepackage[T1]{fontenc}

\usepackage{anyfontsize} 
\DeclareRobustCommand{\VAN}[3]{#2}
\let\VANthebibliography\thebibliography
\def\thebibliography{\DeclareRobustCommand{\VAN}[3]{##3}\VANthebibliography}

\usepackage{graphicx}	
\usepackage{amsmath}	
\usepackage{multicol}
\usepackage{tipx}

\def\afe{$[\alpha/\mathrm{Fe}]$}
\def\feh{$[\mathrm{Fe}/\mathrm{H}]$}
\def\bpass{{\sc{bpass}}}
\def\msun{M$_\odot$}

\def\yg{\textbenttailyogh}

\defcitealias{Grevesse93}{GN93}
\defcitealias{Hurley00}{HPT00}
\defcitealias{Byrne25}{B25}

\def\hpt{\citetalias{Hurley00}}
\def\cmb25{\citetalias{Byrne25}}

\title[Re-calibrating rapid stellar evolution]{Re-calibrated analytic stellar evolution formulae\ for the main sequence:\\Extension to massive stars and variations in $[\alpha/\mathrm{Fe}]$}

\author[C. M. Byrne et al.]{
Conor M. Byrne,$^{1}$\thanks{E-mail: conor.byrne@warwick.ac.uk (CMB)}
Elizabeth R. Stanway,$^{1}$
and Jan J. Eldridge$^{2,1}$
\\
$^{1}$Department of Physics, University of Warwick, Gibbet Hill Road, Coventry CV4 7AL, UK\\
$^{2}$Department of Physics, University of Auckland, Private Bag 92019, Auckland, New Zealand
}

\date{Accepted XXX. Received YYY; in original form ZZZ}

\pubyear{\the\year{}}

\begin{document}
\label{firstpage}
\pagerange{\pageref{firstpage}--\pageref{lastpage}}
\maketitle

\begin{abstract}
Analytic formulae are an efficient way to approximate the evolution of a star of a given mass and chemical composition, and are widely used in the literature to compute large grids of evolutionary models for stellar population synthesis. Such formulae rely on calibration to detailed stellar evolution models, and existing versions are based on a sparse, limited grid in comparison to modern simulations.  By calibrating against the latest generation of single-star evolution models calculated within the Binary Population and Spectral Synthesis (\bpass) framework, we revise and extend pre-existing equations to include the evolution of massive ($M\gtrsim50\,$\msun) stars and changes in chemical abundances in the form of variations in \afe. We introduce additional terms which allow for a better approximation of the main-sequence lifetimes, and the luminosities and radii at the beginning and end of the main sequence. These revised equations describe the main-sequence age, zero-age and terminal-age radius and luminosity with mean errors of less than 10 per cent or 0.05 dex, for all except the terminal age radius, which shows RMS errors of 20 per cent and 0.1 dex. For \afe\ enriched stellar populations, we advise using a matched-iron value for metallicity in the equations, consistent with the expectation that iron is the dominant opacity source affecting the evolution of main-sequence stars.
\end{abstract}




\section{Introduction}

The theory of stellar evolution underpins almost all areas of modern astrophysics. The timescales of evolution dictate the speed of nucleosynthesis and the occurrence rates and times of energetic events such as supernovae and gravitational wave emission from compact object mergers. Nucleosynthetic yields and their release in supernovae control the chemical enrichment of the Universe \citep[{e.g.}][]{Hoyle56,Burbidge57,Kobayashi20}, influencing subsequent generations of star formation and provide energetic feedback which impacts on the evolution of galaxies \citep[{e.g.}][]{Larson74,Hopkins12}. The evolution of stellar luminosity and surface temperatures over time changes the colour and quantity of light that is observed.

Stellar evolution theory is a mature field of study, with the general principles and fundamental behaviour being well-understood \citep[e.g.][]{KippenhahnBook,EldridgeToutBook}. Nonetheless many open questions and uncertainties remain. These restrict the accuracy of simulations of stellar evolution \citep[see e.g.][for reviews of uncertainties in the evolution of massive stars and binary stars respectively]{Langer12,deMarcoIzzard17}. The opaque nature of stellar interiors (and the computational limitations which often restrict simulations to 1-dimensional or short-timescale calculations)  means that numerical approximations and analytic prescriptions are often employed in order to simulate the evolution of a star from its birth through to its death, while accounting for all the microphysical and macrophysical processes taking place in the interior. Some of these can be derived from laboratory measurements \citep[such as atomic and molecular opacity data, e.g.][]{Iglesias96,OP1,OP2}, while others are inferred through calibration of simulations to observed stellar systems \citep[e.g. the treatment of convective mixing,][]{Stancliffe16}.

There are many applications where the calculation of a large ensemble of stars is needed, where different masses, chemical composition or indeed the nature of the numerical physics prescriptions are varied to evaluate the impact across a diverse range of stellar types. This can increase the number of calculations required by several orders of magnitude, if one wishes to evolve each star in detail.

Examples of this kind generally involve stellar population synthesis, which is used to infer global properties of a collection of stars, such as the age of a stellar cluster or the mass of stellar material in a distant galaxy where individual stars cannot be resolved \citep{Tinsley68}. This can be computationally expensive if the desired number of models is of the order $10^6$ or more.

In this situation, {\emph{rapid}} population synthesis is a commonly used technique. The detailed calculation of the stellar structure and evolution are replaced with a set of analytic formulae, which have been calibrated against a sparser set of detailed stellar structure and evolution models. These equations can be used to describe changes in luminosity, radius and temperature of the star as a function of stellar mass, metallicity and evolutionary age. This approach implicitly assumes that the nature of stellar evolution varies smoothly with mass and composition and thus can be well approximated by a series of polynomial equations. One such set of equations is that introduced by \citet{Hurley00}, (hereinafter HPT00). The equations of \citetalias{Hurley00} were derived based on a total of 168 stellar evolution  models, with 24 models at each of 7 different metallicities, spanning from 0.5 to 50 \msun\ and $0.0001\le Z\le0.03$.

In this paper, we present revised and extended analytic formulae for main sequence stellar evolution, calibrated against the Binary Population and Spectral Synthesis (\bpass) v2.4.0 single-star models of \citet{Byrne25}. This grid provides 351 models per composition, spanning masses from 0.1 to 316 \msun\ across 74 distinct combinations of metallicity and \afe. This offers a substantially richer dataset than was available to \citetalias{Hurley00}, extending to higher initial stellar masses and an additional composition dimension in the form of \afe. We describe the \bpass\ framework and the relevant model grid in Section~\ref{Sec:BPASS}, before outlining our fitting methodology in Section~\ref{sec:methods}. In Section~\ref{sec:fitresults} we present revised analytic expressions for the zero-age and terminal-age main sequence properties. The overall performance of the new equations is assessed and compared to the \citetalias{Hurley00} expressions, including a demonstration of their impact on binary population synthesis outcomes using the rapid population synthesis code BSE \citep{Hurley02} in Section~\ref{sec:discussion}. We summarise our conclusions in Section~\ref{sec:conclusions}. The revised equations extend the domain of reliable analytic stellar evolution to massive stars above {100} \msun\ and to non-Solar-scaled abundance patterns{, while improving their performance against a set of modern stellar evolution calculations}. This addresses limitations which are increasingly consequential for studies of gravitational wave progenitors, high-redshift stellar populations, and other regimes in which the \citetalias{Hurley00} equations are unreliable.

\section{Motivation}

As with any analytic model, the equations only provide an accurate approximation of behaviour within the range of parameters for which they are calibrated.The equations presented by \citetalias{Hurley00} are based on a grid of stellar evolution models calculated by \cite{Pols98}. These models were calculated at 7 metallicities in the range $0.0001\le Z\le0.03$, with 24 models at each composition, spanning initial masses of $0.5\le M/\mathrm{M}_\odot\le50$, giving a total of 168 individual stellar evolution models. This presents limitations if one uses these expressions in the study of young stellar populations or populations of stars in the distant Universe, where massive stars greater than 50\,\msun\ make a non-negligible contribution to the stellar population. As noted by \citet{Banerjee20}, the equations of \citetalias{Hurley00} are known to be unreliable for extrapolation, particularly above 100\msun.

The detection of about $30$\,\msun\, black holes in gravitational wave events such as GW150914 \citep{Abbott16} implies progenitors above the mass range in which the \citetalias{Hurley00} equations are calibrated. As a consequence, any population synthesis used to explain and interpret gravitational wave catalogues such as GWTC4 \citep{LIGO-GWTC4} need to consider such progenitors. 
Massive stars show a preference to evolve in binary and multiple star systems \citep{2012Sci...337..444S}. If one considers binary stellar populations, the problem is inherently more complex owing to interactions. As a result, small uncertainties or inconsistencies in the approximations for single star evolution can alter the interaction regimes, binary evolution outcomes and the distribution and population of binary compact object systems one would expect. This emphasises the need for reliable models of stellar evolution - preferably detailed models, but in their absence, well-calibrated analytic approximations.

The analytic equations of \citetalias{Hurley00} are widely used in a number of population synthesis codes such as SSE (\citetalias{Hurley00}), BSE \citep{Hurley02}, binary\_c \citep{Izzard23}, Cosmic \citep{Breivik20}, COMPAS \citep{COMPAS22}, and STARTRACK \citep{Belczynski08}. They also lie at the foundation of emulators and interpolators such as METISSE \citep{Agrawal20}. These tools provide the crucial foundations for interpreting transient event rates, the nature of unresolved stellar populations, providing predictions and estimates for observable quantities such as supernova rates, ionising photon fluxes and gravitational wave merger rates across cosmic time \citep[e.g.][]{Mapelli17,Tang20}. As observational astronomy pushes into more extreme regimes and the computational power available to run stellar evolution simulations increases, it is worth undertaking an assessment of the validity of these analytic equations to describe stellar evolution across a wide range of masses and compositions and make adjustments to them where required. 

Examples of such studies include that of \citet{Stevenson17}, who used COMPAS models to consider the progenitor systems of the first gravitational wave detections from merging binary black holes, \citet{Bruel24}, who used Cosmic models within a stellar cluster simulation to model black-hole mergers in galaxy-scale simulations, and \citet{Mandel21} who used COMPAS models to investigate properties such as the black hole mass distribution and the period-eccentricity distribution of double neutron stars. These studies rely to some extent on analytic prescriptions for stellar evolution, for which extrapolation may not always be reliable. 

An additional consideration in the distant Universe is the presence of non-uniformly scaled abundance patterns, of which $\alpha$-enhancement is the most commonly encountered. This arises from the enrichment timescales for specific elements, with $\alpha$-process elements undergoing rapid enrichment as a product of massive star evolution, while the enrichment of iron-group elements proceeds more slowly, being predominantly released as an end product of low-mass stellar evolution. These changes in the relative abundances of chemical elements are observed in young, low-metallicity stellar populations, such as at high-redshift. It has been shown that $\alpha$-enhancement has an impact on the lifetimes, radii and effective temperatures of main sequence stars \citep{Byrne25}. This additional dimension in chemical composition is not considered in the analytic equations of \citetalias{Hurley00}, nor in population synthesis more widely.

\begin{figure}
    \centering
    \includegraphics[width=0.481\textwidth]{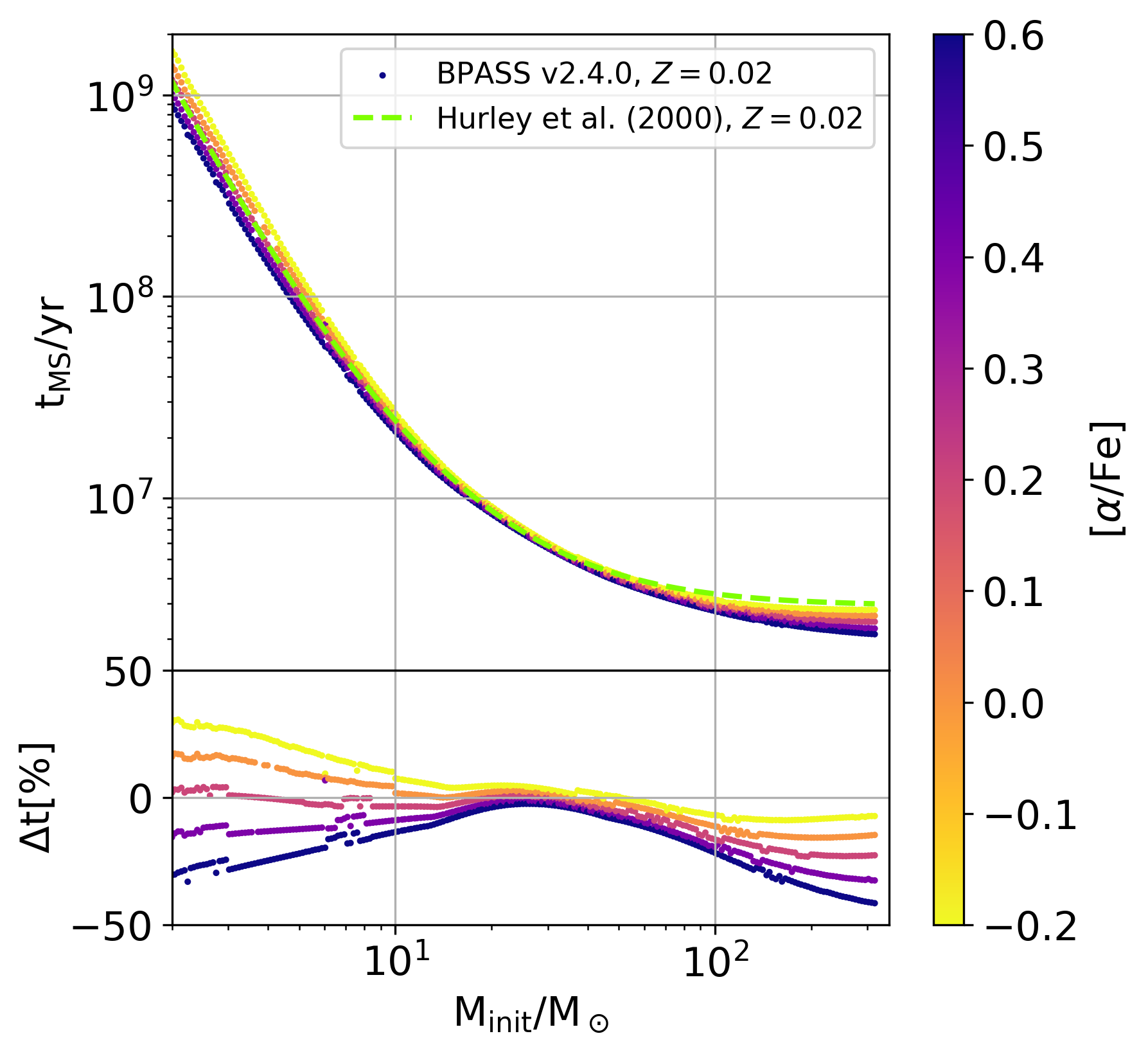}
    \caption{Main-sequence (core-hydrogen-burning) lifetimes as a function of initial stellar mass. The points indicate \citet{Byrne25} models from \bpass\ v2.4.0 with a total metallicity mass fraction of $Z=0.020$,while \afe\ is varied, as indicated by the colour scale. The chartreuse dashed line indicates the analytic lifetimes predicted using the equations of \citetalias{Hurley00} for a metallicity of $Z=0.020$}
    \label{fig:lifetimes}
\end{figure}

Fig.~\ref{fig:lifetimes} illustrates how the main-sequence lifetime is affected by changes in \afe. In this instance, the end of the main sequence is defined as the age at which the star's central hydrogen mass fraction has decreased to $10^{-12}$. In addition to the variation in low-mass star lifetimes with \afe\, this figure highlights how the \citetalias{Hurley00} equations consistently overestimate the main-sequence lifetimes of massive stars ($M\gtrsim80$\msun), which are also \afe-dependent. This behaviour provides clear motivation to revisit these analytic equations and adjust them to i) better capture stellar evolution for stars with masses greater than 50 \msun\ and ii) account for variations in stellar properties arising from different \afe. This will enable the analytic approach to remain relevant in the study of stellar populations dominated by massive stars and/or non-Solar scaled chemical compositions, as commonly found in the distant Universe.

\section{BPASS}
\label{Sec:BPASS}

Binary Population and Spectral Synthesis \citep[\bpass,][]{2017PASA...34...58E,2018MNRAS.479...75S,Byrne22} is a comprehensive framework capable of end-to-end modelling of simple stellar populations (SSPs), combining detailed custom 1-D stellar evolution models with population synthesis weightings and libraries of theoretical stellar spectra.

As part of the latest generation of \bpass\ models, a grid of single-star evolution models (v2.4.0) with variation in $Z$ and \afe\ were calculated and presented by \cite{Byrne25}, hereinafter B25. This comprises 351 stellar models per composition, logarithmically spaced, spanning masses from 0.1 to 316 \msun. This provides a dense set of models with which to evaluate, extend and re-calibrate the \citetalias{Hurley00} semi-analytic expressions used for rapid stellar evolution calculations. 

In the \bpass\ v2.4.0 models, the base elemental abundances used are those of \citep[][hereinafter GN93]{Grevesse93}, with a Solar metallicity mass fraction of $Z_\odot=0.020$. When considering changes in \afe, the standard \citetalias{Grevesse93} mixture is labelled as \afe$=+0.2$ to better reflect more recent Solar abundance determinations \citep[e.g.][]{Asplund09,Asplund21}. Thus, the \bpass\ models labelled as $Z=0.020, [\alpha/\mathrm{Fe}]=+0.2$ are the models which correspond most directly to the underlying \citetalias{Grevesse93} composition, while models labelled $Z=0.014, [\alpha/\mathrm{Fe}]=+0.0$ closely correspond to the \citet{Asplund21} Solar abundance. 

The underlying opacity grids employed are those originally computed by \citet{Eldridge04a}. These tables are built from Type~2 format OPAL \citep{Iglesias96} opacity tables, which account for changes in H, He, C and O. Since oxygen is the $\alpha$-process element with the strongest contribution to stellar opacity, this flexibility provides appropriate, tabulated opacity data to be used for $\alpha$-enhanced compositions.

Because iron is the dominant metal contributing to stellar envelope opacity, and following the results of \citetalias{Byrne25}, which showed that most of the variation in stellar evolution arises from changes in \feh\ rather than \afe, we define an iron-equivalent metallicity $Z'$. This represents the total metallicity mass fraction of a composition with the \citetalias{Grevesse93} abundance that has the same iron mass fraction in the chosen composition. As we have re-defined the standard \citetalias{Grevesse93} to have \afe$=+0.2$, by definition $Z=Z'$ for compositions with \afe=+0.2, while for \afe$>+0.2$, $Z>Z'$ (and vice versa), because at a given $Z$, an $\alpha$-rich abundance pattern necessarily has a lower iron abundance. We also define a log-normalised quantity \yg$=\log(Z'/0.02)$, which proves useful to describe the metallicity dependence of the parameters.

{During stellar evolution, the \bpass\ v2.4.0 models apply main sequence mass-loss prescriptions based on the effective temperature. For temperatures below 20\,000\,K and above 60\,000\,K, the mass-loss prescription of \citet{deJager88} is applied, while for stars with effective temperatures above 25\,000\,K and below 50\,000\,K, the mass-loss prescription of \citet{Vink01} was applied, with linear interpolation used in the intermediate regimes. For Wolf-Rayet stars, stripped of hydrogen on the main sequence, the mass-loss rates of \citet{Nugis00}, as modified by \citet{Dray03} are applied, which scale based on surface composition and temperature. Where these mass-loss rates scale with stellar metallicity, in this implementation this is taken to be the iron-equivalent metallicity, $Z'$, because iron is the primary contributor to stellar wind rates.}

In this work, we restrict ourselves to the considering the early stages of stellar evolution from stellar birth on the main sequence, through to the core hydrogen exhaustion. The post-main sequence evolution is more strongly dominated by stellar winds than composition, see Section~\ref{sec:limits} for discussion. In addition, we only consider the evolution of single stars, because the effects of binary evolution are considerably more complex and are not the focus of this study.

\section{Methods}
\label{sec:methods}
The origin of analytic formulae or interpolation equations as an approximation for stellar evolution calculations can be traced back at least as far as the appendix of \citet{Eggleton89}. 38 stellar evolution models with $Z=0.020$, spanning from 0.1 to 100\,\msun\, were calculated with the Cambridge STARS evolution code \citep{Eggleton71,Eggleton72}. These models were used to construct equations describing $L_0$ and $R_0$, the zero-age main-sequence (ZAMS) luminosity and radius, along with $t_{\mathrm{MS}}$, the main-sequence lifetime. In addition quadratic and cubic equations were used to describe the main-sequence evolution in luminosity and radius respectively.

Using a revised version of the STARS code, \citet{Pols95} calculated 600 stellar structure models for ZAMS stars between 0.1 to 100\,\msun\, for each of 7 different metallicities ranging from $Z=10^{-4}$ to $Z=0.030$. These were used by \citet{Tout96} to construct new analytic formulae describing the ZAMS luminosity and radius of stars as a function of both mass and metallicity for the first time{, and applied in an early form of binary population synthesis model by \citet{Tout97}}. 

\subsection{Methods of Hurley et al.}
\label{sec:HPT_methods}
A further grid of detailed stellar evolution models was calculated by \citet{Pols98} comprising models at 7 different metallicities between $Z=10^{-4}$ to $Z=0.030$ and stellar masses between 0.5 to 50\,\msun\, in steps of 0.1 in $\log(M)$. These models were used as the foundation to produce the comprehensive set of analytic stellar evolution equations outlined in \citetalias{Hurley00}. These analytic equations were used to describe the evolution of luminosity, temperature and radius of stars as a function of their age, mass and metallicity mass fraction, covering both the main sequence and post-main sequence phases of evolution, such as the Giant Branch and the Horizontal Branch.

For the equations governing main-sequence evolution, the original fitting by \citetalias{Hurley00} was made with least-squares fits to their stellar models. The functional form of each equation (the number of terms and the choice of exponents) was guided by visual inspection of the data. These are typically {rational polynomials} of the form 
$\sum\limits_{n} a_nM^{\beta_n} / \sum\limits_{m} a_mM^{\beta_m} $
 where $n$ and $m$ can represent any number of terms, and are typically rational polynomials.

The exponents $\beta_n$ are usually fixed based on the observed mass dependence, while the coefficients $a_n$ provide the scaling. The choice of $\beta$ together with whether a term appears in the numerator or denominator, determines the mass range over which it influences the fit.
To incorporate metallicity dependence, each coefficient $a_n$ is usually parametrised a function of metallicity by a fourth-order polynomial $a_n=c_0+c_1\zeta+c_2\zeta^2+c_3\zeta^3+c_4\zeta^4$, where $\zeta=\log(Z/0.02)$.

The different evolutionary stages are defined in terms of key transitions in stellar structure (e.g. core exhaustion) and surface composition (e.g. hydrogen or helium-stars). The underlying \citet{Pols98} stellar models used for these fits were computed assuming no mass loss by stellar winds. {This is a contrast to the approach of the \bpass\, models, where mass loss is routinely applied to detailed stellar evolution models. Mass loss is treated as an additional post-processing step in the \citet{Pols98} models. Further discussion of the impacts of the treatment of stellar mass-loss is presented in Section~\ref{sec:massloss}}

\begin{figure}
    \centering
    \includegraphics[width=0.481\textwidth]{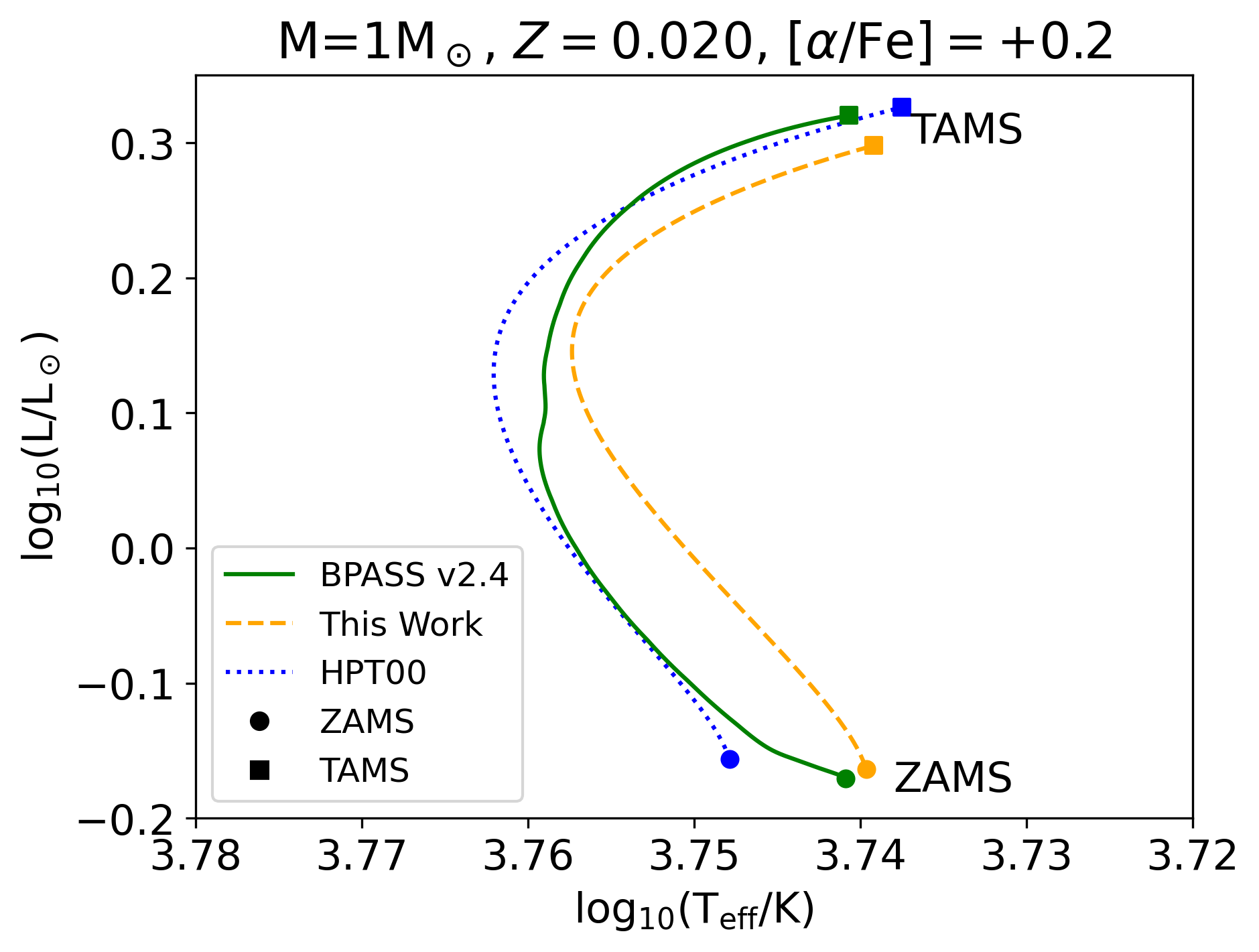}
    \includegraphics[width=0.481\textwidth]{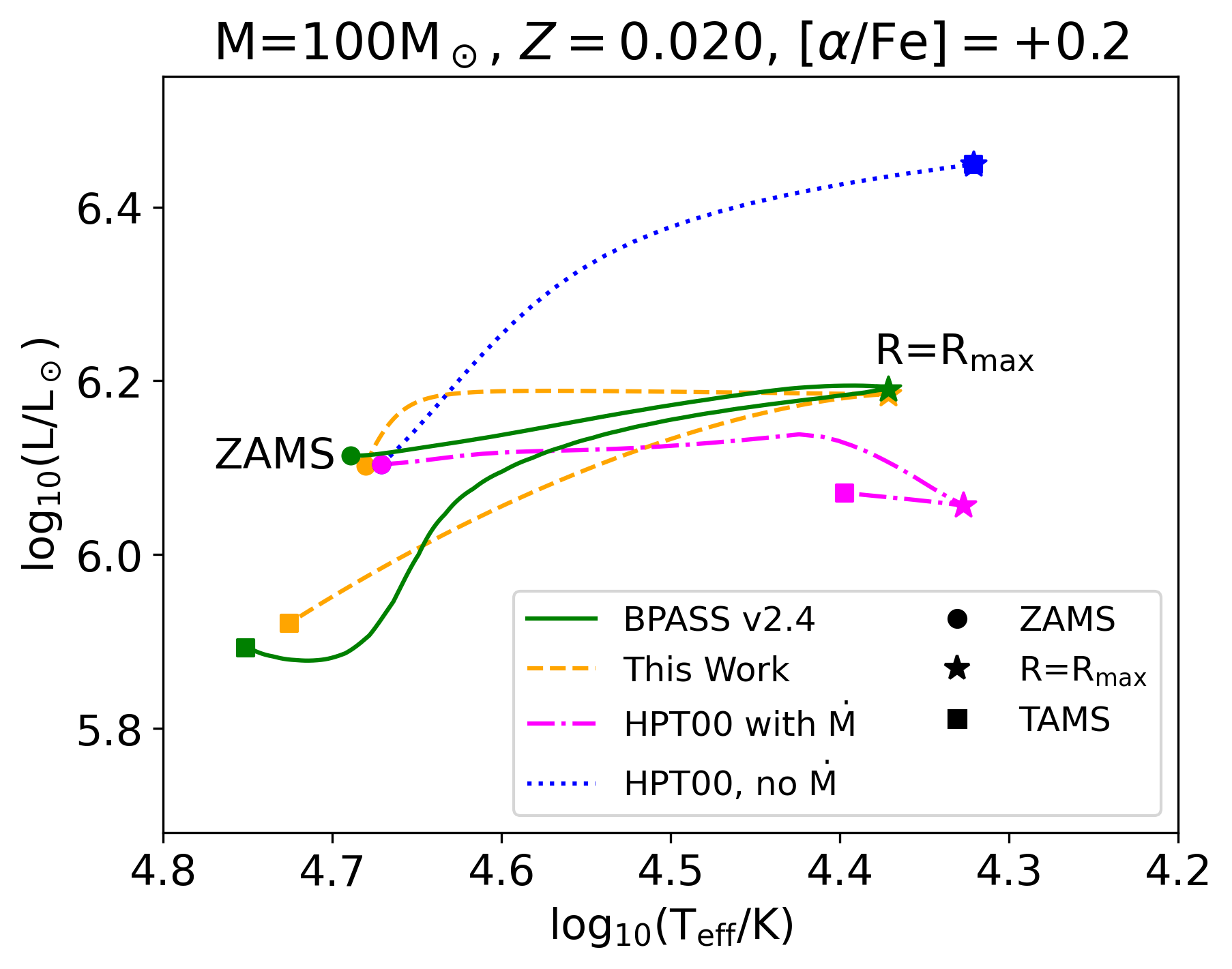}
    \caption{Core-hydrogen-burning evolution of a 1\msun\ (upper panel) and a 100\msun\ star (lower panel). The detailed \bpass\ v2.4.0 evolution tracks are shown by solid green lines, while the analytic formulae {with zero mass-loss} of \citetalias{Hurley00} are shown as dotted blue lines and the new analytic fits presented in this work are plotted as orange dashed lines. {In the lower panel, an additional \citetalias{Hurley00} evolution track incorporating stellar mass-loss using the same prescriptions as in the detailed \bpass\ models is indicated by a magenta dot-dashed line}. The Zero Age main sequence is further indicated by a round point on each track, while the Terminal Age main sequence (defined as the point of core hydrogen exhaustion) is shown by square symbols. In the 100\msun\ tracks, the point at which the models reach maximum main sequence radius is indicated by the star-shaped symbols.}
    \label{fig:HRD1-100}
\end{figure}

Fig.~\ref{fig:HRD1-100} illustrates the evolution of stars with Solar metallicity ($Z=0.020$) and masses of 1\msun\ and 100\msun. A detailed model from \citetalias{Byrne25} is shown by a solid line, while the dotted line illustrates the approximate evolution track as determined by the \citetalias{Hurley00} equations, calculated using the SSE code they were devised for\footnote{{Caution should be exercised when transcribing the \citetalias{Hurley00} equations, in particular equation 21, where there is a typo in the preprint version ($a_{59}M^{a_{61}}$) which is corrected in the final published version ($a_{59}+M^{a_{61}}$).}}. A dashed line is used to show the updated approximate evolution tracks implemented later in this work. For the 1\msun\ model, it can be seen that the \citetalias{Hurley00} approximation works well across the entire hydrogen-burning evolution, with only small offsets in effective temperature. {The new fits presented in this work perform marginally less well in this specific example, with offsets less than about 0.01 dex in effective temperature, but as will be shown in Section~\ref{sec:recalibrate} below, the fitting undertaken was a global optimisation across all masses, metallicities and combinations, with overall performance being better on average}. For the 100\msun\ model, the \citetalias{Hurley00} evolution track evolves in a manner contrary to the detailed model, resulting in substantially incorrect temperature and luminosity estimates. {This is true regardless of whether the \citetalias{Hurley00} SSE evolution model is computed in its simplest form with zero mass loss (blue dotted line) or if the same mass-loss laws are applied to it as used in the detailed \bpass\ stellar evolution model (magenta dot-dashed line)}. This demonstrates the inadvisability of extrapolating the \citetalias{Hurley00} equations into a regime well beyond that in which they were calibrated.

It is worth highlighting at this point that the stellar evolution models of \citet{Pols98} used to calibrate the \citetalias{Hurley00} equations and the evolution models of \citetalias{Byrne25} used to calibrate the equations presented in this work are based on the same underlying stellar evolution framework, both having been calculated with derivatives of the earlier versions of the Cambridge STARS code \citep{Eggleton72}. Notable differences include the choices of opacity tables, nuclear burning rates, and {as noted above,} the inclusion of wind mass-loss in the \citetalias{Byrne25} models, which was not used in the \citet{Pols98} models.

\subsection{Recalibration Methodology}
\label{sec:recalibrate}
For the stellar evolution grids at each of the 74 compositions, the age of every model when it reaches the {end of the main sequence, $t_\mathrm{MS}$}, was determined. In \citetalias{Hurley00}, this was defined differently {with the age of the star at the base of the giant branch, $t_\mathrm{BGB}$ calculated based on certain criteria, and then a small correction applied (Equations 5, 6 and 7 of \citetalias{Hurley00}) to translate this into $t_\mathrm{MS}$}. Different criteria were applied in different mass ranges, related to the mass of the convective envelope or the helium-burning luminosity. In this work we apply a simpler metric which is {to define the end of the main sequence, $t_\mathrm{MS}$ as the} time at which the central hydrogen mass fraction drops below $10^{-12}$. For massive stars, {when applying the definitions of \citetalias{Hurley00} to determine $t_\mathrm{MS}$} the time differs by a systematic offset of approximately 1 per cent of the main sequence lifetime. Below 2\,\msun\ the offset between applying the \hpt\ approach and our definition is described by a random scatter with an rms of approximately two per cent.

The closest model timestep to $t_\mathrm{MS}$ was extracted, and the luminosity and radius ($L_\mathrm{t_{MS}}$ and $R_\mathrm{t_{MS}}$) were recorded for comparison to Equations 8 and 9 of \citetalias{Hurley00}. Models with an initial mass less than about 0.7\,\msun\ have main sequence lifetimes which exceed the Hubble time and the arbitrary 30\,Gyr time limit imposed on the \bpass\ v2.4.0 models. Consequently, these low-mass models were excluded from the fitting at this stage. 

Tabulated data were compiled, containing $Z$, \afe, $Z'$ $M_\mathrm{init}$, $L_0$, $R_0$, $t_\mathrm{MS}$, $t_\mathrm{BGB}$, $L_\mathrm{t_{MS}}$ and $R_\mathrm{t_{MS}}$ for each single star evolution model in \bpass\ v2.4.0 with an initial mass greater than 0.7\msun. This comprises 19\,287 models in total, a much larger data set than the total of 168 models used by HPT00. Due to difficulties extending the metallicity evolution of some coefficients to very low metallicities, compositions with $Z'<0.001$ were excluded, leaving a total of 14\,587 models for the fitting process. 

An optimisation parameter was defined as the one minimizing the log-space loss function across the full dataset. The loss function was quantified in two ways.  
\begin{enumerate}
    \item Root-mean-square (RMS) percentage error,
    \begin{equation}
        \Delta_\% = \sqrt{ \left\langle 
        \left(\frac{f_{\mathrm{analytic}} - f_{\mathrm{BPASS}}}{f_{\mathrm{BPASS}}}\right)^2 
        \right\rangle } \times 100 ,
    \label{eq:deltapct}
    \end{equation}
    which gives a measure of fractional error across the dataset.  
    \item Log-space RMS error,
    \begin{equation}
        \Delta_{\log} = \sqrt{ \left\langle 
        \left[ \log(f_{\mathrm{analytic}}) -\log\left(f_{\mathrm{BPASS}}\right)\right]^2 
        \right\rangle } ,
    \label{eq:deltalog}
    \end{equation}
    which does the comparison in log space, a meaningful comparator when the properties involved (mass, age, radius and luminosity) span multiple orders of magnitude.  
\end{enumerate}

Because the parameters being fit each span several orders of magnitude, in most cases we optimized the fit in log-space, minimizing the mean-squared deviation between $\log f_{\mathrm{analytic}}$ and $\log f_{\mathrm{BPASS}}$, where $f$ indicates the parameter being fit in each instance. This approach prevents the fit from being dominated by the highest values of $f$ and ensuring the fit gives similar importance to data points across the full range.  

The optimisation was carried out using gradient-based methods implemented in \texttt{PyTorch} \citep{Pytorch}. Specifically, we treated the model parameters as differentiable variables and applied the limited-memory Broyden-Fletcher-Goldfarb-Shanno \citep[L-BFGS][]{LBFGS} optimiser, which is well-suited for problems with many correlated parameters. Given the number of free parameters (up to 10 coefficients, each described by equations extending in some cases up to a 4th order polynomial in $Z$ and a 2nd order polynomial in [$\alpha$/Fe]) fitting with a Markov-chain Monte Carlo approach was inappropriate owing to the vast volume of parameter space to be searched.

An iterative approach was taken, with initial fit optimisation done with the \citetalias{Hurley00} equations on the $Z=0.020, [\alpha/\mathrm{Fe}]=0.2$ models with terms added and removed to better match the functional form of the \bpass-derived parameters, until the overall linear and/or log-space errors were reduced to an acceptable level (typically less than 5 per cent) and the mass-dependence of the function matched the data through visual inspection. Dependence of the coefficients on $Z$ and [$\alpha$/Fe] was then determined through gradual addition of more grids of data, iteratively refining the fit parameters until an acceptable fit that worked globally across all compositions was obtained, and the addition of further, higher order terms made minimal difference to the goodness of fit. To account for  differences in the overall shape of curves, some of the fixed exponents in the \citetalias{Hurley00} equations were allowed to vary and, in some mass regimes, entirely new terms were added. In this work we replace the $a_n$ and $\beta_n$ terms used by \citetalias{Hurley00} with terms $b_n$, which are allowed to vary both in numerical value and in composition dependence. It is worth remembering that the choice of $\beta$ by \citetalias{Hurley00} does not always have a clear physical motivation, acting only as a descriptor of the slope of the relationship, so making adjustments to these values is not implying any fundamental change in the physics of stellar interiors or evolution.

As well as the overall metallicity dependence, some coefficients show a dependence on \afe. Examples of this would be the coefficients governing $t_\mathrm{MS}$ at both high and low masses, as seen in Fig.~\ref{fig:lifetimes}. To account for this, additional terms were added to the typical fourth-order polynomial adopted for most coefficients: a dependence on \afe, \afe$^2$, and in a couple of instances, the product of $Z$ and \afe\ were included to better capture the composition-dependence.

\begin{figure*}
    \centering
    \includegraphics[width=\linewidth]{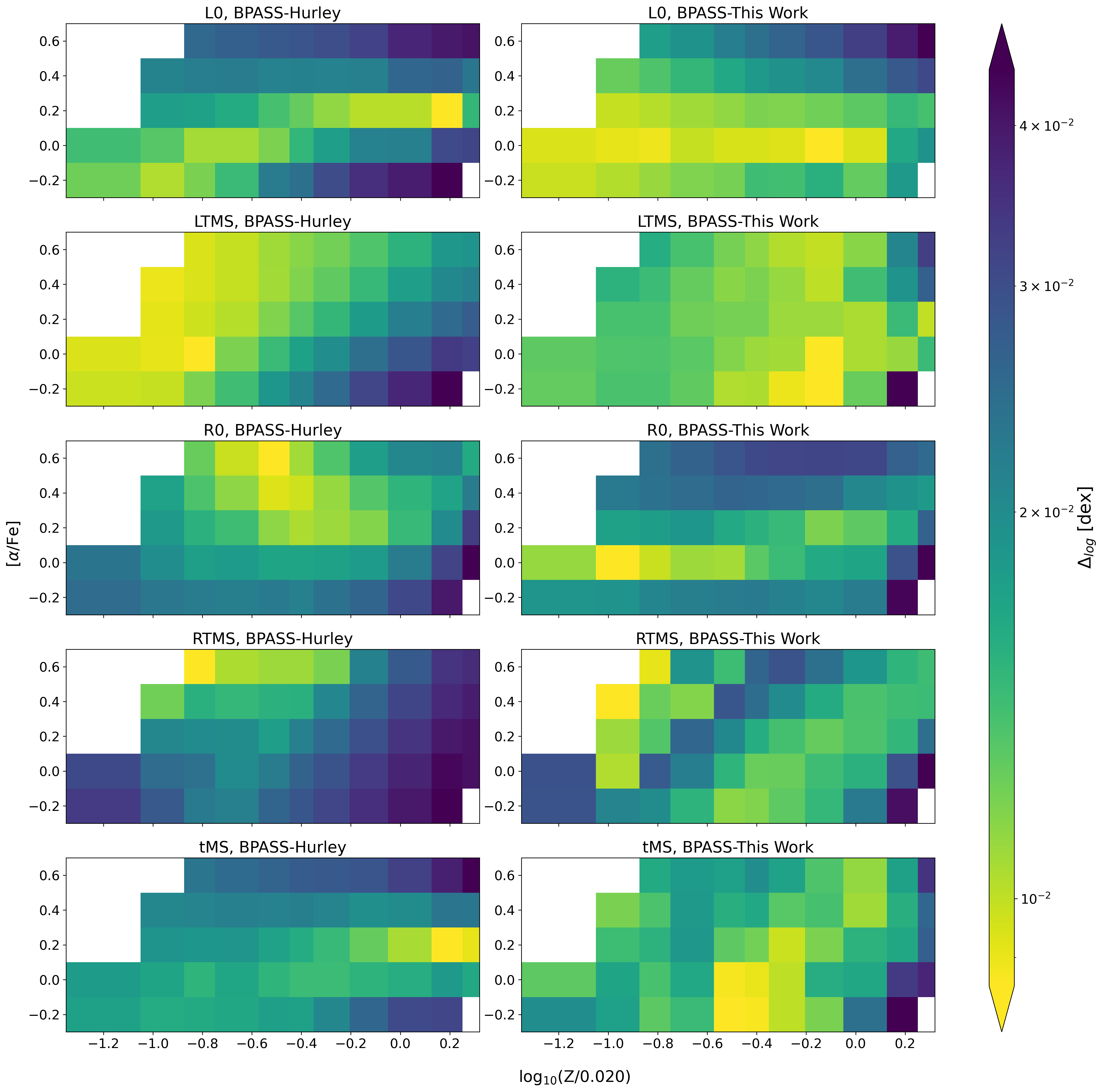}
    \caption{The RMS loss function for each of the fitted quantities in log-space at each value of $\log(Z)$ and \afe. $\Delta_\mathrm{log}$ is calculated as per Equation~\ref{eq:deltalog}. The left-hand panels compare the \citetalias{Hurley00} equations to \citetalias{Byrne25} model data, while the right-hand panels compare the analytic fits derived in this work to the \citetalias{Byrne25} models. There is no data in the upper left corner of the panels because these combinations of $Z$ and \afe\ yield $Z'<0.001$, and are thus not included in the fitting process. The empty data point at $Z=0.04$ and \afe$=-0.2$ is a composition not calculated in the \citetalias{Byrne25} as the iron content is higher than any of the available opacity tables.}
    \label{fig:goodfitlog}
\end{figure*}

Fig.~\ref{fig:goodfitlog} shows the goodness of fit (the value of the loss function) of each of the five stellar parameters considered as a function of $Z$ and [$\alpha$/Fe], comparing the original \citetalias{Hurley00} form of the equations to the revised fits derived here. In general, the revised fits represent a slight improvement ($R_0$) a moderate improvement ($R_\mathrm{t_{MS}}$) or a significant improvement over the \citetalias{Hurley00} equations when applied to the \bpass\ detailed models. 

Finding a global fit which works well over so many parameters ($M$, $Z'$, \afe) without over-fitting or including excessive free parameters is challenging, and some of the corners (e.g. high $Z'$, low \afe\ and vice versa) show larger errors in the fit as a result. {This emphasises the point that extrapolation of analytic equations in these contexts should be avoided.} In our fitting, we aimed to follow the general style and structure of equations as \citetalias{Hurley00} to best facilitate introduction into existing rapid population synthesis codes. 

All numerical results, and the values and composition dependence of the fitted parameters are tabulated in Appendix~\ref{app:append} and key results are presented in the following section.

\section{Fitting Formulae Results}
\label{sec:fitresults}

{The equations can be broadly separated into three categories: the zero-age properties (Section~\ref{sec:zams}), the main-sequence lifetime (Section~\ref{sec:tms}) and the terminal-age properties (Section~\ref{sec:tams}).}

\subsection{Zero-age main sequence properties}
\label{sec:zams}
The equations for the zero-age main sequence properties were originally formulated by \citet{Tout96}, but continued to form a part of the overall analytic description presented by \citetalias{Hurley00}. In the below section, references to \citet{Tout96} and \citetalias{Hurley00} are made interchangeably.

\subsubsection{ZAMS radii}

The location of the zero-age main-sequence (ZAMS) is an important anchor point for the analytic description of stellar evolution.

The ZAMS stellar radii were fit with an equation of the form 
\begin{equation}
R_0=
        \begin{cases}
        \dfrac{b_0 + b_1 M^{b_2}+b_3 M^{b_4}+b_5 M^{b_6}}{b_7 M^{b_8}+b_9 M^{b_{10}}} & M/M_\odot \le 3.4, \\ \\
        b_{11}+b_{12}M^{b_{13}}+b_{14}M^{b_{15}} & M/M_\odot \ge 3.5.
    \end{cases}
\label{eq:R0}
\end{equation}

The form of this equation is similar to that of \citet{Tout96}, but it has been split in two parts to better follow the kink in $R_0$ seen between 1-3\msun, and the exponents have been allowed to vary. {To ensure continuity, a simple interpolation can be applied to connect these two regimes. This is similar to the approach of \citetalias{Hurley00} for $R_\mathrm{TMS}$, where there is a similar gap of 0.1\,\msun\ between the two parts of the equation.} Coefficients $b_0$ to $b_{10}$ are either constant (the exponents) or vary linearly with \yg\ $=\log(Z'/0.02)$ (the scaling factors), with the exception of $b_3$, $b_5$ and $b_9$, which additionally depend on \yg$^2\alpha'$, $\alpha'$ and \yg$^2$ respectively, where $\alpha' = [\alpha/\mathrm{Fe}]$. Coefficients $b_{11}$ to $b_{16}$ variously include dependence on \yg$\alpha'$, \yg$^2$, \yg$^2\alpha'^2$, \yg$^3\alpha'$ and \yg$^3\alpha'^2$, as indicated in Table~\ref{tab:L0} in the Appendix. 

Fitting was done by minimising $\Delta_\%$, with a reduced weighting of 0.7 applied to masses $7\le M/$\msun$\le 40$. Comparing the results of this fit to the \citetalias{Byrne25} models, $\Delta_\%=5.8$ per cent, while in log-space, $\Delta_\mathrm{log}=0.027$ dex. This compares favourably to making the same comparison between the \citetalias{Hurley00} equations and the \citetalias{Byrne25} models, which yield $\Delta_\%=8.4$ per cent, and $\Delta_\mathrm{log}=0.041$ dex.

\begin{figure*}
    \includegraphics[width=0.95\textwidth]{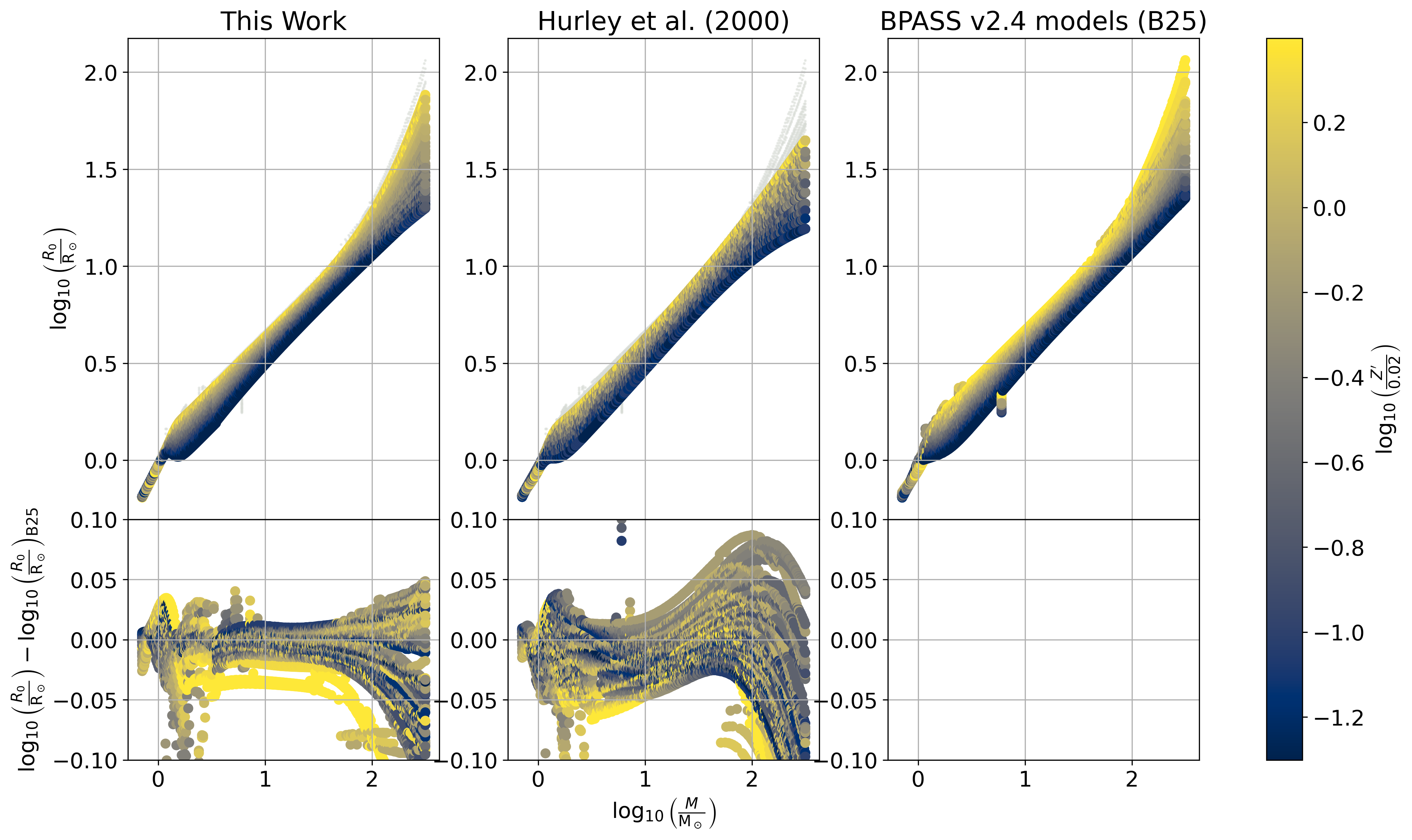}
    \caption{Dependence of Zero Age main sequence (ZAMS) stellar radii on initial mass and metallicity. The left-hand panel illustrates the derived analytic fits, the centre panel shows the analytic fits of \citetalias{Hurley00}, and the right-hand panel illustrates the actual values measured in the grid of detailed \bpass\ stellar evolution models. The data from the detailed \bpass\ models are also underplotted as light grey points in the centre and left-hand panels. The lower section of each panel shows the difference in \protect{$\log(R_0)$} between the {indicated analytic fitting formula and the detailed evolution models of \citetalias{Byrne25}}.}
    \label{fig:RZAMS}
\end{figure*}

These equations, the \citet{Tout96} ZAMS relations, and the ZAMS properties of the \bpass\ models are compared in Fig.~\ref{fig:RZAMS} ($R_0$) and Fig.~\ref{fig:LZAMS} ($L_0$). Models with metallicities between $Z'=0.001$ and $Z'=0.030$, with \afe=0.0 are plotted. The lower section of each panel showing the residual (in log space) shows the difference {between the analytic fits and the \citetalias{Byrne25} models}. In the left-hand and centre panels, the \bpass\ model data points are plotted in light grey. 

For $R_0$, it can be seen that both the existing and revised equations agree well with one another and the models, up to around 100 \msun, at which point the \citet{Tout96} relation departs to noticeably larger radii than seen in the \bpass\ v2.4.0 models. The new fits still slightly over-estimate the radii, but to a lesser extent than the original equations. In all instances, the RMS errors in $\log(Q)$ (where $Q$ is any of the quantities considered) are lower for the new equations than using the \citetalias{Hurley00} equations, so this still represents an overall improvement to the treatment of this parameter.

\subsubsection{ZAMS luminosities}
The ZAMS stellar luminosities were fit using the equation
\begin{equation}
    L_0 = \dfrac{b_{16} M^{5.5}+b_{17}M^{11}}{b_{18} + M^{3}+b_{19} M^{5}+b_{20} M^{7}+b_{21} M^{8}+b_{22} M^{9.5}}
\end{equation}

This equation has the same form as that of \citet{Tout96}, with only the coefficients ($b_{16}$ to $b_{22}$) and their composition dependence  being changed. The exponents in the equation are unchanged being kept the same. All of these coefficients are a fourth-order polynomial in \yg\ with additional terms for $\alpha'$ and $\alpha'^2$. 

\begin{figure*}
    \includegraphics[width=0.95\textwidth]{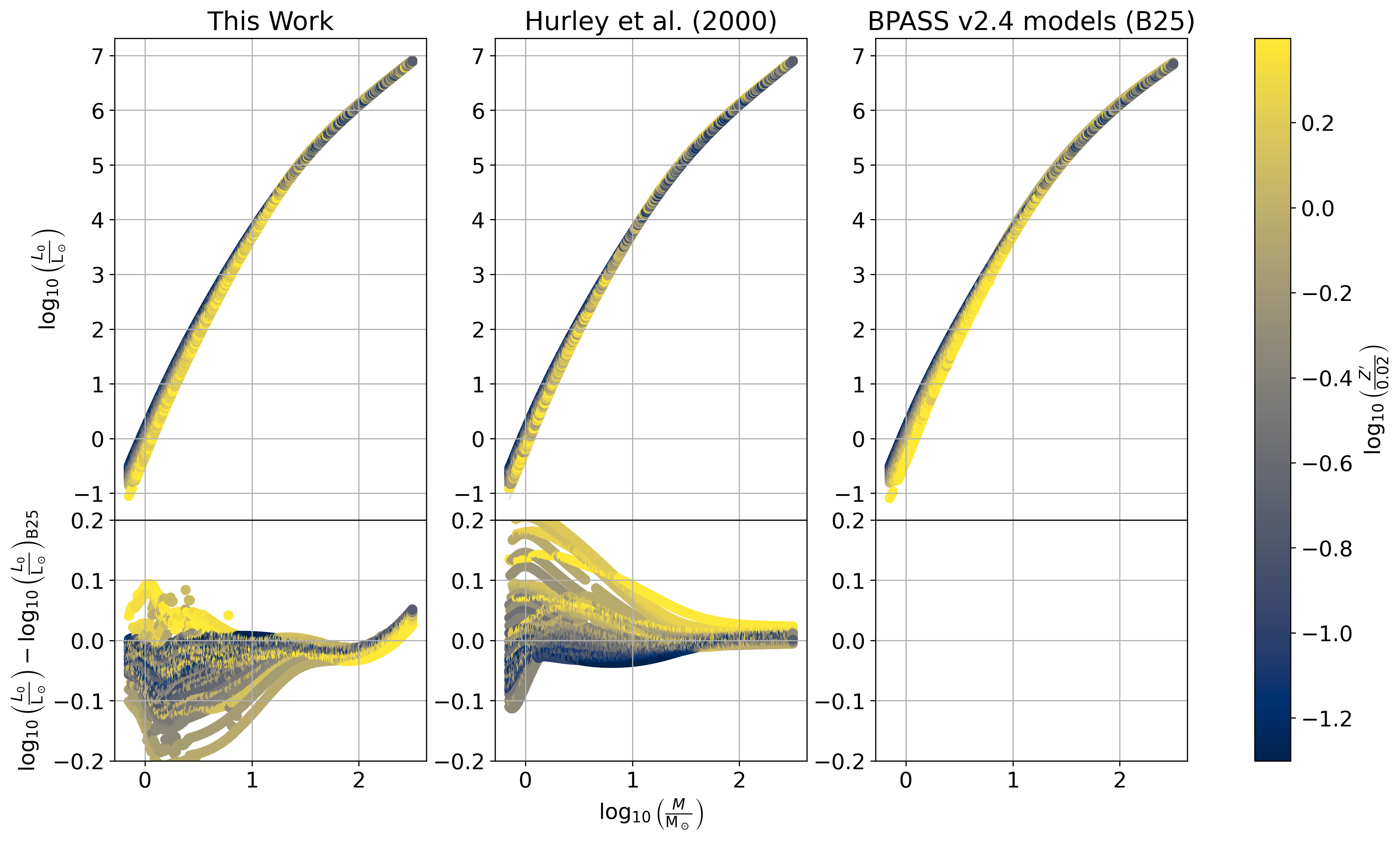}
    \caption{Comparison of zero-age main sequence luminosities, \protect{$\log(L_0)$}, as a function of mass and metallicity, with the panels following the same layout as Fig.~\protect\ref{fig:RZAMS}.}
    \label{fig:LZAMS}
\end{figure*}

In terms of fitting, the goodness of fit was evaluated based on minimising $\Delta_{\log}$. In this case, there are only minor differences between the new fits and those of \citetalias{Hurley00}, but the new equations do account for variation in \afe, which has a small effect. When all compositions are considered together, the overall RMS $\Delta_{\log}=0.039$ dex, while $\Delta_{\%}=8.12$ per cent. Once again, this compares favourably to the equivalent comparison with the \citet{Tout96} equation, which yields $\Delta_{\log}=0.039$ and $\Delta_{\%}=9.40$ per cent. From Fig.~\ref{fig:LZAMS}, it can be seen that the residuals for the models at \afe$=0.0$, are similar, albeit with slightly larger errors in the new fits for moderate masses at moderate metallicities and a slight under-prediction of luminosities for stars with masses above 100\msun.

\begin{figure*}
    \includegraphics[width=0.95\textwidth]{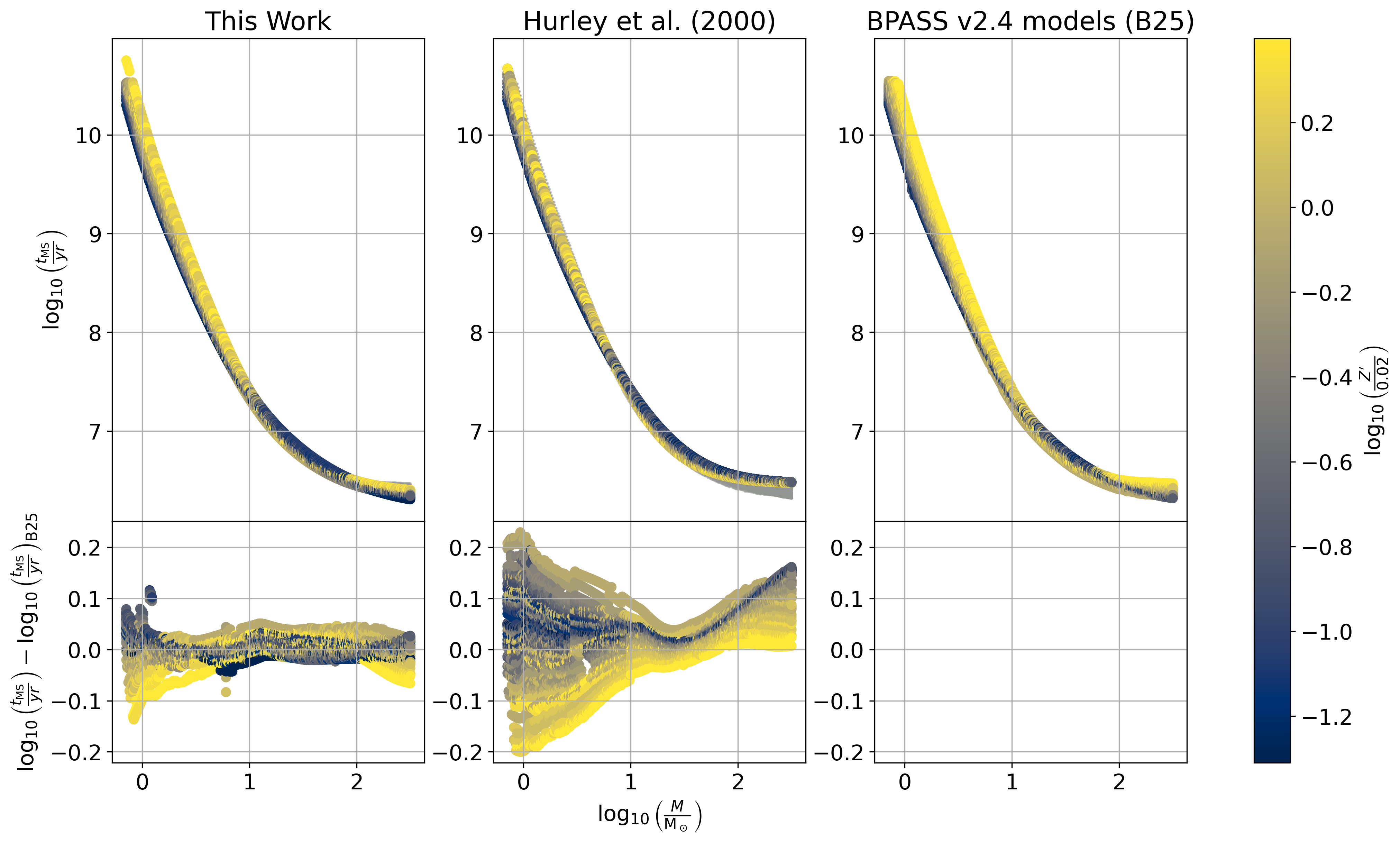}
    \caption{Comparison of main sequence lifetimes, \protect{$t_\mathrm{MS}$} as a function of mass and metallicity, with the panels following the same layout as Fig.~\protect\ref{fig:RZAMS}.}
    \label{fig:tMS}
\end{figure*}

\subsection{Main-sequence lifetimes}
\label{sec:tms}
The terminal-age main-sequence (TAMS) marks the end of a star's core-hydrogen burning phase. This represents roughly 90 per cent of the total lifetime of a given star, and thus the evolution between ZAMS and TAMS brackets the typical range of surface properties for stars of a given mass for the majority of the time they are contributing to the light of a stellar population. The time taken to reach this point is crucial, as the main sequence turn-off is an important metric for determining the age of stellar populations. \citetalias{Hurley00} derived an expression for this, which we amend to
\begin{equation}
    t_\mathrm{BGB}=\dfrac{b_{23}+b_{24} M^{4+b_{25}}+b_{26} M^{5.5+{b_{27}}}+b_{28}M^7}{a_{4} M^2+a_{5} M^7}
\end{equation}

\noindent This describes the time taken to reach the base of the giant branch (or the ignition of helium in the core for stars more massive than about $12\,$\msun). A small correction is then applied (Equations 5--7 of \citetalias{Hurley00}) to transform this to the main sequence lifetime, $t_\mathrm{MS}$. Leaving this correction unaltered makes the implicit assumption that the mass- and composition-dependence of the time taken to transition from the end of core-hydrogen burning to the base of the giant branch is unchanged between that determined by \citetalias{Hurley00} and that of the \citetalias{Byrne25} models. While this is unlikely to be strictly true, the differences are not likely to be significant relative to the overall {main sequence} lifetime of a given star. 

{The fitting was carried out by directly comparing the value of $t_\mathrm{MS}$ found in the models to the outcome of using Eqn. 5 above and modifying it using Equations 5, 6 and 7 of \citetalias{Hurley00} to arrive at a resulting analytic value for $t_\mathrm{MS}$, providing a direct fit, rather than a two-step fit of $t_\mathrm{BGB}$ and the modification needed to reach $t_\mathrm{MS}$. This was done as the primary motivation of this work was to calibrate the evolution on the main sequence, and defining the end of the main sequence and revising the procedure to relate $t_\mathrm{BGB}$ to $t_\mathrm{MS}$ akin to that used by \citetalias{Hurley00} while accounting for the extended mass range and $\alpha$-enhancement axis would require a number of additional free parameters which made optimisation challenging, with little to no gain in the best fit performance, and would require more extensive modification to be compatible with population synthesis software which already includes the equations of \citetalias{Hurley00}. Expanding the fitting procedures to cover the entirety of the post-main sequence evolution is beyond the scope of this work.}

The modified form of this equation is very similar to that of \citetalias{Hurley00}, with the addition of coefficients ($b_{25}$ and $b_{27}$ here) which allow the exponents of the middle two terms in the numerator to vary with composition. The terms $b_{23},\,\,b_{24},\,\,b_{25}$ and $b_{27}$ have additional dependence on $\alpha'$ and $\alpha'^2$ to account for the \afe-sensitivity of age at both the high-mass and low-mass ends of the parameter range. On the other hand, the denominator is entirely unchanged, using the same form, metallicity dependence and the $a_4$ and $a_5$ coefficients as defined by \citetalias{Hurley00}. The fit was optimised by minimising $\Delta_\mathrm{log}$, with additional weighting added to high-mass models. All models with a mass less than 10\msun were given a weighting of 1, with a linear ramp increasing the weights to 2 for models with the maximum stellar mass (316 \msun), to ensure that the function accurately reproduces the shape and size of the plateau in lifetimes seen at the high-mass end.

Fig.~\ref{fig:tMS} compares these analytic expressions as a function of mass to the \bpass\ v2.4.0 models for different compositions. 

From this, it can be seen that there are effectively two regimes where the \bpass\ models deviate from the predictions of \citetalias{Hurley00};
\begin{enumerate}
    \item At low masses ($M\lesssim50\,\mathrm{M}_\odot$), lifetimes deviate noticeably when \afe\ is increased or decreased 
    \item At high masses ($M\gtrsim50\,\mathrm{M}_\odot$), the \citetalias{Hurley00} equations consistently over-estimate the lifetimes for all compositions, with the shortest lifetimes seen in metal poor and/or $\alpha$-rich compositions.
\end{enumerate}

With the new equations, the errors remain low ($\lesssim0.1$ dex), and no longer show a large systematic departure at the high mass end. Regarding the overall errors of the fit, $\Delta_{\log}=0.018$ dex and $\Delta_{\%}=4.10$ per cent. This marks a considerable improvement from applying the \citetalias{Hurley00} equation, which yields $\Delta_{\log}=0.062$ dex and $\Delta_{\%}=15.42$ per cent, largely impacted by the overestimation of massive star lifetimes.

\subsection{Terminal Age main sequence properties}
\label{sec:tams}
{To define the surface properties of the star at the end of the main sequence, an analytic description of both the radius and luminoisty at the point of core hydrogen exhaustion is required.}

\subsubsection{TAMS radii}

As discussed above, the end of the main sequence can be difficult to precisely define. As stars evolve rapidly at the point of core hydrogen depletion, slight differences in the chosen age can lead to drastically different radii and luminosities. After experimentation, it was deemed that using the criterion that the central hydrogen mass fraction drops below 10$^{-12}$ is sufficient to identify core hydrogen exhaustion, and typically leads to stellar radii and luminosities which match well with those seen at time $t=t_\mathrm{MS}$. For stars which show a hook at the end of the main sequence, the turning point in the effective-temperature-luminosity plane is used to determine the terminal-age radius. This turning point matches closely to the top of the hook, where the main sequence of the \citetalias{Hurley00} tracks terminate. Applying these criteria to identify $L_\mathrm{t_{MS}}$ and $R_\mathrm{t_{MS}}$ provides a consistent metric for extracting the appropriate model time step, and provides a defined end-point for this phase of evolution which is independent of the conversion from $t_\mathrm{BGB}$ to $t_\mathrm{MS}$ used by \citetalias{Hurley00}.

\begin{figure*}
    \includegraphics[width=0.95\textwidth]{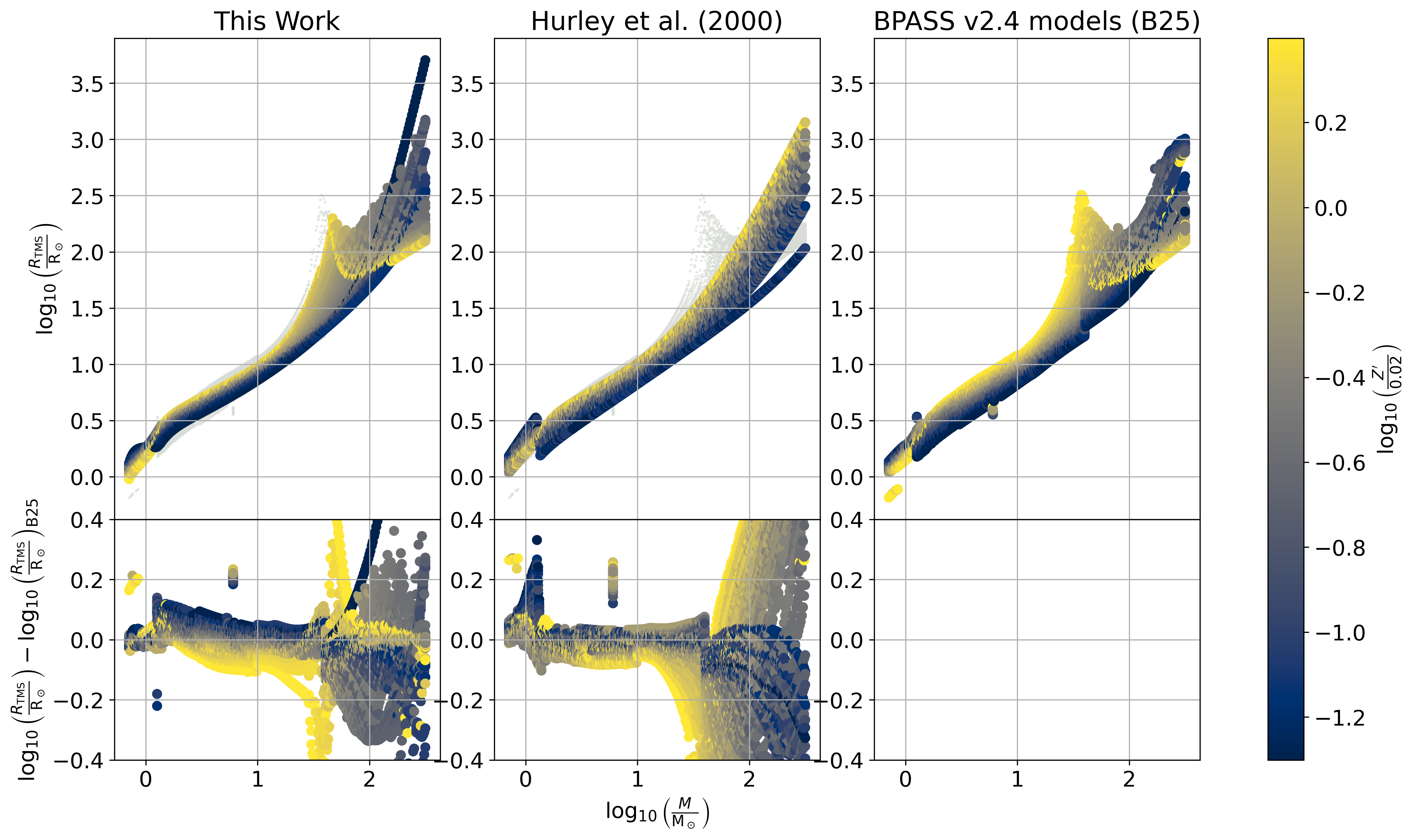}
    \caption{Comparison of Terminal Age main sequence (TAMS) radii, \protect{$\log(R_\mathrm{t_{MS}})$}, as a function of mass and metallicity, with the panels following the same layout as Fig.~\protect\ref{fig:RZAMS}.}
    \label{fig:RTMS}
\end{figure*}

As with the ZAMS properties, fitting the radius and luminosity at the end of the main sequence began from the analytic functions of \citetalias{Hurley00}, with the addition of extra terms and changes to composition dependence as deemed appropriate. The analytic form of the equation is
\begin{equation}
\label{eq:RTMS}
\mathrm{R}_\mathrm{t_{MS}}=
        \begin{cases}
        \dfrac{b_{45}+b_{46}\mathrm{M}^{b_{47}}}{b_{48}+\mathrm{M}^{b_{49}}} & \dfrac{M}{M_\odot} \le a_{17} \\ \\
        \dfrac{b_{50}\mathrm{M}^{b_{51}}+b_{52}\mathrm{M}^{b_{53}}+b_{54}\mathrm{M}^{b_{55}}}{b_{56}+b_{57}\mathrm{M}^{b_{58}}} & a_{17}+0.1 \le \dfrac{M}{M_\odot} \le 0.9M_\mathrm{WR} \\ \\
        10^{b_{59}}M^{b_{60}}+b_{61}M^{b_{62}} & \dfrac{M}{M_\odot} \ge 1.1\mathrm{M}_\mathrm{WR}\\
    \end{cases}
\end{equation}
\noindent and the results of the fitting, for compositions with \afe$=0.0$, shown in Fig.~\ref{fig:RTMS}. For $R_\mathrm{t_{MS}}$, the equation is defined in three parts, an increase from the two-part description in \citetalias{Hurley00}. The first term in Equation~\ref{eq:RTMS} has the same form as the first part from \citetalias{Hurley00}, while the second has an additional term ($b_{54}M^{b_{55}}$) in the numerator. The exponents have been given flexibility to vary, and the mass-dependent term in the denominator is given a scaling factor $b_{57}$. The mass at which the equation transitions from one to another remains the same as in \citetalias{Hurley00}, using $a_{17}$ as originally defined. {As with Equation~\ref{eq:R0}, linear interpolation can be used to ensure continuity across the different mass ranges.}

The third part of Equation~\ref{eq:RTMS} relates to the most massive stars, which undergo envelope stripping before core hydrogen exhaustion. This part of the equation ($10^{b_{59}}M^{b_{60}}+b_{61}M^{b_{62}}$), does not define the true end of the main sequence, more accurately it refers to the point of maximum radius during the main sequence evolution (e.g. the point with lowest effective temperature, labelled in the evolution of the 100 \msun\ model in Fig.~\ref{fig:HRD1-100}). The mass at which local maximum in radius occurs is (iron-scaled) metallicity-dependent, because it is a consequence of massive star mass-loss rates which are scaled according to iron abundance. This mass can be well-described by

\begin{equation}
    M_\mathrm{WR,R} = 2164.9(\text{\yg}+2)^{-7}+64.947(\text{\yg}+2)^{-0.8},
\end{equation}

\noindent where \yg\ continues to mean $\log(Z'/0.02)$. 

In terms of composition dependence of the coefficients, most are simply quadratic or cubic in \yg, which is similar to \citetalias{Hurley00}. The exceptions are $b_{50}$, $b_{51}$, $b_{52}$, $b_{53}$, $b_{54}$, $b_{56}$, $b_{59}$ and $b_{60}$, which show mild dependence on $\alpha'$ or $\alpha'^2$, and $b_{55}$, which depends on \yg$^4$. This three-part function for $R_\mathrm{t_{MS}}$ describes the model radii reasonably well, with $\Delta_\%=17.8$ per cent, while $\Delta_\mathrm{log}=0.09$ dex. This compares well when we consider that fitting the \citetalias{Hurley00} equation to the \bpass\ models gives $\Delta_\%=127.3$ per cent and $\Delta_\mathrm{log}=0.25$ dex. As can be seen in Fig.~\ref{fig:RTMS}, the \citetalias{Hurley00} equation underestimates the final radii for most intermediate masses, while it considerably overestimates the radii of the most metal-rich massive stars, which do not achieve such large radii because wind stripping takes effect before they can expand so far. It should of course be noted that a different definition for the end of the main sequence was used. Stars expand considerably at the end of core-hydrogen burning. Nonetheless, comparing the radii of the \bpass\ models at core-hydrogen exhaustion to the radii at time $t_\textrm{MS}$ deduced from following the same methods as \citetalias{Hurley00} produces only minimal differences in the \bpass\ models and does not explain the offsets between the \citetalias{Hurley00} equations and the equation used here.

\begin{figure*}
    \includegraphics[width=0.95\textwidth]{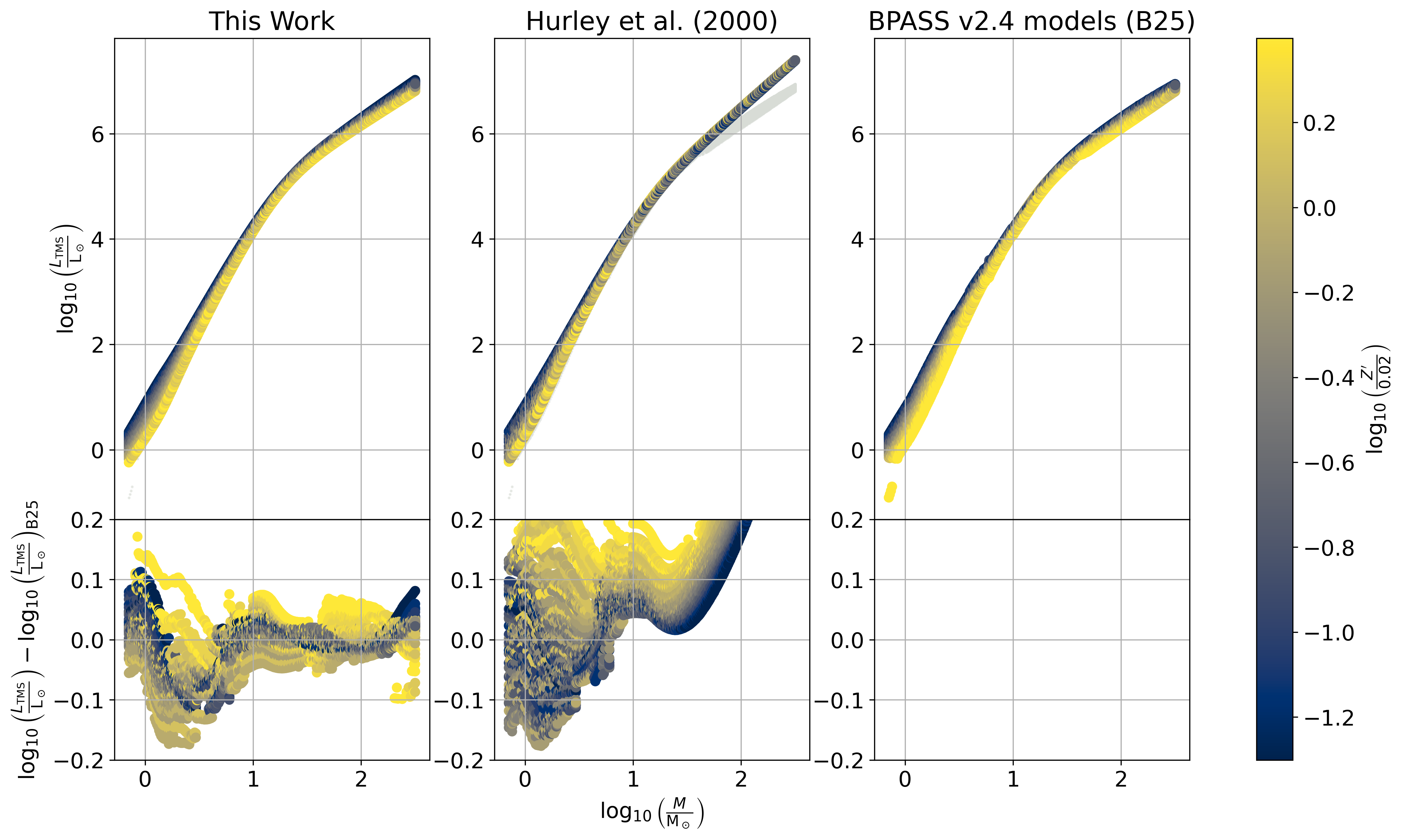}
    \caption{Comparison of TAMS luminosities, \protect{$\log(L_\mathrm{t_{MS}})$}, as a function of mass and metallicity, with the panels following the same layout as Fig.~\protect\ref{fig:RZAMS}.}
    \label{fig:LTMS}
\end{figure*}

\subsubsection{TAMS Luminosities}

For the terminal age luminosities, the equation used is very similar to \citetalias{Hurley00}. 

\begin{equation}
\label{eq:LTMS}
L_\mathrm{t_{MS}}= \dfrac{b_{29}M^3+b_{30}M^4+b_{31}M^{b_{32}+b_{33}}}{b_{34}+b_{35}M^5+b_{36}M^{b_{32}}+b_{37}M^{b_{38}}}
\end{equation}

The only differences are that the exponent $b_{33}$ is allowed to vary, (in contrast to the fixed exponent of 1.8 in equation 8 of \citetalias{Hurley00}), and an additional term ($b_{37}M^{b_{38}}$) was added to the denominator to provide additional curvature to the function at high masses. Similar to $R_\mathrm{t_{MS}}$, most coefficients are cubic in \yg, while a few need dependence on \yg$^4$. The coefficient $b_{32}$ additionally has dependence on the product \yg$\alpha'$. Once again, optimisation was done by minimising $\Delta_\mathrm{log}$.

Similar to $R_\mathrm{t_{MS}}$, for the high-mass stars, there is a change in final luminosity as a result of envelope stripping. This does not appear in this equation because the luminosities of the stripped stars at maximum radius are a smooth continuation of this function. Nonetheless, for the additional steps needed to follow the stripping process through to core hydrogen exhaustion, the definition of the mass at which the change in luminosity behaviour is seen can be stated here, as 

\begin{equation}
    M_\mathrm{WR,L} = 694.252(\text{\yg}+2)^{-3}.
\end{equation}

\noindent this is subtly different from the mass threshold for a change in radius evolution defined in the previous section, with $M_\mathrm{WR,L}$ being around 70 per cent of $M_\mathrm{WR,R}$ for high and moderate metallicities, increasing to parity or slightly exceeding it at low metallicities. 

Overall, the fitting of $L_\mathrm{t_{MS}}$, again optimised using $\Delta_\mathrm{log}$, provides a fit with $\Delta_\%=10.4$ per cent, while $\Delta_\mathrm{log}=0.04$ dex. This is a marked improvement on \citetalias{Hurley00}, where $\Delta_\%=73.8$ per cent and $\Delta_\mathrm{log}=0.20$ dex. From Fig.~\ref{fig:LTMS}, it can be seen that most of the error comes from the overestimated luminosities above about 50\,\msun.

\begin{table}
\centering
\caption{Comparison of $\Delta_{\%}$ and $\Delta_{\log}$ for each of the properties considered, showing the difference between the \citet{Byrne25} models and the analytic equations of \citet{Hurley00} (using the iron-scaled metallicity equivalent $Z'$ and the true metallicity mass fraction $Z$) and this paper respectively.}
\begin{tabular}{lrrrrrr}
\hline
Property & \multicolumn{2}{c}{\citetalias{Byrne25}-\citetalias{Hurley00}($Z'$)}  & \multicolumn{2}{c}{\citetalias{Byrne25}-\citetalias{Hurley00}($Z$)} & \multicolumn{2}{c}{\citetalias{Byrne25}-This Work} \\
 & $\Delta_{\%}$ & $\Delta_{\log}$ & $\Delta_{\%}$ & $\Delta_{\log}$ & $\Delta_{\%}$ & $\Delta_{\log}$ \\
\hline
$L_0$            & 9.4006 & 0.0387 & 12.219 & 0.0516 & 8.1162 & 0.0389 \\
$R_0$            & 8.3543 & 0.0410 & 9.2368 & 0.0449 & 5.7963 & 0.0268 \\
$t_\mathrm{MS}$  & 15.420 & 0.0614 & 17.103 & 0.0678 & 4.0963 & 0.0179 \\
$R_\mathrm{t_{MS}}$ & 128.44 & 0.2517 & 127.31 & 0.2500 & 20.504 & 0.1026 \\
$L_\mathrm{t_{MS}}$ & 73.775 & 0.1948 & 74.289 & 0.1982 & 10.424 & 0.0429 \\
$R_\mathrm{WR}$  & ------ & ------ & ------ & ------ & 8.5469 & 0.0451 \\
$L_\mathrm{WR}$  & ------ & ------ & ------ & ------ & 5.9166 & 0.0271 \\
\hline
\end{tabular}
\end{table}

\subsection{Additional step for massive stripped stars}

As massive stars shed their outer envelopes through strong winds, there arises a (metallicity-dependent) critical mass whereby the star strips its hydrogen rich envelope before reaching the end of the main sequence/core-hydrogen exhaustion. The consequence of this is that the star initially expands and becomes redder, before contracting and evolving blue-wards as the hydrogen is stripped. An example evolutionary track for a 100\msun\ star was shown above in Fig.~\ref{fig:HRD1-100}.

In order to capture this behaviour in the analytic equations, a two-step approach is taken to the main sequence evolution. In the first step, the star evolves from (R$_0$, L$_0$) to (R$_{\mathrm{Rmax}}$,L$_{\mathrm{Rmax}}$), where R$_{\mathrm{Rmax}}$ and L$_{\mathrm{Rmax}}$ are R and L derived from Equations~\ref{eq:RTMS} and~\ref{eq:LTMS}, respectively, the same method as used for lower mass stars. This point in the evolution occurs prior to reaching the end of core-hydrogen exhaustion. The time taken for a star to reach this point can be described by $t_\mathrm{Rmax}=t_\mathrm{MS}-\Delta t_\mathrm{strip}$, where

\begin{equation}
    \Delta t_\mathrm{strip} = \max(0,b_{67}+M b_{68}).
\end{equation}

\noindent with $b_{67}$ and $b_{68}$ being polynomials with terms up to \yg$^3$ and $\alpha'$. Stars with a mass above the critical thresholds should then use $t_\mathrm{R_{max}}$ as the timescale to evolve from ZAMS to the point of maximum radius. After this, an interpolation is applied so that between $t_\mathrm{R_{max}}$ and $t_\mathrm{MS}$, the radius and the luminosity of the massive star evolves from 
(R$_{\mathrm{Rmax}}$,L$_{\mathrm{Rmax}}$) to new values that describe the loci of such stars on the HRD at the point of core-hydrogen exhaustion ($R_\mathrm{WR}$, $L_\mathrm{WR}$). These quantities can be well described by 
\begin{equation}
    R_\mathrm{WR} = 10^{b_{63}}M^{b_{64}}+b_{65}M^{b_{66}}
\end{equation}
\noindent for the radius and 
\begin{equation}
    L_\mathrm{WR} = {b_{39}}M^{b_{40}}+b_{41}M^{b_{42}}+b_{43}M^{b_{44}}
\end{equation}

\noindent for the luminosity. When fitting $R_\mathrm{WR}$ and $L_\mathrm{WR}$ to all data points with initial masses above the relevant thresholds, optimising for $\Delta_\mathrm{log}$, we get total errors of $\Delta_\mathrm{log}=0.0451$ dex and $0.0271$ dex respectively and $\Delta_\%=8.55$ per cent and $5.92$ per cent respectively. Applying these additional steps, the new equations produce the orange dashed line in Fig.~\ref{fig:HRD1-100} for the evolution of an 100\msun\ star at Solar metallicity, which is a reasonable approximation of the detailed \bpass\ model shown with the green solid line.

Accurately capturing the expansion to maximum radius, rather than simply evolving from the ZAMS to TAMS location is important in the context of binary evolution. Without doing so, the analytic evolution would never approach the maximum radius for these stars. This would impact the predictions of whether such stars would fill their Roche lobes and undergo interactions with their companion. When considering composite spectra of stellar populations, omitting this step could also introduce a small bias towards bluer temperatures for populations dominated by massive stars.

\subsection{Overall performance of the equations}

\begin{figure}
    \centering
    \includegraphics[width=0.481\textwidth]{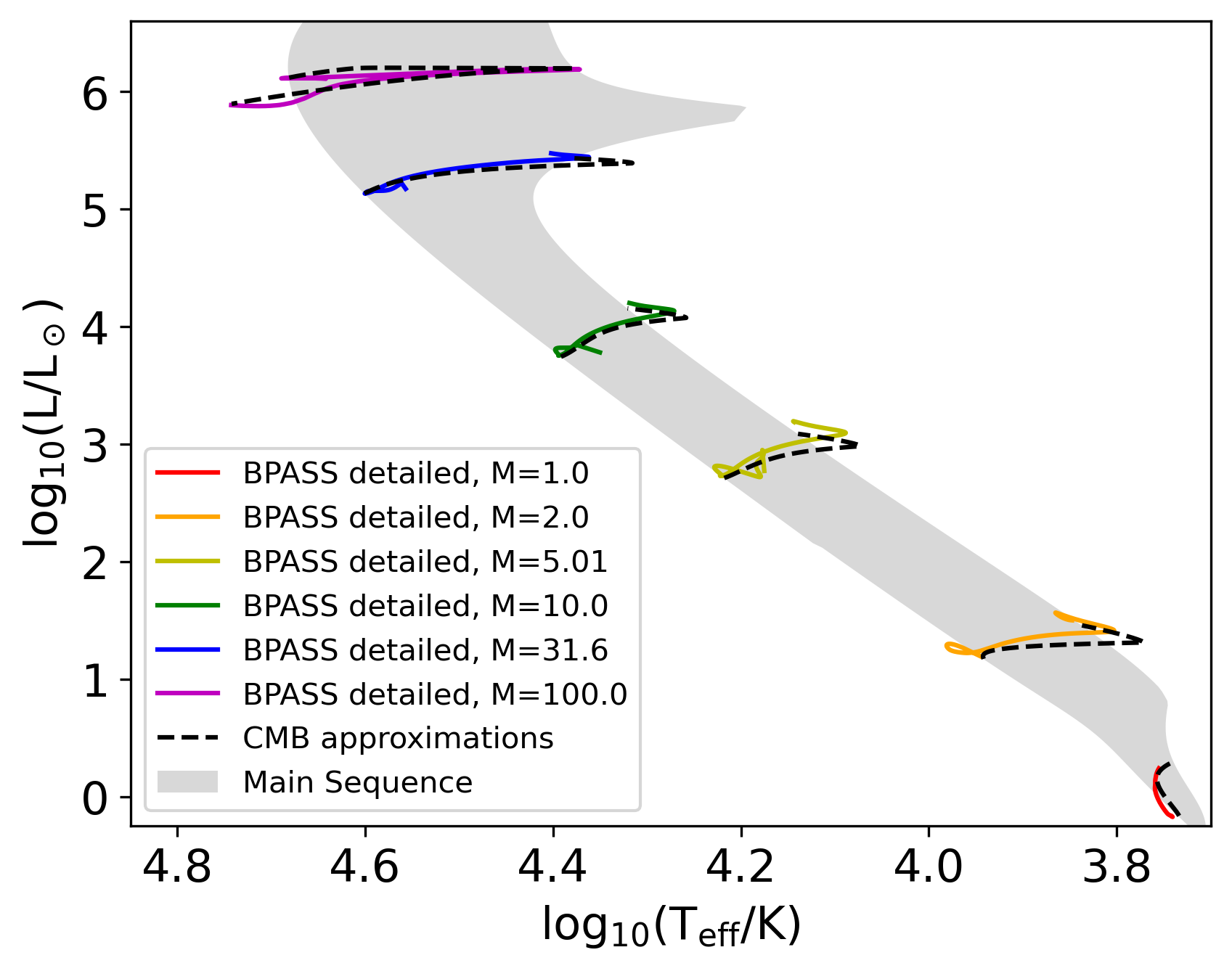}
    \includegraphics[width=0.481\textwidth]{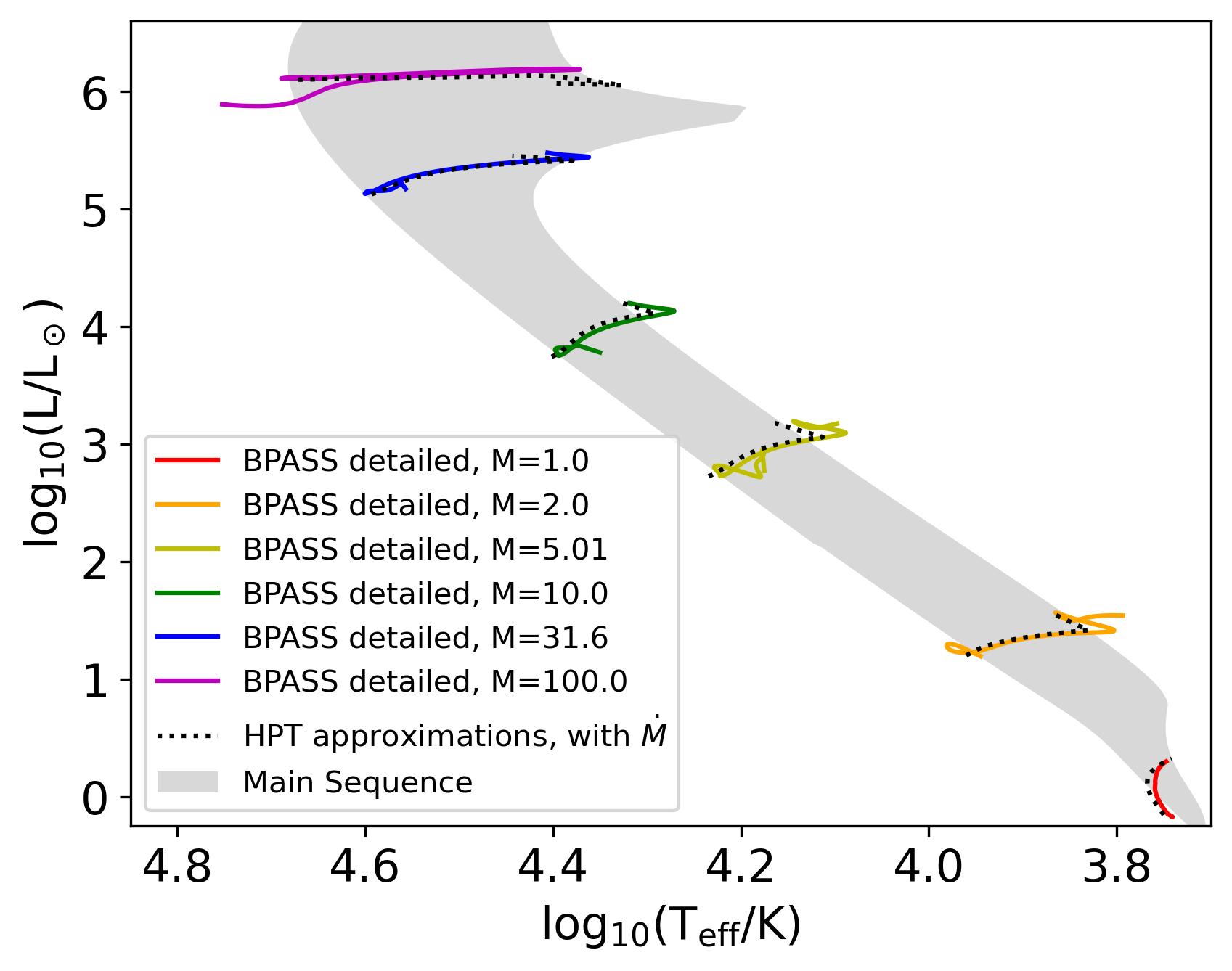}
    \caption{Luminosity-Temperature diagrams illustrating the main sequence evolution of stars of selected masses, comparing detailed evolution calculations from \bpass\ to the analytic prescriptions derived in this work (black, dashed lines, upper panel) and the analytic prescriptions of \citetalias{Hurley00} (black, dotted lines, lower panel). These comparisons are being made at a composition of $Z=0.020$, \afe=$+0.2$.The region between the ZAMS and TAMS (or point of maximum radius, for stars which move bluewards before core hydrogen exhaustion) is shaded in grey.}
    \label{fig:multiHRD}
\end{figure}

\begin{figure*}
    \centering
    \includegraphics[width=0.95\linewidth]{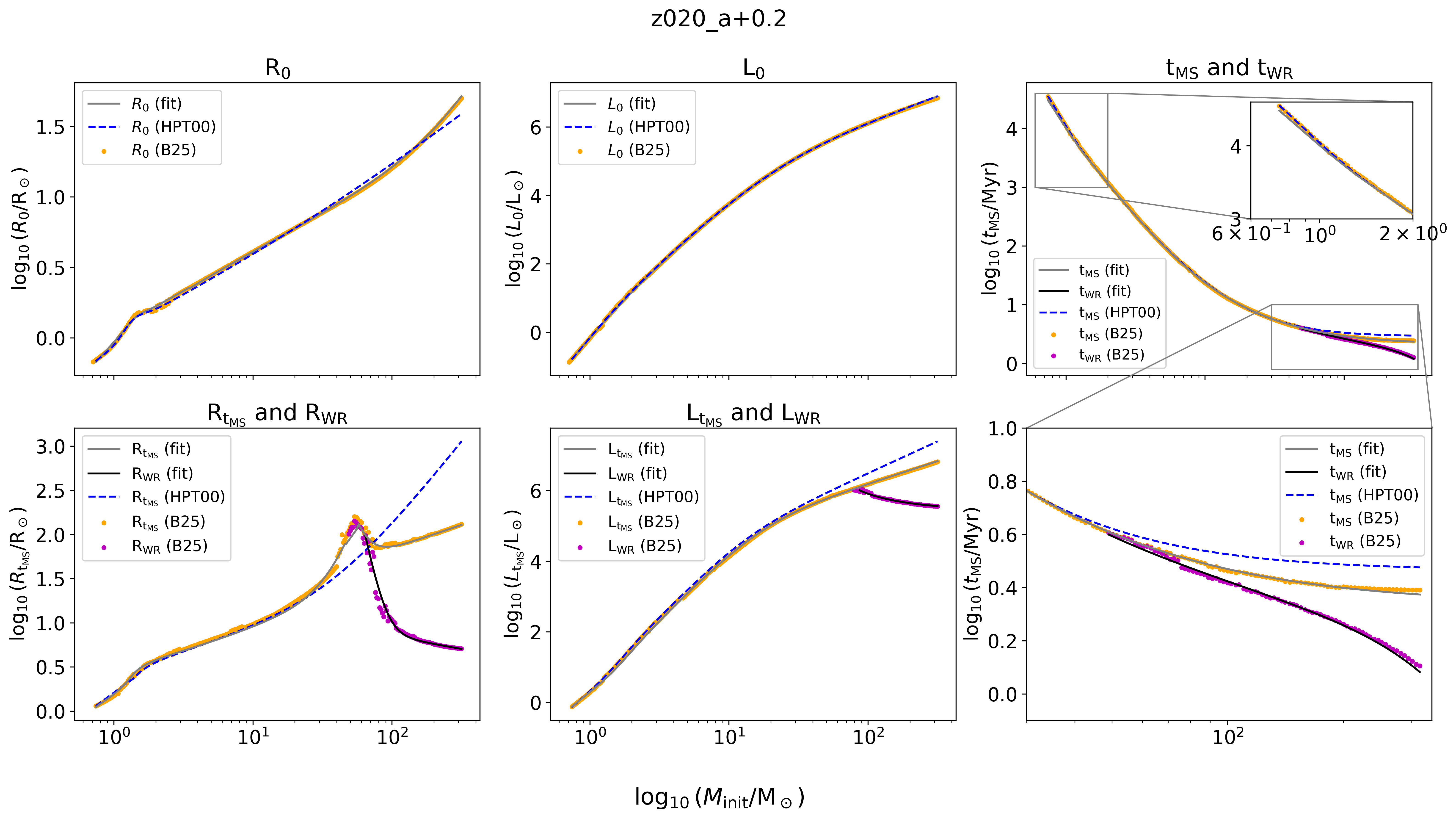}
    \caption{Mass evolution for each of the five parameters being fit in this work, shown at a single composition ($Z=0.020$, \afe$=+0.2$). From left to right, top to bottom these are: i) zero-age main-sequence radii, ii) zero-age main-sequence luminosities, iii) main-sequence evolutionary timescales, iv) Radii at the Terminal Age main sequence (and at the point of maximum radius for massive stars), v) Luminosities at the terminal-age mai sequence (and at the point of maximum radius for massive stars), and vi) an expanded view of the main sequence lifetimes of stars with initial masses between 30 and 300\msun. The blue dashed lines show the analytic results from \citetalias{Hurley00}, while the black and grey solid lines show the fits introduced in this paper. The \bpass\ v2.4.0 model data from \citetalias{Byrne25} are shown as magenta and orange points, with the orange points indicating the properties at the time of core-hydrogen exhaustion for non-stripped stars, and for the point of maximum radius for stars which become stripped, while the purple points indicate the properties of the stripped stars at the point of core-hydrogen exhaustion.}
    \label{fig:fits_z020a02}
\end{figure*}

Figs~\ref{fig:multiHRD} and~\ref{fig:fits_z020a02} show a comparison between the \citetalias{Hurley00} equations, the equations presented here, and the detailed \bpass\ v2.4.0 models of \citetalias{Byrne25} with a composition of $Z=0.020,\,\,[\alpha/\mathrm{Fe}]=+0.2$, showing only the evolution from the zero-age main sequence to the exhaustion of core hydrogen. Fig.bb~\ref{fig:multiHRD} considers the evolution now predicted in the HR diagram for key stellar masses. The grey shaded region indicates the full range of models calculated by \cmb25, while coloured tracks are compared to the \hpt\ equations in the lower panel, and our new analytic equations in the upper panel. The performance of the new analytic formalism in reproducing the detailed stellar evolution model is comparable to the \hpt\ equations for most masses. At masses of around 2-5$\,$\msun\, the \hpt\ models produce a slightly better approximation. However at $M>50$\,\msun\ the new formalism provides a substantially better match to the detailed model{, accurately showing the departure toward higher effective temperatures as a result of envelope stripping}.

In Fig.~\ref{fig:fits_z020a02} we show the fit to individual parameters across the full mass range at the same composition. Noticeable improvements can be seen for masses greater than $10\,$\msun\ in all parameters, with particularly striking improvements in the TAMS parameters. It is interesting to note that the R$_\mathrm{t_{MS}}$ parameter is also better fitted by our new analytic prescription at around 2\,\msun, despite the less impressive performance in the H-R diagram. This is likely attributable to the difference in definition of the end of the main sequence lifetime, and the very rapid transition of these stars through a significant temperature change en route to the base of the giant branch.

We note that these figures only present one composition, while the fitting algorithm optimises across the full range in total metallicity and alpha-enhancement. Additional figures showing the same information for a selection of other compositions can be found in Appendix~\ref{app:fitfigures}

\section{Discussion}
\label{sec:discussion}
\subsection{Composition dependence in rapid population synthesis}

As shown both here and in \citetalias{Byrne25}, the stellar evolution is primarily sensitive to the iron abundance, rather than the total metallicity. This is not a surprise. \citet{Vink01} demonstrated that CNO-cycle products such $\alpha$-elements are unlikely to dominate the opacity in the stellar envelope until metallicities as low as $Z=7\times10^{-4}$. They nonetheless contribute to the wind mass loss at higher metallicities. 

The majority of the parameters in the analytic formalism presented here are primarily dependent on \yg, with \afe-dependent terms typically an order of magnitude or more weaker in their scaling factors. For the ZAMS radius, only the most massive stars are \afe-dependent, while the ZAMS luminosity shows a more general influence of \afe\ on the fit. This arises due to the effects of \afe\ on the opacity of the stellar envelope, and hence on the surface temperature.  

\begin{figure*}
    \centering
    \includegraphics[width=\linewidth]{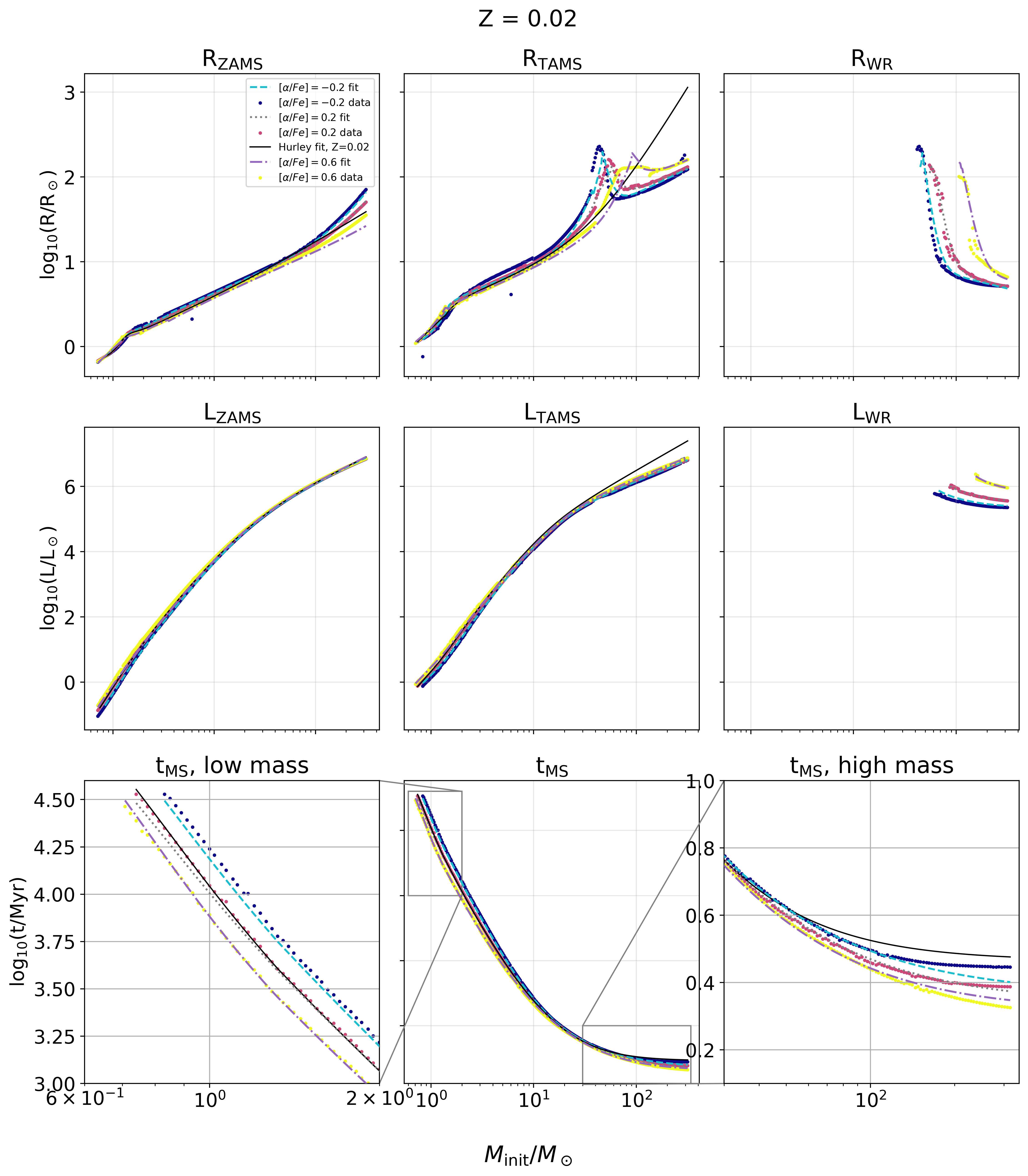}
    \caption{Fitting of radius, luminosity and main sequence lifetime as a function of mass and $\alpha$-enhancement. BPASS v2.4 model data is shown by dark blue (\afe=-0.2), magenta (\afe=0.2) and yellow (\afe=0.6) points, while the corresponding best fitting analytic expressions are shown by the cyan dashed, grey dotted and purple dot-dashed lines respectively. Where relevant, the  values derived using the \protect\citetalias{Hurley00} equations are shown by the black dash-double dot line for comparison. Upper panels: $R_0$ (left), $R_\mathrm{t_{MS}}$ (centre) and $R_\mathrm{WR}$, the main sequence radius post-stripping for massive stars (right). Middle panels: $L_0$ (left), $L_\mathrm{t_{MS}}$ (centre) and $L_\mathrm{WR}$, the main sequence luminosity post-stripping for massive stars (right). Lower panels: main sequence lifetimes for stars <=2\msun\ (left), the entire mass range (centre) and for stars >=30\msun\ (right).}
    \label{fig:alphafit}
\end{figure*}

Fig.~\ref{fig:alphafit} shows the performance of the analytic equations (cyan dashed, grey dotted and purple dot-dashed lines) compared to the data at $Z=0.020$ for three different values of \afe (blue, magenta and yellow points), with the nominal best \citetalias{Hurley00} data shown by the black solid line. Note once again that the equations take the iron-scaled effective metallicity $Z'$, so the \citetalias{Hurley00} line on these plots only serves as a good comparison to the \afe=0.2 data and fits. For $R_\mathrm{0}$, the \afe=0.6 fit slightly under-predicts the radius at high mass. Similarly, the shape and location of the peak value of $R_\mathrm{t_{MS}}$ in the $\alpha$-rich case differs slightly from the data, but captures the general behaviour much better than the functional form of the \citetalias{Hurley00} track.

For the massive stars, post-stripping, the fits to $R_\mathrm{WR}$ perform well at all \afe. Luminosity is largely insensitive to \afe, with the exception of $L_\mathrm{WR}$, where the fit performs well, with a slight over-prediction of luminosities for the $\alpha$-poor (\afe=-0.2) composition. For main sequence lifetimes, the $\alpha$-dependence in the parameters yield a good fit throughout. At low-mass, where there is noticeable $\alpha$-dependence, the fits match the data. At the highest masses, the lifetimes seem to be more strongly $\alpha$-dependent than the fits, but in all cases the predicted main-sequence lifetimes for the stars with masses greater than about $\,$50\msun, are closer to the data points than would be obtained by applying the \citetalias{Hurley00} equations.

It must be borne in mind that this figure represents just 3 of the 56 distinct compositions used to determine the global best fit and that, at other compositions the fits do better and worse at different masses in the various properties (see e.g. the figures in Appendix~\ref{app:fitfigures} for examples of other comparisons between the models and the data).

The main sequence lifetime shows \afe\ dependence at all ages, with significant modification to the shape of the $t_\mathrm{MS}$-Mass relation. The L$_\mathrm{t_{MS}}$ relation to mass has very little dependence on \afe. However the final luminosity resulting from the additional step which accounts for stripping through wind mass loss is \afe-dependent. This change in L$_\mathrm{t_{MS}}$ arises primarily through temperature rather than radius. Because the mass-loss rate prescription assumed by \cmb25\ stellar evolution models depended only on iron abundance, this cannot be due to the presence of additional stripping, but rather reflects the impact of stellar atmosphere opacity on the final temperature after stripping has occurred.

The iron-dependence of modelled mass loss rates in the massive OB star regime is based on the prescription of \citet{Vink01}, which scales as 0.85\yg. In recent years different mass loss prescriptions for core-hydrogen burning stars approaching their Eddington limit have been proposed for the massive star regime by authors such as \citet{Bestenlehner20Mdot} or \citet{Figueroa26}. However these works do not explore metallicity-dependence in any detail, instead relying on metallicity prescriptions contemporary with \citet{Vink01}. There has been no detailed study of the impact of composition changes at low metallicity on these winds.  We hence argue that in the massive star regime, in which stellar wind stripping of the stellar envelope can be significant, the incorporation of winds and the response of the stellar structure to their impact is essential in analytic formalisms. The approach of \hpt, in which wind mass loss is applied as a correction at a later stage of the evolution does not capture the true response of the stars, or correctly predict their binary interactions. This is demonstrated by our need to introduce a strongly metallicity- and composition-dependent additional step in the main sequence evolution. 

\subsection{Limits to applicability}
\label{sec:limits}
In this paper our analytic prescriptions have been derived from models which extend to 300\,\msun, and are valid in the very massive star (VMS, M>100\msun) regime for the assumptions made by \cmb25. However we note that significant uncertainties remain in the physics of VMS, particularly in the area of relevant mass loss prescriptions, and the possible impact of rotation and convective mixing. 

We have limited our analysis in this paper to the main sequence, the core hydrogen-burning stage of evolution, from hydrogen ignition to hydrogen depletion. As we have already discussed, the rate of mass loss from the stellar surface in winds can have a significant effect at high masses even in this regime. The later stages of stellar evolution are still more dominated by the extreme mass loss experienced by stars on the giant branch, the prescription for which do not currently incorporate $\alpha$-dependence. These evolved stars are also subject to processes such as dredge-up and convective mixing which transport the products of nuclear burning to the other layers of the star. As a result, the composition of the envelope may differ from that at ZAMS, and particularly so in the case of stripped Wolf-Rayet stars. We hesitate to suggest substantial changes to the post-main sequence without future, more detailed studies of these evolutionary stages across the full mass range. As outlined when describing the \bpass\ v2.4.0 models above, the maximum age for which stars were allowed to evolve was approximately 30 Gyr, meaning that stars with masses less than about $\,0.7$\msun\ were not included in our fitting. Consequently, extrapolation of the equations presented here to stars of lower mass is not advised. 

We also note that our prescriptions have been derived from a grid of single star evolution models, in which there can be no interaction (i.e. mass loss or gain) owing to a binary partner. The majority of binary stars which undergo interaction do so either very early in their evolution (effectively creating a more massive zero-age star) or in the post-main sequence phase (at which time they expand to fill and overflow their Roche Lobe). As a result, the fraction of stars for which the main sequence properties are modified by binary interactions is small.  
The radius and luminosity at the end of the main sequence, and the evolutionary timescale of the main sequence for the most massive stars nonetheless have knock-on effects for later binary interactions and hence on the end state of stars as a function of mass.

\subsection{{Mass loss}}
\label{sec:massloss}
{As mentioned in Section~\ref{sec:HPT_methods} above, the \citetalias{Hurley00} equations were constructed with models with zero mass loss, with the effects of mass change being a post-processing step in the evolution calculations, to enable flexibility. This is the method implemented in the SSE and BSE population synthesis codes. This is an important difference from the method we adopted, where stellar wind prescriptions were applied to the detailed models. To account for mass changes on the main sequence, \citetalias{Hurley00} preserves the fractional lifetime of the star between time increments but calculates the subsequent luminosity and radius from the evolution of a star with effectively a lower mass but at the same fractional evolution state. This has the effect of translating the evolution from a point on one stellar evolution track to a point on a lower mass evolution track at the same fractional timescale. As a result, the direct comparisons between the \citetalias{Hurley00} equations and the new equations presented here do not represent a like-for-like comparison.}

\begin{figure}
    \centering
    \includegraphics[width=0.48\textwidth]{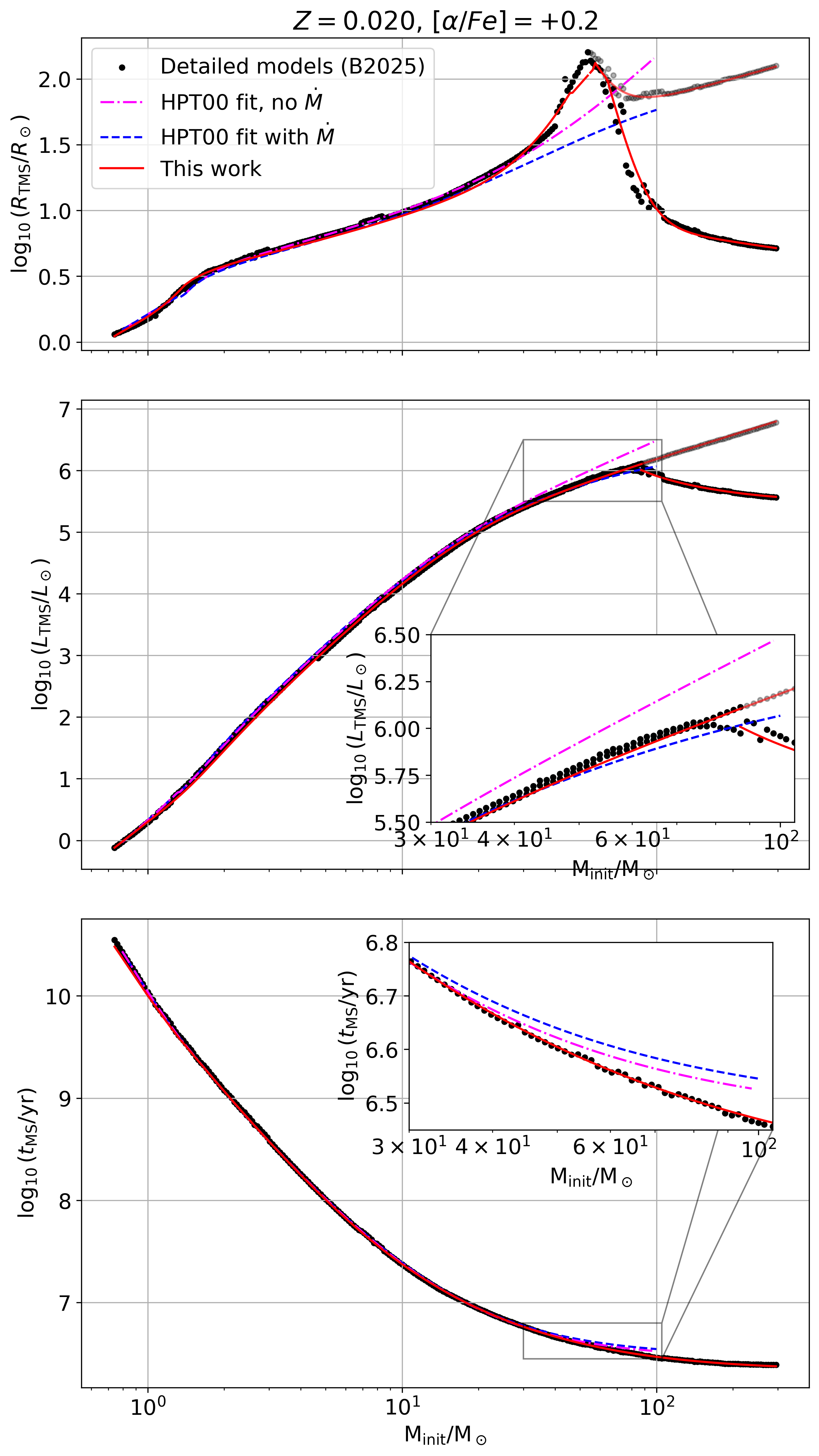}
    \caption{{$R_\mathrm{TMS}$ (upper panel), $L_\mathrm{TMS}$ (middle panel), and $t_\mathrm{MS}$ (lower panel) as a function of initial stellar mass, comparing the \citetalias{Byrne25} detailed stellar evolution models (black points), the new analytic fits derived here (red solid line), the \citetalias{Hurley00} analytic equations where zero mass loss is assumed (magenta dash-dot line) and SSE models where the \citetalias{Hurley00} equations are augmented by the same prescriptions for mass loss as in the \citetalias{Byrne25} models. This comparison has been done at $Z=0.020$ and $[\alpha/$Fe$]=+0.2$. At the highest masses, where stripping occurs before the end of core hydrogen burning, the pale red points and grey line indicate the radius and luminosity at the point of maximum radius while the solid colours indicate the values at core hydrogen exhaustion.}}
    \label{fig:ssewind}
\end{figure}

{Fig~\ref{fig:ssewind} provides a comparison at $Z=0.020$, $[\alpha$/Fe]$=+0.2$ between the detailed evolution models of \citetalias{Byrne25}, the fits derived in this work and SSE models using the \citetalias{Hurley00} equations with and without mass-loss. In the case where mass-loss is calculated, the same mass-loss prescriptions are applied as in \bpass\ namely \citet{deJager88} for cool stars and \citet{Vink01} for hot stars above 20\,000\,K. From this, it can be seen that while this adaptive approach with mass loss reproduces well the terminal-age luminosities, it does not fully encapsulate the full range of behaviour seen in terminal-age radii. This can also be seen in Fig~\ref{fig:HRD1-100}, where the evolution of a 100 solar mass star using mass loss gives a reasonable representation of the evolution, but does not quite reach the expected location in the luminosity-effective temperature plane at the end of the main sequence.}

{A more substantial point to highlight is the impact this has on the estimated total main sequence lifetime. Following the approach of \citetalias{Hurley00} a massive star which loses mass on the main sequence will take longer to reach the end of the main sequence, owing to its evolution as a lower mass star, which has an inherently longer main sequence lifetime. This exacerbates the discrepancy between the lifetimes of massive stars predicted by \citetalias{Hurley00} and those observed. The new analytic fitting carried out here provides a significant improvement in this regard, with the lifetime predictions for massive stars in much better agreement with the detailed evolution models. Even if a mass-loss responsive implementation of the \citetalias{Hurley00} equations yields luminosities and temperature evolution which is in better agreement with the detailed models, the time over which the star evolves along the track can vary substantially, which is important for stellar population considerations such as supernova delay time distributions.}

{The main drawback of the approach we adopted is that the current construction relies on the assumed mass-loss prescriptions in order to produce the correct evolution tracks, with both the zero-age and terminal-age properties being tied to the initial mass of the star, making it less adaptive and responsive to varying the physics (e.g. if one were to implement new mass-loss prescriptions). In addition, by tying the equations to the initial mass of the star, further consideration is needed for how the evolution is impacted by mass changes other than stellar winds. A na\"{i}ve implementation similar to that of \citetalias{Hurley00}, where an effective mass is calculated may be a reasonable starting point. In this case, the effective mass would be determined by taking the initial mass of the star and adding or subtracting any mass lost or gained through mass transfer, but ignoring mass lost through winds, since this is already accounted for in the equations. Further work will be required in future to test the validity of this approach using the forthcoming \bpass\ v2.4 binary star models.}

\subsection{Implementation in BSE}
Many of the most common uses for rapid population synthesis (i.e. the construction of stellar populations based on analytic models rather than detailed stellar structure models) are in the regime of binary evolution, and particular the final merger statistics of compact objects. 

To demonstrate the impact of our revised analytic formalism on these applications, we construct a modified version of the population synthesis code, BSE \citep{Hurley02}. BSE is an open-source rapid population binary synthesis code, where the evolution of the individual stars is approximated with the \hpt\ equations and additional equations governing binary interactions. To accommodate our revised formalism, the original single star evolution equations determining $R_0$, $L_0$, $t_\mathrm{BGB}$, $R_\mathrm{t_{MS}}$ and $L_\mathrm{t_{MS}}$ were replaced with the versions described in Section~\ref{sec:fitresults} above. In addition, the relevant functions in BSE were modified such that the mass used in each time-step in these equations is the initial mass, rather than the time-step mass. 

We have not changed the prescription for the end state of the stellar evolution in BSE, which determines the mass and type of compact remnant based on analytic prescriptions for the core mass, total mass, luminosity and radius. We note that these prescriptions were themselves derived based on the same mass-limited grid of models used by \hpt. While our detailed \cmb25 BPASS v2.4.0 single star models contain detailed structure information and hence would enable us to make more precise estimates of the eventual remnant mass for each model, we do not attempt to implement these here, because they are also subject to significant post-main-sequence mass loss and mass transfer effects.

\begin{figure}
    \centering
    \includegraphics[width=0.481\textwidth]{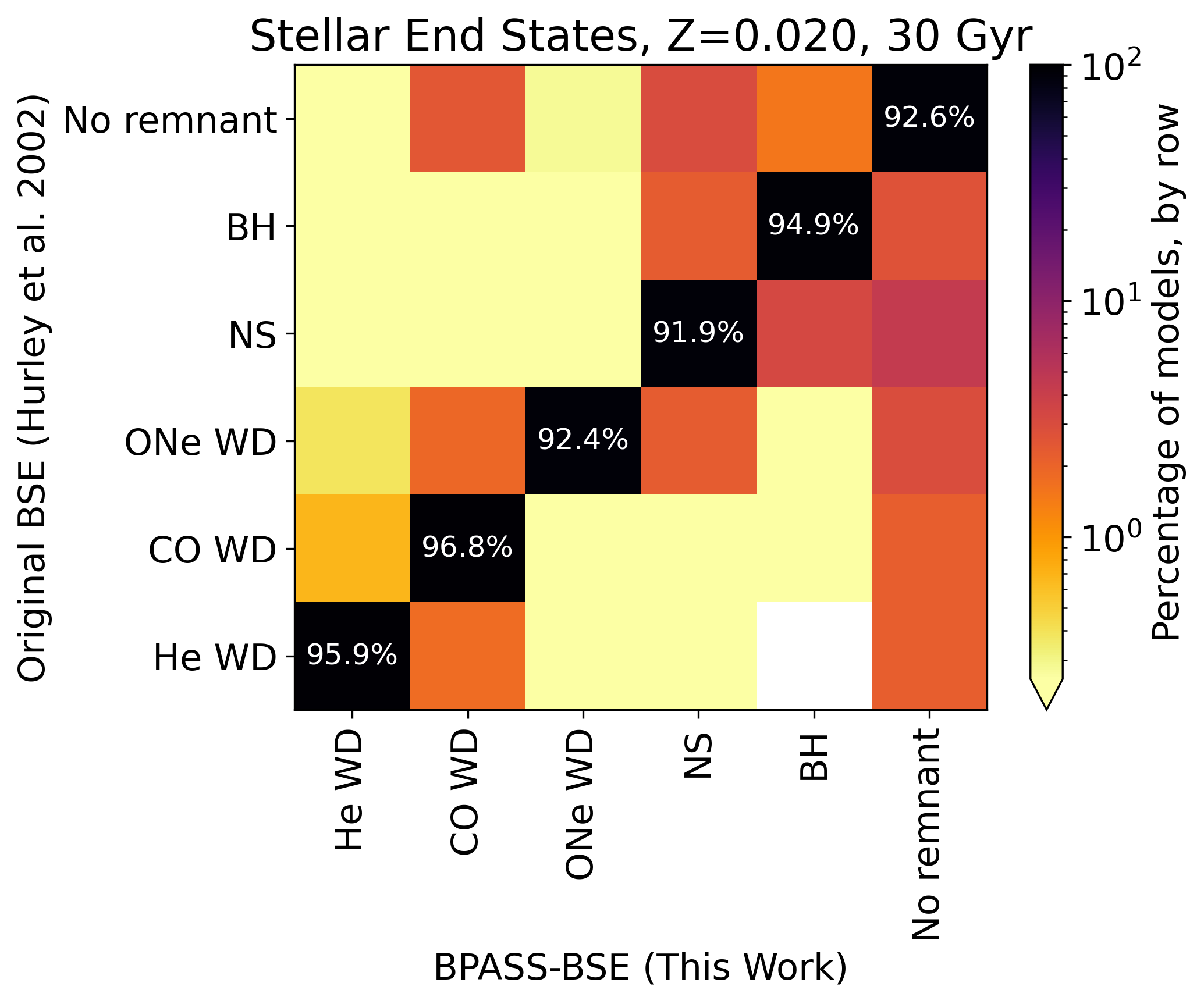}
    \includegraphics[width=0.481\textwidth]{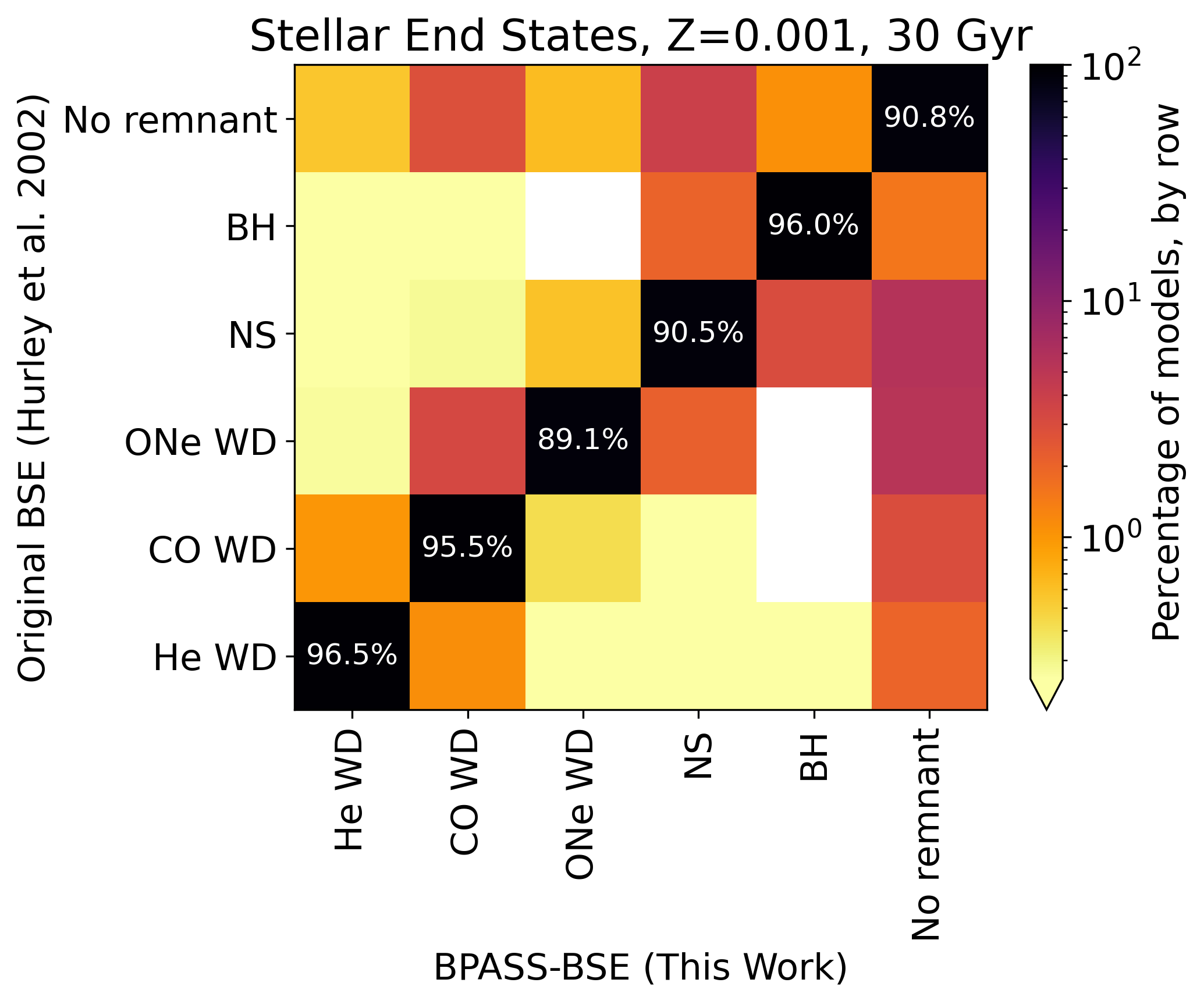}    
    \caption{Comparison of stellar end states for the original, unmodified BSE code, and a modified version incorporating the new analytic equations. Each row of the histogram is normalised to sum to 100 per cent, and the percentage of models which have the same end state in both versions of the simulation are indicated along the diagonal. Comparisons are made for systems which have completed their evolution in 30 Gyr, and are made at $Z=0.020$ (upper panel) and $Z=0.001$ (lower panel)}
    \label{fig:bsecom}
\end{figure}

Fig.~\ref{fig:bsecom} provides a comparison between models with the original BSE code and a modified version with the new analytic equations for the main-sequence evolution. The population generated was idealised, with models spaced on the regular grid in $M_1$, $q$ and $\log_{10}(P$/d) defined for the detailed \citetalias{Byrne25} BPASS v2.4.0 models. No attempt is made to weight the population with any assumptions of initial mass function or binary parameter distributions. Populations with metallicities of $Z=0.020$ and $Z=0.001$ are considered. In general, the overall results are qualitatively similar between the two metallicities, with the end state of around 90 per cent of the models yielding the same category of stellar remnant. In general most of the movement is to adjacent categories of stellar remnant (e.g. from neutron star to black hole, or from He white dwarf to CO white dwarf). The implication of this is that 10 per cent of the uniformly sampled parameter space produces a \textit{different} end state when these new equations are applied. For the purposes of this demonstration, we have not examined the details of the outcome, for example the mass of individual compact remnants, as these are poorly defined in rapid population synthesis codes, and should ideally be addressed with detailed stellar structure models rather than approximated. We would, however, expect that the mass distributions would be modified by changes in the main-sequence properties of stars, and the same subsequent effects which lead to reclassification of ten per cent or more of the overall population.

We note that in the case of main-sequence mass transfer, more accurate behaviour may be obtained by adapting the current initial mass-based prescription to reflect the mass changes owing to binary mass transfer, but ignoring changes in mass owing to stellar winds as these are already accounted for in the new equations. For post-main sequence mass-transfer, the binary interaction is accounted for in the post-main sequence evolution.  To evaluate the impact of this change, we also considered a modified variant of BSE in which the main sequence mass adapted in line with any mass change, and found a qualitatively similar result, with around 10 to 20 per cent of models again reclassified. 

As Fig.~\ref{fig:fits_z020a02} demonstrates, the radii and luminosity of massive stars are both overestimated by the \hpt\ equations towards the end of the main-sequence {in the na\"{i}ve case where zero-mass loss is included. When mass-loss is included in models using the \citetalias{Hurley00} equations (Section~\ref{sec:massloss}), terminal-age main sequence luminosities may be broadly correct, but the radius may be underestimated.}. Given most of the commonly-used massive star mass-loss prescriptions \citep[e.g. see][]{2022ARA&A..60..203V} {are very sensitive to proximity to the Eddington limit, and by extension, radius and luminosity, this can have significant implications for the inferred final black hole masses predicted by evolutionary models}. The importance of the mass loss on the main sequence has been recognised for some time, e.g. as discussed with reference to the StarTrack rapid population synthesis code in \citet{Belczynski10}. However the approach usually adopted in rapid population synthesis such as BSE or StarTrack applies a correction for mass loss effectively as a post-processing of the original \hpt\ prescription, rather than accounting for the effects of mass loss on the stellar structure, nuclear burning and hence stellar evolution.  This provides flexibility to adjust the mass-loss rate rapidly during the population synthesis, and so is valuable for exploring large parameter spaces. However the handling of mass loss does have important consequences for the stellar evolution, perhaps most significantly in evaluating the fraction of the initial mass which becomes bound to any compact remnant formed at the end of the stellar life. 

While we would advise the community studying binary stellar populations, including compact mergers and their mass distributions, to use detailed modelling where possible, we propose that the revised analytic prescriptions for main sequence evolution presented in this paper are a useful first step towards improving current rapid population synthesis efforts.

\section{Conclusions}
\label{sec:conclusions}
In this work, we have presented a new set of analytic expressions which provide a good description of the main sequence stellar parameters of the \bpass\ v2.4.0 single star stellar evolution models. We summarise our results as follows.

\begin{enumerate}
    \item We have calculated revised analytic formulae to describe the main sequence evolution of stars from M=0.7\msun to 300\msun, and spanning metallicities from $Z=0.001$ to $Z=0.040$, with compositions with varying \afe\ in the range -0.2 to +0.6.

    \item Our revised formalism accounts for the impact of main-sequence mass loss on the properties of massive stars at the end of their main-sequence lifetimes, significantly revising estimates for $t_\mathrm{MS}$, $R_\mathrm{t_{MS}}$ and L$_\mathrm{t_{MS}}$ extrapolated from the equations of \hpt.

    \item Four of the five main sequence properties have a global fit with RMS errors of less than 0.05 dex, or 10 per cent. The exception is the terminal age main sequence which shows a slightly larger RMS error of 0.1 dex and 20 per cent. Care should be taken not to extend these new equations beyond their regime of validity. Their performance below $Z'=0.001$ and for initial stellar masses less than 0.7\msun\ is not always trustworthy.

    \item We find that all parameters show some degree of dependence on \afe, with particular impacts on the evolution of the most stars due to the balance between opacity and wind mass-loss. 

    \item Using the rapid population synthesis code BSE, we demonstrate that the end states of approximately 10\,per cent of binary stellar models would be reclassified as a result of changes to the analytic main sequence evolution. 

    \item {By treating massive stars with high mass-loss rates as evolving like stars with a lower effective mass, the equations of \citetalias{Hurley00} provide a closer match to the main sequence evolution of the \citetalias{Byrne25} detailed models than when zero mass loss is applied. While this can produce a similar evolution track in terms of luminosity and effective temperature, it overestimates the lifetimes of stars with masses greater than about 50 solar masses, and this approach to mass loss exacerbates this discrepancy.}

    \item We caution strongly against extrapolation of the widely-used analytic formalisms of \hpt\ for stars above initial stellar masses of 50\,\msun, and in regions of parameter space in which they were not originally calibrated (e.g. $\alpha$-enhanced environments). In such regimes, the new equations presented in this work represent a more reliable set of values.

\end{enumerate}

\section*{Acknowledgements}

{We are very grateful to the referee for their thorough review, which has led to numerous improvements to the manuscript.} We thank members of the \bpass\ team for providing helpful discussions and feedback while carrying out this research. ERS and CMB acknowledge funding from the UK Science and Technology Facilities Council (STFC) through Consolidated Grants ST/T000406/1 and ST/X001121/1. JJE acknowledges funding from the Royal Society Te Apar\=angi of New Zealand Marsden Grant Scheme. Computing facilities were provided by the Scientific Computing Research Technology Platform (SCRTP) of the University of Warwick. Data processing made use of Astropy\footnote{\url{https://www.astropy.org/}}, a community-developed core Python package for Astronomy \citep{astropy:2013,astropy:2018}.

\section*{Data Availability}
 
The stellar evolution models underlying this work are the \bpass\ v2.4.0 single-star models, available on the \bpass\ websites\footnote{\url{https://warwick.ac.uk/bpass}}\footnote{\url{https://bpass.auckland.ac.nz}}. {Python and FORTRAN-compatible functions for computing all of the stellar properties ($R_0$, $L_0$, $t_\mathrm{MS}$,$t_\mathrm{WR}$, $R_\mathrm{TMS}$,$R_\mathrm{WR}$,$L_\mathrm{TMS}$ and $L_\mathrm{WR}$) as a function of mass and composition will be made available on the \bpass\ websites.} Tabulated data showing the stellar properties at zero-age main sequence, terminal-age main sequence and the base of the giant branch are also made available on the \bpass\ websites. All other data can be shared upon reasonable request to the author at the end of the peer-review process.



\bibliographystyle{mnras}
\bibliography{mybib} 

\bsp	




\appendix

\cleardoublepage

\section{Metallicity dependence of coefficients}
\label{app:append}
The composition dependence of the coefficients, $b_n$, and the selected $a_n$ values taken from \citet{Hurley00} are given here. The functional form of each set of coefficients is outlined above each table, typically of the form:
\begin{equation}
\mathrm{\textbf{[Column\,1]}} + \mathrm{\textbf{[Column\,2]}}\text{\yg} + ...
\end{equation}
Note that :
\[
\text{\yg}=\log(Z'/0.02),
\]
\[
\sigma=\log(Z'),
\]
\noindent and
\[
\alpha'=[\alpha/\mathrm{Fe}].
\]
$Z'$ is the iron-equivalent metallicity mass fraction as described in Section~\ref{Sec:BPASS}. The equations from the main paper associated with each set of coefficients are duplicated at the top of each table for ease of use.

\begin{table*}
\label{tab:R0}
\caption{Coefficients $b_i$ for indices 0--15, used for calculating expressions for $R_0$, the radius of a star at the zero-age main sequence (ZAMS).}
    
\begin{equation*}
\hspace{4.5cm}
R_0=
        \begin{cases}
        \dfrac{b_0 + b_1 M^{b_2}+b_3 M^{b_4}+b_5 M^{b_6}}{b_7 M^{b_8}+b_9 M^{b_{10}}} & M/M_\odot \le 3.4 \\
        \\
        b_{11}+b_{12}M^{b_{13}}+b_{14}M^{b_{15}} & M/M_\odot \ge 3.5 
    \end{cases}
\end{equation*}

\hspace{-6.5cm} where $\alpha' = [\alpha/\mathrm{Fe}]$, \yg\ $=\log(Z'/0.02)$ and $b_n$ are of the form

\begin{equation*}
\hspace{1.1cm}
    b_{n} = \textbf{[1]}+\textbf{[2]}\text{\yg}+\textbf{[3]}\alpha'+\textbf{\textbf{[4]}}\text{\yg}\alpha'+\textbf{[5]}\text{\yg}^2+\textbf{[6]}\text{\yg}^2\alpha'+\textbf{[7]}\text{\yg}^2\alpha'^2+\textbf{[8]}\text{\yg}^3\alpha'+\textbf{[9]}\text{\yg}^3\alpha'^2
\end{equation*}

\setlength{\tabcolsep}{3pt}
\begin{tabular}{lllllllllll}
\hline
$b_n$   & \textbf{[1]}         & \textbf{[2]}       & \textbf{[3]}       & \textbf{\textbf{[4]}}       & \textbf{[5]}       & \textbf{[6]}       & \textbf{[7]}       & \textbf{[8]}       & \textbf{[9]}\\ \hline
$b_0$   & +8.97539\,E+0 &-----------&-----------&-----------&-----------&-----------&-----------&-----------&-----------\\
$b_1$   & +5.33807\,E+2 &-----------&-----------&-----------&-----------&-----------&-----------&-----------&-----------\\
$b_2$   & -3.24394\,E+0 &-----------&-----------&-----------&-----------&-----------&-----------&-----------&-----------\\
$b_3$   & +2.60967\,E+1 &+1.54326\,E+0&-----------&-----------&-----------&+1.05658\,E-1&-----------&-----------&-----------\\
$b_4$   & +1.00967\,E+1 &-----------&-----------&-----------&-----------&-----------&-----------&-----------&-----------\\
$b_5$   & -2.18839\,E+1 &-7.27841\,E+0&+3.26805\,E-1&-----------&-----------&-----------&-----------&-----------&-----------\\
$b_6$   & +1.00965\,E+1 &-----------&-----------&-----------&-----------&-----------&-----------&-----------&-----------\\
$b_7$   & +5.80086\,E+2 &+4.96437\,E+0&-----------&-----------&-----------&-----------&-----------&-----------&-----------\\
$b_8$   & -4.29736\,E+0 &-----------&-----------&-----------&-----------&-----------&-----------&-----------&-----------\\
$b_9$   & +3.74971\,E+0 &-6.36795\,E+0&-----------&-----------&+1.79430\,E+0&-----------&-----------&-----------&-----------\\
$b_{10}$& +9.54465\,E+0 &-----------&-----------&-----------&-----------&-----------&-----------&-----------&-----------\\
\\
$b_{11}$& -5.63050\,E-1 &+4.15280\,E-1&-----------&-----------&+3.08940\,E-1&-----------&-----------&-----------&-----------\\
$b_{12}$& +1.55030\,E+0 &+3.14690\,E-1&-----------&-----------&-----------&-----------&+3.20190\,E-2&-----------&-1.04840\,E-3\\
$b_{13}$& +4.73720\,E-1 &-2.71010\,E-2&-----------&-----------&-----------&-----------&+1.16970\,E-2&-----------&-----------\\
$b_{14}$& +4.27090\,E-4 &+4.05530\,E-5&-----------&+1.21580\,E-3&-----------&-----------&-----------&+4.95840\,E-4&-----------\\
$b_{15}$& +1.93210\,E+0 &+4.80180\,E-1&-----------&+4.70410\,E-2&-----------&-----------&-----------&-----------&-----------\\
\hline
\end{tabular}

\end{table*}

\begin{table*}
\centering
\label{tab:L0}
\caption{Coefficients $b_i$ for indices 16--22, used for calculating expressions for $L_0$, the luminosity of a star at the zero-age main sequence (ZAMS).}

\begin{equation*}
\hspace{4.5cm}
    L_0 = \dfrac{b_{16} M^{5.5}+b_{17}M^{11}}{b_{18} + M^{3}+b_{19} M^{5}+b_{20} M^{7}+b_{21} M^{8}+b_{22} M^{9.5}}
\end{equation*}

\hspace{-9.5cm} where $b_n$ are of the form

\begin{equation*}
\hspace{2.5cm}
    b_{n} = \textbf{[1]}+\textbf{[2]}\text{\yg}+\textbf{[3]}\text{\yg}^2+\textbf{\textbf{[4]}}\text{\yg}^3+\textbf{[5]}\text{\yg}^4+\textbf{[6]}\alpha'+\textbf{[7]}\alpha'^2
\end{equation*}
\vspace{0pt}
\begin{tabular}{llllllll}
\hline
$b_n$    & \textbf{[1]}        & \textbf{[2]}       & \textbf{[3]}       & \textbf{\textbf{[4]}}       & \textbf{[5]}       & \textbf{[6]}        & \textbf{[7]}\\ \hline
$b_{16}$ & +1.7047\,E+0 &-8.4516\,E+0 &+1.3890\,E+1 &-9.8473\,E+0 &+1.3510\,E+1 &+5.76784\,E-1 &+1.40516\,E+1\\
$b_{17}$ & +3.6911\,E+1 &-7.6272\,E+1 &+9.9889\,E+1 &-2.4105\,E+1 &+2.4279\,E+1 &+2.83841\,E+1 &+4.53143\,E+1\\
$b_{18}$ & +6.9398\,E-1 &-5.0715\,E+1 &+5.2277\,E+0 &-8.3235\,E+0 &+7.9322\,E-1 &-1.25317\,E-1 &+1.39085\,E+1\\
$b_{19}$ & +2.8543\,E+1 &-2.8499\,E+1 &+2.7650\,E+1 &+2.7932\,E+1 &+2.8138\,E+1 &+1.92527\,E+0 &+1.23523\,E+1\\
$b_{20}$ & +2.5531\,E+1 &-2.1217\,E+1 &+2.5429\,E+1 &+8.6413\,E+0 &+1.5680\,E+1 &+1.71860\,E+1 &+9.70280\,E+1\\
$b_{21}$ & +3.3885\,E+0 &-8.9833\,E+0 &+1.1551\,E+1 &-4.6203\,E+0 &+3.1806\,E-1 &+9.96020\,E-1 &+2.67763\,E+0\\
$b_{22}$ & +2.6312\,E-2 &-5.4357\,E-2 &+6.9621\,E-2 &-1.6750\,E-2 &+1.8662\,E-2 &+1.85043\,E-2 &+2.70854\,E-2\\
\hline
\end{tabular}
\end{table*}

\begin{table*}
\label{tab:tBGB}
\centering
\caption{Coefficients $b_i$ for indices 23--28, used for calculating expressions for $t_\mathrm{BGB}$, the time taken for stars to reach the base of the giant branch. These values are then modified as per \protect\citet{Hurley00} to derive $t_\mathrm{MS}$, the main sequence lifetime.}

\begin{equation*}
    \hspace{4.5cm}
    t_\mathrm{BGB}=\dfrac{b_{23}+b_{24} M^{4+b_{25}}+b_{26} M^{5.5+{b_{27}}}+b_{28}M^7}{a_{4} M^2+a_{5} M^7}
\end{equation*}

\hspace{-9.5cm} where $b_n$ are of the form (unless otherwise stated)

\begin{equation*}
\hspace{3.0cm}
    b_{n} = \textbf{[1]}+\textbf{[2]}\text{\yg}+\textbf{[3]}\text{\yg}^2+\textbf{\textbf{[4]}}\text{\yg}^3+\textbf{[5]}\alpha'+\textbf{[6]}\alpha'^2
\end{equation*}

\begin{tabular}{lllllll}
\hline
$b_n$     &\textbf{[1]}&\textbf{[2]}&\textbf{[3]}&\textbf{[4]}&\textbf{[5]}&\textbf{[6]}\\ \hline
$b_{23}'$ &+9.02878\,E-1 &------------&------------&------------&------------&------------\\
$b_{24}'$ &+6.25289\,E-2 &------------&------------&------------&------------&------------\\
$b_{23}''$&+6.51418\,E+0 &------------&------------&------------&-1.25468\,E-1 &+1.55575\,E-2 \\
$b_{24}''$&+4.22274\,E+0 &------------&------------&------------&-1.04958\,E+0 &-3.02614\,E+0 \\
$b_{25}$  &+6.17196\,E-2 &-2.04701\,E-1 &-4.87759\,E-2 &------------&-4.12218\,E-2 &+4.38856\,E-2 \\
$b_{26}$  &+6.87345\,E+1 &-7.13544\,E+1 &-7.35245\,E+1 &-4.71213\,E+0 &------------&------------\\
$b_{27}$  &+3.01514\,E-1 &+2.02105\,E-1 &+2.62031\,E-1 &------------&-4.92494\,E-2 &-2.35150\,E-2 \\
$b_{28}$  &+7.45886\,E-1 &+1.91214\,E-1 &+3.04747\,E-1 &+2.02850\,E-1 &------------&------------\\
\\
$a_{4}$  &+4.14196\,E-2 &+4.56489\,E-2 &+2.95854\,E-2 &+5.57148\,E-3 &------------&------------\\
$a_{5}$  &+3.42635\,E-1 &------------&------------&------------&------------&------------\\
\hline
\end{tabular}

\begin{equation*}
\hspace{3cm}
b_{23}= {645.547(\text{\yg}+2)^{b_{23}'} + 0.953718(\text{\yg}+2)^{b_{23}''}}
\end{equation*}
\begin{equation*}
\hspace{3cm}
b_{24} = 1539.73(\text{\yg}+2)^{b_{24}'} + 80.7146(\text{\yg}+2)^{b_{24}''}
\end{equation*}

\end{table*}

\begin{table*}
\label{tab:LTMS}
\centering
\caption{Coefficients $b_i$ for indices 29--44, used to derive the luminosity of a star at the Terminal Age main sequence (TAMS), immediately before core hydrogen exhaustion. In the case of the most massive stars, the luminosity of the star at the point of maximum radius is given by $L_\mathrm{t_{MS}}$, and at the point of central hydrogen exhaustion by $L_\mathrm{WR,L}$}

\begin{equation*}
\hspace{4.5cm}
L_{\mathrm{t_{MS}}}= \dfrac{b_{29}M^3+b_{30}M^4+b_{31}M^{b_{32}+b_{33}}}{b_{34}+b_{35}M^5+b_{36}M^{b_{32}}+b_{37}M^{b_{38}}}
\end{equation*}\\
\begin{equation*}
\hspace{4.5cm}
    L_{\mathrm{WR,L}} = {b_{39}}M^{b_{40}}+b_{41}M^{b_{42}}+b_{43}M^{b_{44}}
\end{equation*}

\hspace{-9.5cm} where $b_n$ are of the form (unless otherwise indicated)

\begin{equation*}
\hspace{1.85cm}
    b_{n} = \textbf{[1]}+\textbf{[2]}\text{\yg}+\textbf{[3]}\text{\yg}^2+\textbf{\textbf{[4]}}\text{\yg}^3+\textbf{[5]}\text{\yg}^4+\textbf{[6]}\alpha'+\textbf{[7]}\text{\yg}\alpha'
\end{equation*}

\begin{tabular}{lllllllll}
\hline
$b_n$    & \textbf{[1]}        & \textbf{[2]}        & \textbf{[3]}        & \textbf{\textbf{[4]}}        & \textbf{[5]}        & \textbf{[6]}        & \textbf{[7]}\\ \hline
$b_{29}'$&+1.40136\,E+0 &-2.33766\,E-1 &+2.32278\,E+0 &+1.06889\,E+0 &------------&------------&------------\\
$b_{30}'$&+6.71190\,E-1 &-1.90987\,E+0 &-1.39804\,E+0 &-2.60405\,E+0 &-8.73777\,E-1 &------------&------------\\
$b_{31}$ &+6.63154\,E+2 &-6.72684\,E+1 &+5.35932\,E-1 &-1.00923\,E+1 &------------&------------&------------\\
$b_{32}$ &+8.78872\,E+0 &-1.27360\,E+0 &+6.95135\,E-1 &+4.56210\,E-2 &------------&------------&+7.96724\,E-3 \\
$b_{33}$ &+1.28464\,E+0 &+5.15597\,E-2 &+1.23244\,E-1 &------------&------------&------------&------------\\
$b_{34}$ &+3.87440\,E+3 &-4.01073\,E-4 &-7.63010\,E+1 &-3.48610\,E+2 &-4.86180\,E+1 &------------&------------\\
$b_{35}$ &+3.68307\,E+2 &+4.54792\,E+2 &+1.11320\,E+2 &+4.13110\,E-1 &------------&------------&------------\\
$b_{36}$ &+1.57492\,E-1 &+8.86317\,E-2 &+1.12835\,E-1 &+4.03034\,E-3 &+1.63748\,E-2 &------------&------------\\
$b_{37}$ &+1.35375\,E+2 &+1.99643\,E+0 &-1.02517\,E+0 &+1.10427\,E+0 &-1.24576\,E+0 &------------&------------\\
$b_{38}$ &+6.48550\,E+0 &-1.37040\,E+0 &+6.56024\,E-1 &------------&------------&------------&------------\\
\\
$b_{29}$ & $b_{29}'\times b_{34}$ \\
$b_{30}$ & $b_{30}'\times b_{34}$ \\
\\
$b_{39}$ &+1.32434\,E+16&------------&------------&------------&------------&------------&------------\\ 
$b_{40}$ &-5.56835\,E+0 &-2.83833\,E+0 &-2.59356\,E-1 &------------&------------&-1.17485\,E-1 &------------\\
$b_{41}$ &+2.29996\,E+8 &------------&------------&------------&------------&------------&------------\\ 
$b_{42}$ &-1.30895\,E+0 &-8.99849\,E-1 &-2.77480\,E-1 &------------&------------&-4.34803\,E-2 &------------\\
$b_{43}$ &+7.86430\,E+4 &------------&------------&------------&------------&------------&------------\\ 
$b_{44}$ &+1.95814\,E-1 &-1.37772\,E-1 &+1.06041\,E-2 &------------&------------&+2.41298\,E-2 &------------\\
\\
\hline
\end{tabular}
\end{table*}

\begin{table*}
\label{tab:RTMS}
\centering
\caption{Coefficients $b_i$ for indices 45--66, used to derive the radius of a star at the TAMS, immediately before core hydrogen exhaustion. In the case of the most massive stars, the luminosity of the star at the point of maximum radius is given by $R_\mathrm{t_{MS}}$, and at the point of central hydrogen exhaustion by $R_\mathrm{WR,R}$}

\begin{equation*}
\hspace{4.5cm}
R_\mathrm{t_{MS}}=
        \begin{cases}
        \dfrac{b_{45}+b_{46}M^{b_{47}}}{b_{48}+M^{b_{49}}} & M/M_\odot \le a_{17} \\ \\
        \dfrac{b_{50}M^{b_{51}}+b_{52}M^{b_{53}}+b_{54}M^{b_{55}}}{b_{56}+b_{57}M^{b_{58}}} & a_{17} \le M/M_\odot \le 0.90M_\mathrm{WR,R} \\ \\
        10^{b_{59}}M^{b_{60}}+b_{61}M^{b_{62}} & M/M_\odot \ge 1.1M_\mathrm{WR,R}\\
    \end{cases}
\end{equation*}
\\
\begin{equation*}
\hspace{4.5cm}
    R_\mathrm{WR,R} = 10^{b_{63}}M^{b_{64}}+b_{65}M^{b_{66}}
\end{equation*}
\\

\hspace{-9.5cm} where $b_n$ are of the form (unless otherwise indicated)

\begin{equation*}
\hspace{1.9cm}
    b_{n} = \textbf{[1]}+\textbf{[2]}\text{\yg}+\textbf{[3]}\text{\yg}^2+\textbf{\textbf{[4]}}\text{\yg}^3+\textbf{[5]}\text{\yg}^4+\textbf{[6]}\alpha'+\textbf{[7]}\alpha'^2
\end{equation*}

\begin{tabular}{llllllll}
\hline
$b_n$    & \textbf{[1]}        & \textbf{[2]}        & \textbf{[3]}        & \textbf{\textbf{[4]}}        & \textbf{[5]}        & \textbf{[6]}        & \textbf{[7]} \\ \hline
\hline

$b_{45}'$ & +5.4227\,E-1 &-5.79154\,E-1 &+4.35132\,E-1 &------------&------------&------------&------------\\
$b_{46}'$ &+4.93212\,E-1 &-3.74170\,E-1 &+2.84740\,E-1 &------------&------------&------------&------------\\
$b_{47}$  &+3.51950\,E+0 &+4.26910\,E-1 &-1.76940\,E+0 &------------&------------&------------&------------\\
$b_{48}$  &+3.33410\,E+1 &+6.99640\,E+1 &+1.22530\,E+2 &------------&------------&------------&------------\\
$b_{49}$  &-5.57740\,E+0 &-8.67930\,E+0 &-1.03770\,E+1 &------------&------------&------------&------------\\
\\
$b_{45}$  & $b_{45}'\times b_{48}$ \\
$b_{46}$  & $b_{46}'\times b_{48}$ \\
\\

$b_{50}$ &+3.54080\,E+0 &+7.07320\,E-1 &--------------&--------------&--------------&--------------&--------------\\
$b_{51}$ &+5.39990\,E+0 &--------------&--------------&--------------&--------------&--------------&--------------\\
$b_{52}$ &+1.96210\,E-1 &+2.26140\,E-5 &--------------&--------------&--------------&--------------&--------------\\
$b_{53}$ &+6.19300\,E+0 &--------------&--------------&--------------&--------------&--------------&--------------\\
$b_{54}$ &+5.75789\,E-6 &+2.70629\,E-5 &+4.73741\,E-5 &+3.62919\,E-5 &+1.02411\,E-5 &+1.64681\,E-9 &--------------\\
$b_{55}$ &+9.26580\,E+0 &+2.67420\,E-3 &-2.08340\,E-3 &+1.59660\,E-3 &-1.93750\,E-9 &--------------&--------------\\
$b_{56}$ &+1.47630\,E+0 &-1.34180\,E-3 &--------------&--------------&--------------&--------------&--------------\\
$b_{57}$ &+1.30210\,E+0 &-7.52720\,E-4 &--------------&--------------&--------------&--------------&--------------\\
$b_{58}$ &+5.00750\,E+0 &-2.70760\,E-3 &-5.00400\,E-8 &-1.66280\,E-8 &--------------&--------------&--------------\\
\\
$b_{59}$ &+1.13330\,E+1 &+9.38750\,E+0 &+7.66066\,E+0 &+1.02193\,E+0 &--------------&--------------&-1.35006\,E-1 \\
$b_{60}$ &-5.31271\,E+0 &-6.62489\,E+0 &-4.74497\,E+0 &-2.97616\,E+0 &--------------&-2.05366\,E-1 &--------------\\
$b_{61}$ &+5.81595\,E+0 &-8.69598\,E+0 &-1.07450\,E+0 &-2.06903\,E-1 &--------------&--------------&--------------\\
$b_{62}$ &+5.38100\,E-1 &+1.84685\,E-1 &+4.47447\,E-1 &+5.76257\,E-1 &--------------&--------------&--------------\\
\\
$b_{63}$ &+1.54080\,E+1 &+2.21700\,E+0 &+8.20890\,E+0 &--------------&--------------&--------------&--------------\\
$b_{64}$ &-7.44410\,E+0 &-3.90170\,E+0 &-3.96250\,E+0 &--------------&--------------&--------------&--------------\\
$b_{65}$ &+2.56010\,E+1 &+3.91800\,E+0 &+1.15320\,E+0 &--------------&--------------&--------------&--------------\\
$b_{66}$ &-2.81160\,E-1 &-8.91678\,E-2 &+7.88068\,E-2 &--------------&--------------&--------------&--------------\\
\\
\multicolumn{8}{l}{$\log(a_{17}) = \mathrm{max}(0.097 - 0.1072 (\sigma + 3), \mathrm{max}(0.097, \mathrm{min}(0.1461, 0.1461 + 0.1237 (\sigma + 2))));\,\,\sigma=\log(Z')$} \\
\hline
\end{tabular}
\end{table*}

\begin{table*}
\label{tab:twr}
\centering
\caption{Coefficients $b_i$ for indices 67--68, used to derive $\Delta t_\mathrm{strip}$, the difference in time between the age of a massive star at the point of maximum radius, and the age at the point of core hydrogen exhaustion, after the star has lost its hydrogen envelope and contracted}

\begin{equation*}
\hspace{4.9cm}
    \Delta t_\mathrm{strip} = b_{67}+M b_{68}
\end{equation*}

\hspace{-12.5cm} where $b_n$ are of the form

\begin{equation*}
\hspace{4.9cm}
    b_{n} = \textbf{[1]}+\textbf{[2]}\text{\yg}+\textbf{[3]}\text{\yg}^2+\textbf{\textbf{[4]}}\text{\yg}^3+\textbf{\textbf{[5]}}\alpha'
\end{equation*}

\begin{tabular}{llllll}
\hline
$b_n$    & \textbf{[1]}        & \textbf{[2]}        & \textbf{[3]}        & \textbf{\textbf{[4]}} & \textbf{[5]}        \\ \hline
\hline
$b_{67}$ &-1.05390\,E-1 &+2.01633\,E-1 &+5.53316\,E-1 &+1.90185\,E-1 &+2.97669\,E-2 \\
$b_{68}$ &+3.89793\,E-3 &+8.72170\,E-3 &+9.33672\,E-3 &+7.09167\,E-3 &+3.89436\,E-4 \\
\hline
\end{tabular}
\end{table*}

\clearpage

\section{Goodness of fit plots}
\label{app:goodfit}

\begin{figure*}
    \centering
    \includegraphics[width=\linewidth]{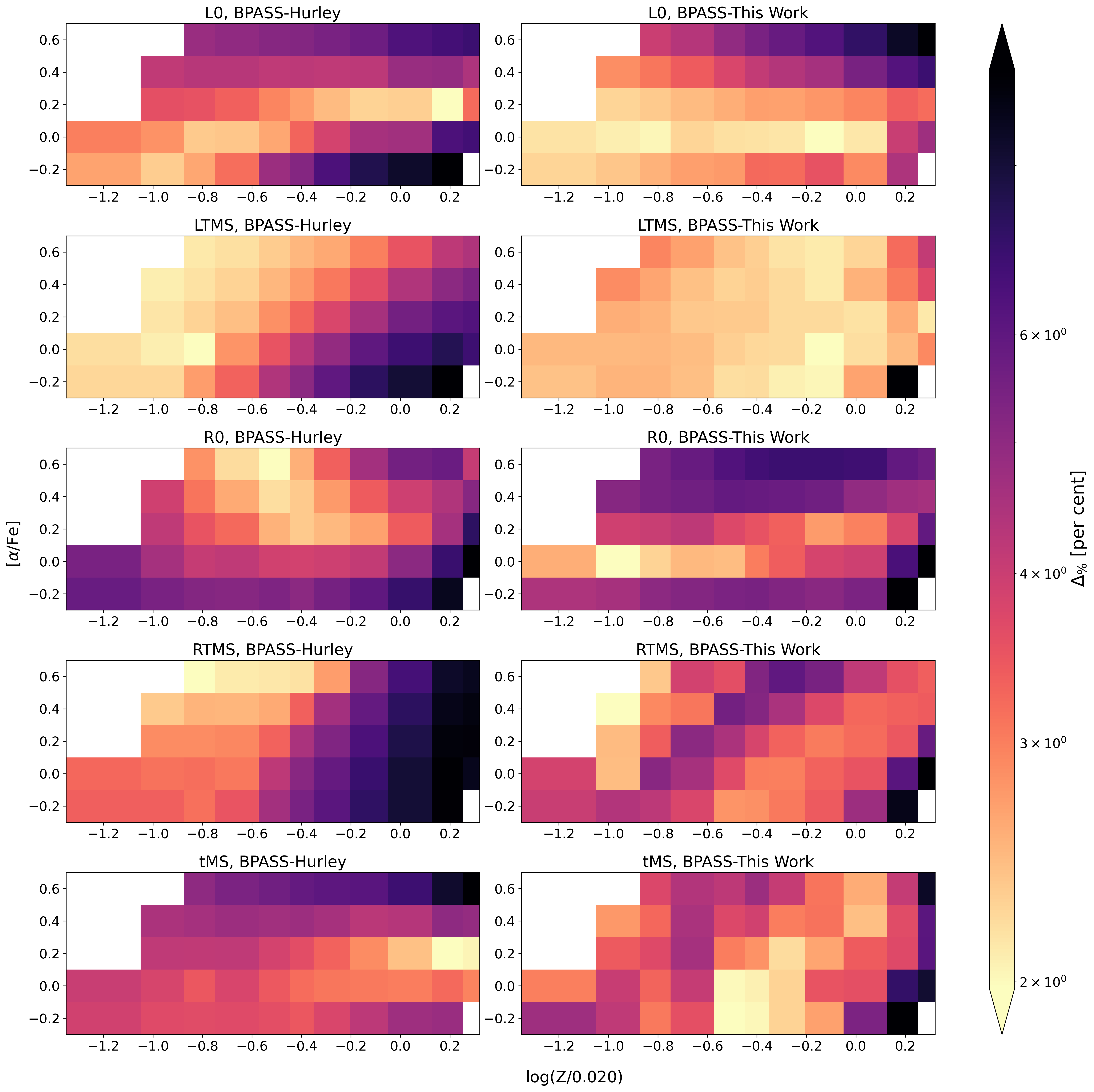}
    \caption{The RMS loss function for each of the fitted quantities expressed as percentage for each value of $\log(Z/0.020)$ and \afe. $\Delta_\mathrm{\%}$ is calculated as per Equation~\ref{eq:deltapct}. The left-hand panels compare the \citetalias{Hurley00} equations to \citetalias{Byrne25} model data, while the right-hand panels compare the analytic fits derived in this work to the \citetalias{Byrne25} models. There is no data in the upper left corner of the panels since these combinations of $Z$ and \afe\ yield values of $Z'<0.001$, and are thus not included in the fitting process. The empty data point at $Z=0.04$ and \afe=-0.2 is a composition not calculated in the \citetalias{Byrne25} as the iron content is higher than any of the available opacity tables.}
    \label{fig:goodfitpct}
\end{figure*}

Fig.~\ref{fig:goodfitpct} show the goodness of fit in terms of $\Delta_\%$ for all of the main sequence evolution parameters fit in this work. Analogously to Fig.~\ref{fig:goodfitlog}, the left-hand panels show the difference between the original \citetalias{Hurley00} and the \citetalias{Byrne25} models, while the right-hand panels shows the comparison between the new equations presented here and the \citetalias{Byrne25} models.

\section{Fits to each parameter at selected metallicities}
\label{app:fitfigures}
\begin{figure*}
    \centering
    \includegraphics[width=0.95\linewidth]{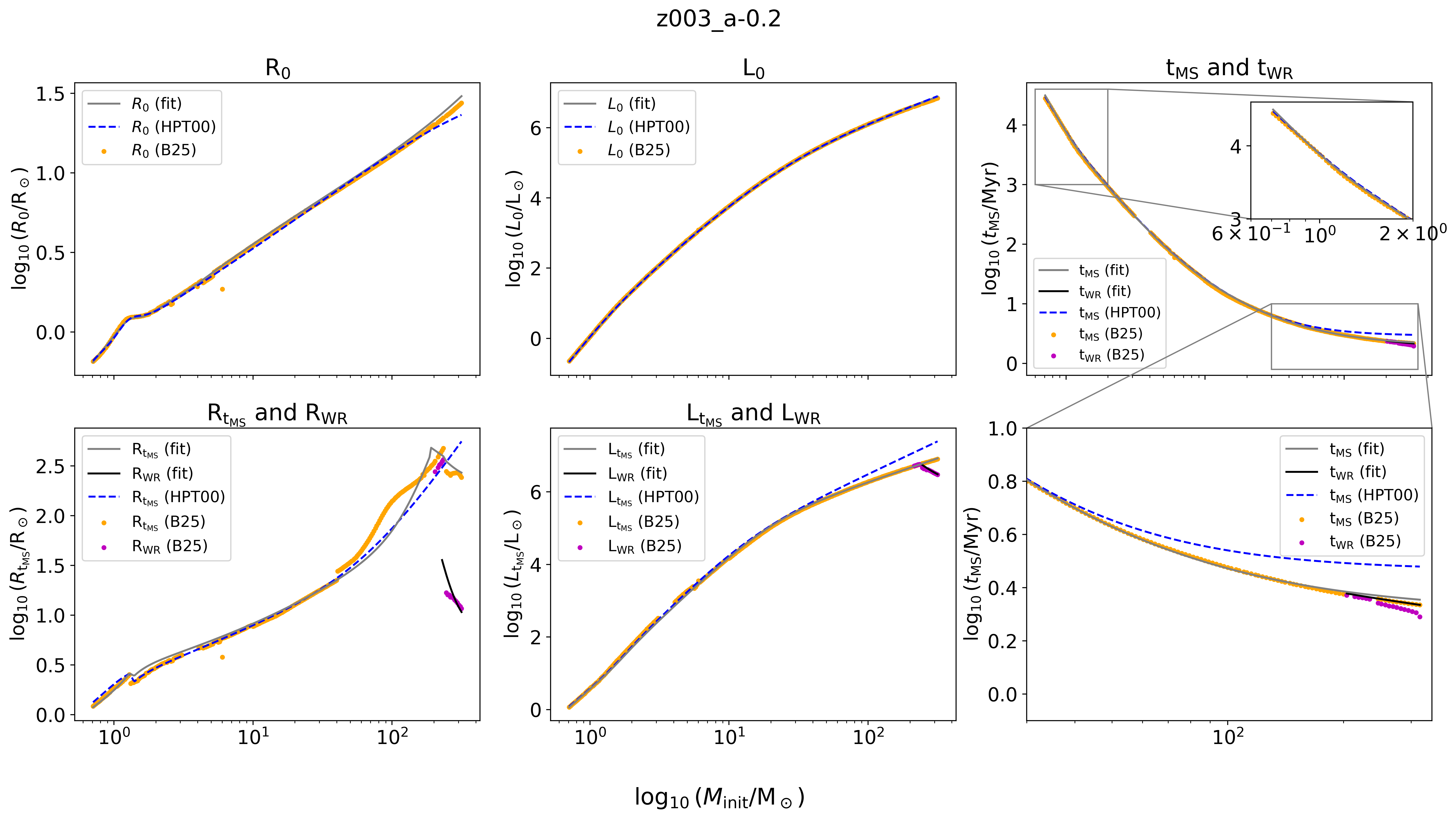}
    \caption{Mass evolution for each of the five parameters being fit in this work, shown at a single composition ($Z=0.003$, \afe$=-0.2$). From left to right, top to bottom these are: i) Zero Age main sequence radii, ii) Zero Age main sequence luminosities, iii) main sequence evolutionary timescales, iv) Radii at the Terminal Age main sequence (and at the point of maximum radius for massive stars), v) Luminosities at the Terminal Age main sequence (and at the point of maximum radius for massive stars), and vi) an expanded view of the main sequence lifetimes of stars with initial masses between 30 and 300\msun. The blue dashed lines show the analytic results from \citetalias{Hurley00}, while the black and grey solid lines show the fits introduced in this paper. The \bpass\ v2.4.0 model data from \citetalias{Byrne25} are shown as magenta and orange points, with the orange points indicating the properties at the time of core-hydrogen exhaustion for non-stripped stars, and for the point of maximum radius for stars which become stripped, while the purple points indicate the properties of the stripped stars at the point of core-hydrogen exhaustion.}
    \label{fig:fits_z003am2}
\end{figure*}

\begin{figure*}
    \centering
    \includegraphics[width=0.95\linewidth]{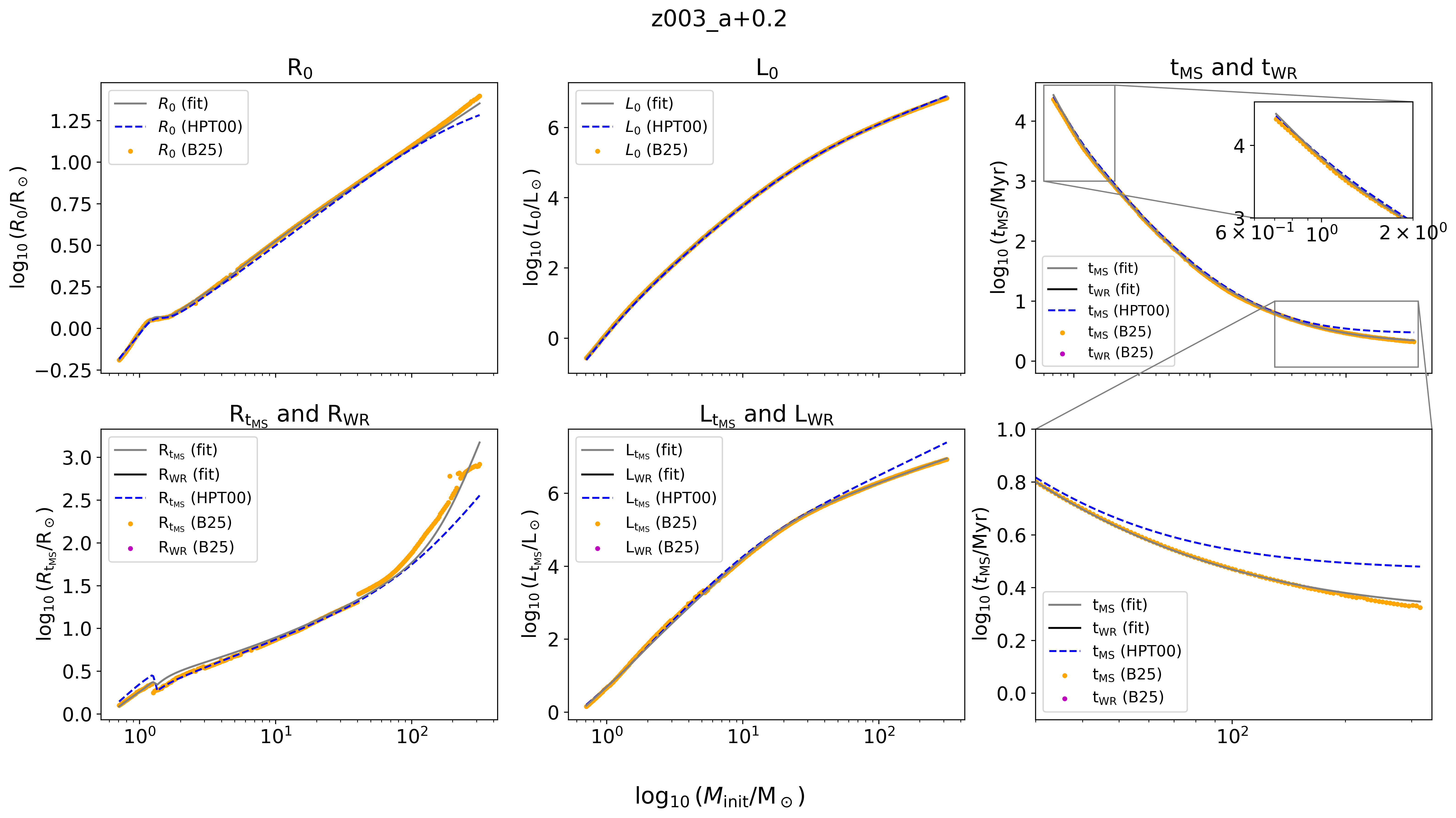}
    \caption{As Fig.~\ref{fig:fits_z003am2}, but for a composition of $Z=0.003$ and \afe=+0.2.}
    \label{fig:fits_z003a02}
\end{figure*}

\begin{figure*}
    \centering
    \includegraphics[width=0.95\linewidth]{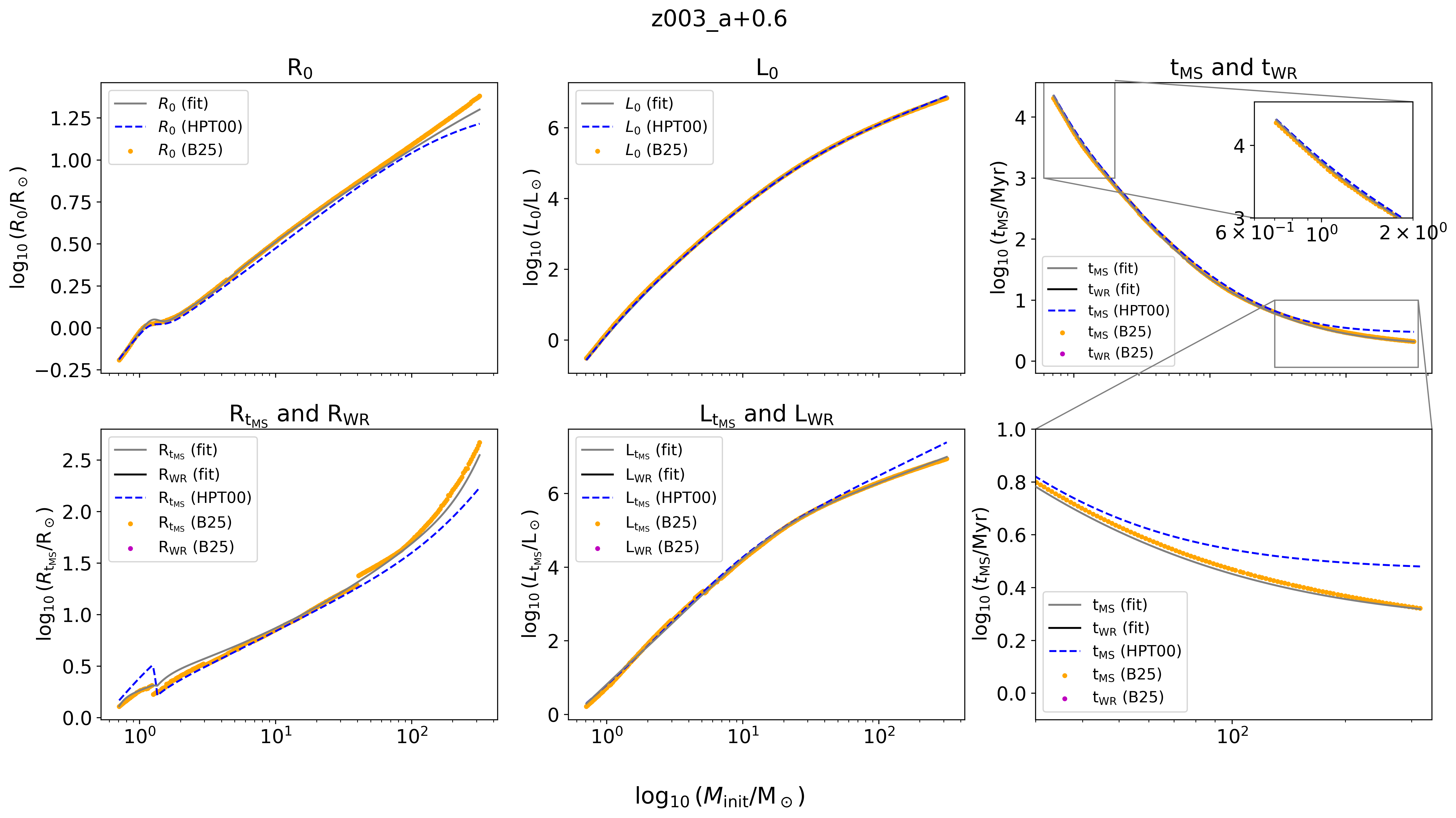}
    \caption{As Fig.~\ref{fig:fits_z003am2}, but for a composition of $Z=0.003$ and \afe=+0.6.}
    \label{fig:fits_z003a06}
\end{figure*}

\begin{figure*}
    \centering
    \includegraphics[width=0.95\linewidth]{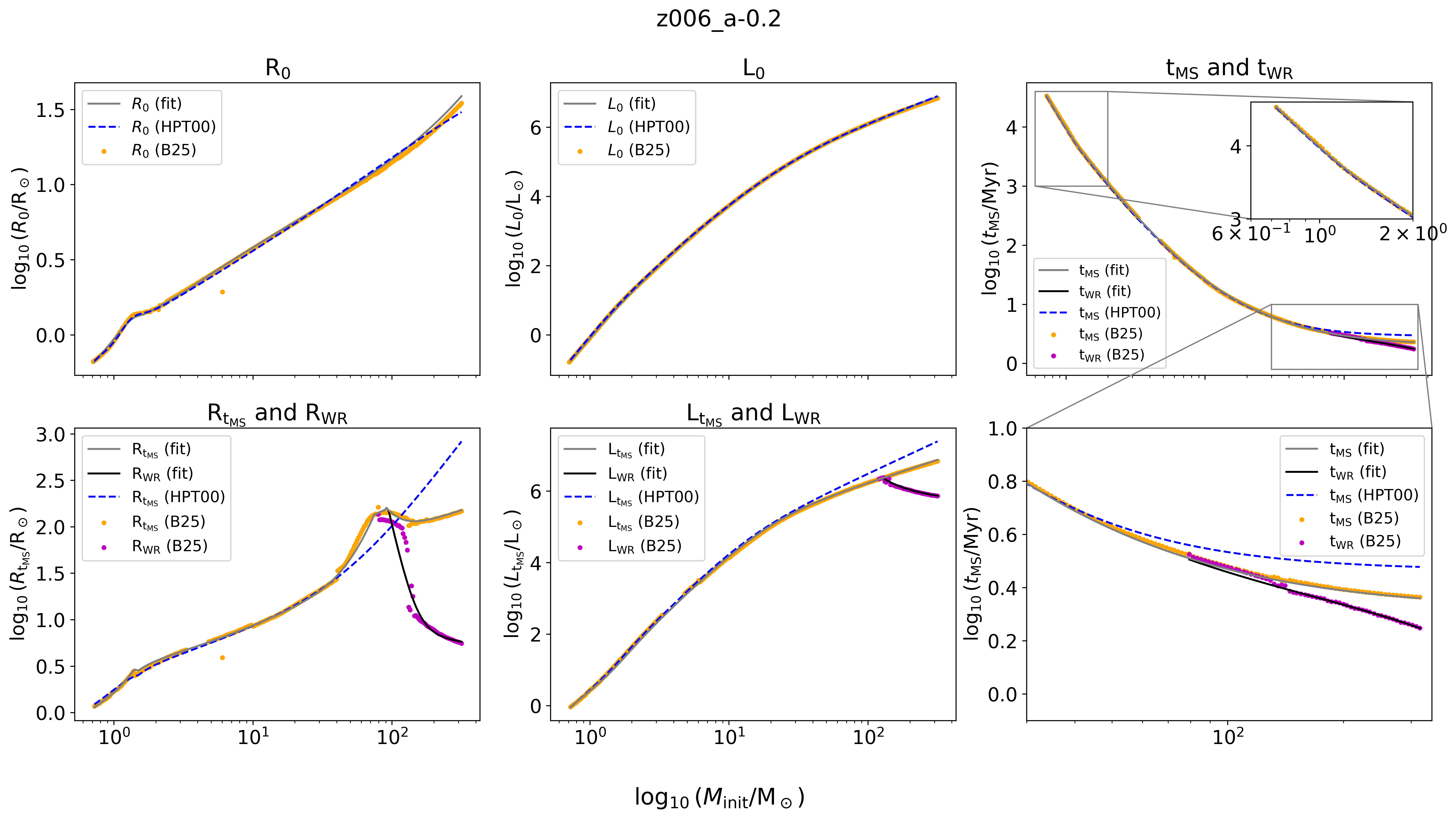}
    \caption{As Fig.~\ref{fig:fits_z003am2}, but for a composition of $Z=0.006$ and \afe=-0.2.}
    \label{fig:fits_z006am2}
\end{figure*}

\begin{figure*}
    \centering
    \includegraphics[width=0.95\linewidth]{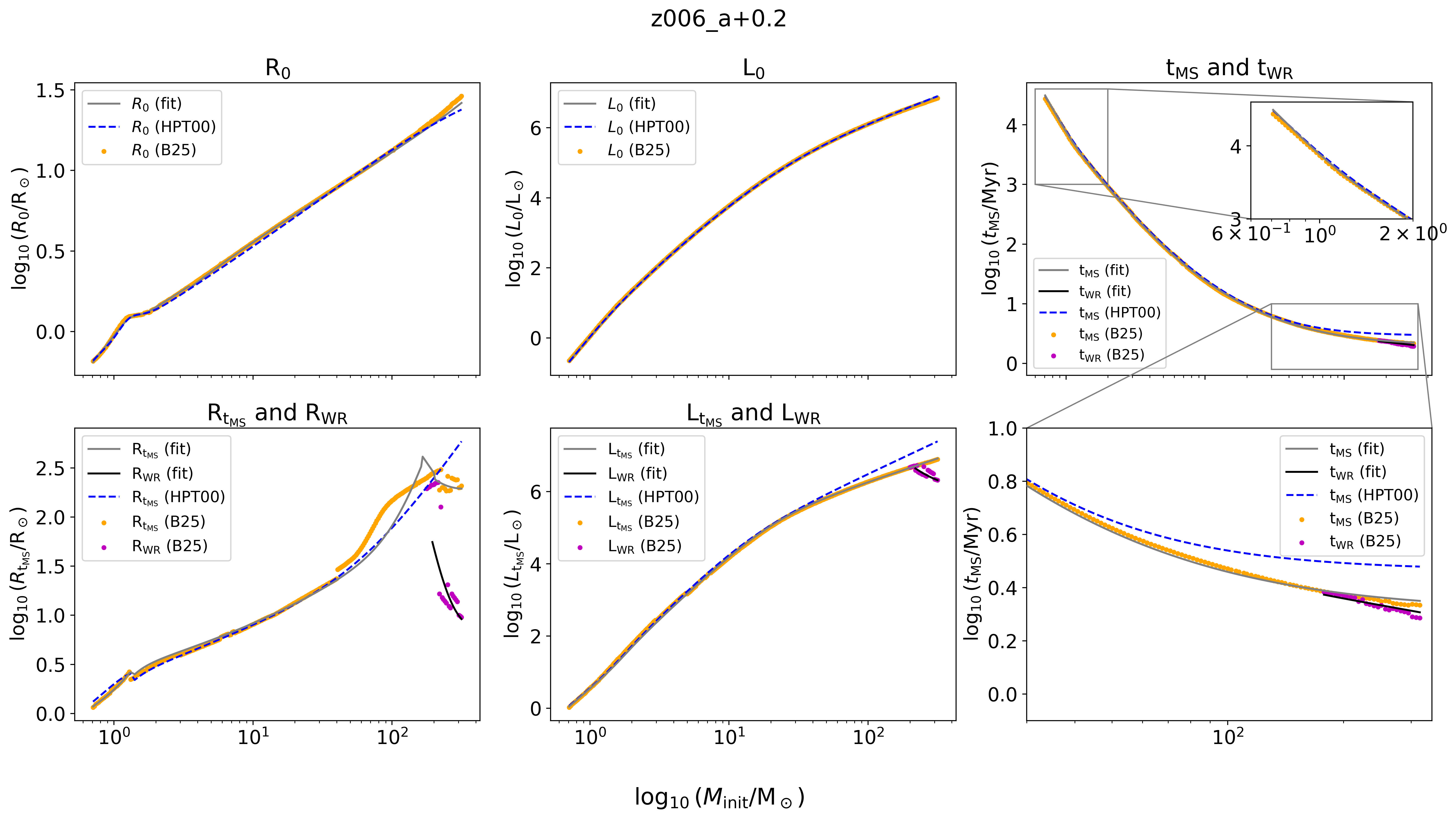}
    \caption{As Fig.~\ref{fig:fits_z003am2}, but for a composition of $Z=0.006$ and \afe=+0.2.}
    \label{fig:fits_z006a02}
\end{figure*}

\begin{figure*}
    \centering
    \includegraphics[width=0.95\linewidth]{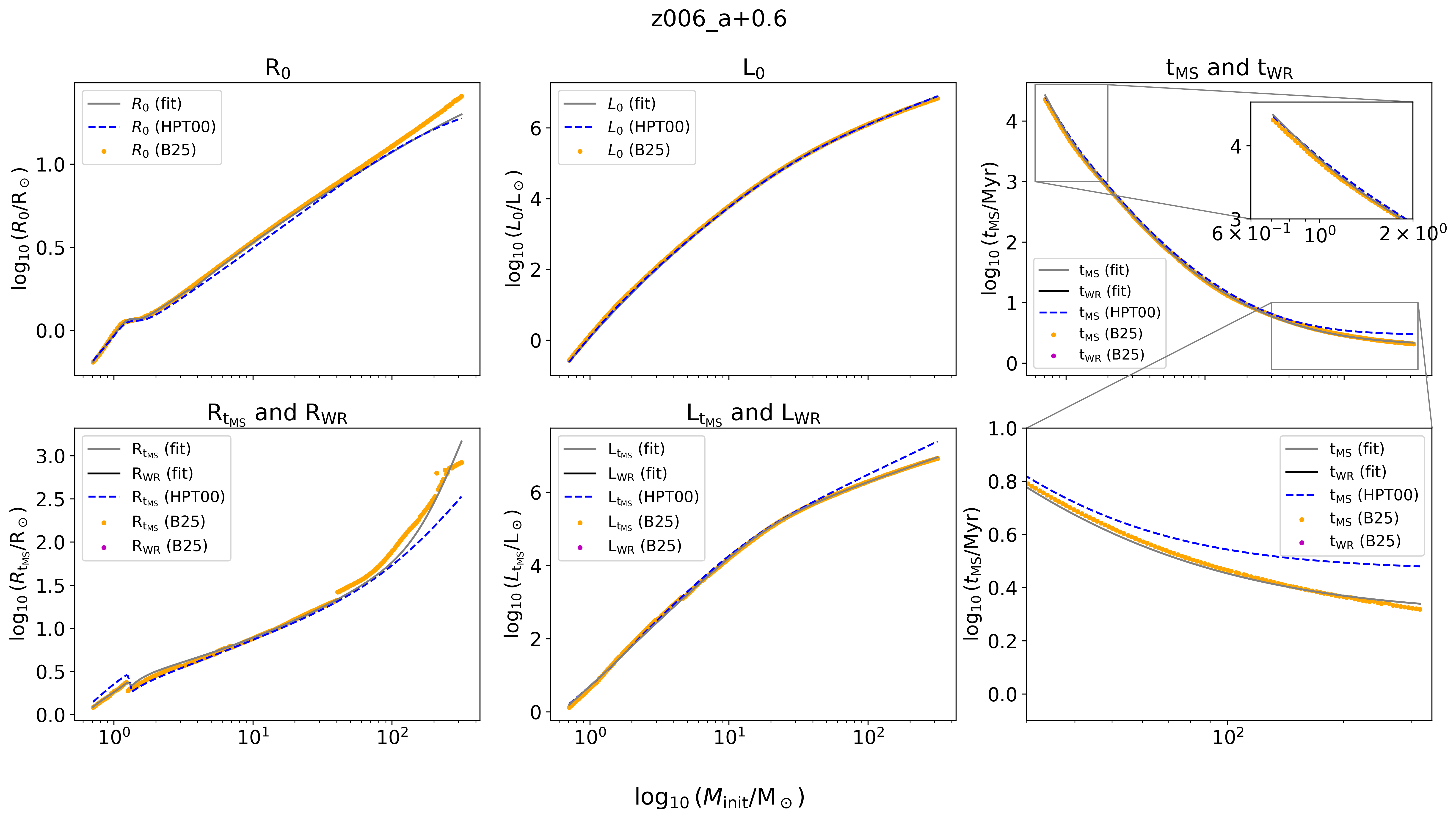}
    \caption{As Fig.~\ref{fig:fits_z003am2}, but for a composition of $Z=0.006$ and \afe=+0.6.}
    \label{fig:fits_z006a06}
\end{figure*}

\begin{figure*}
    \centering
    \includegraphics[width=0.95\linewidth]{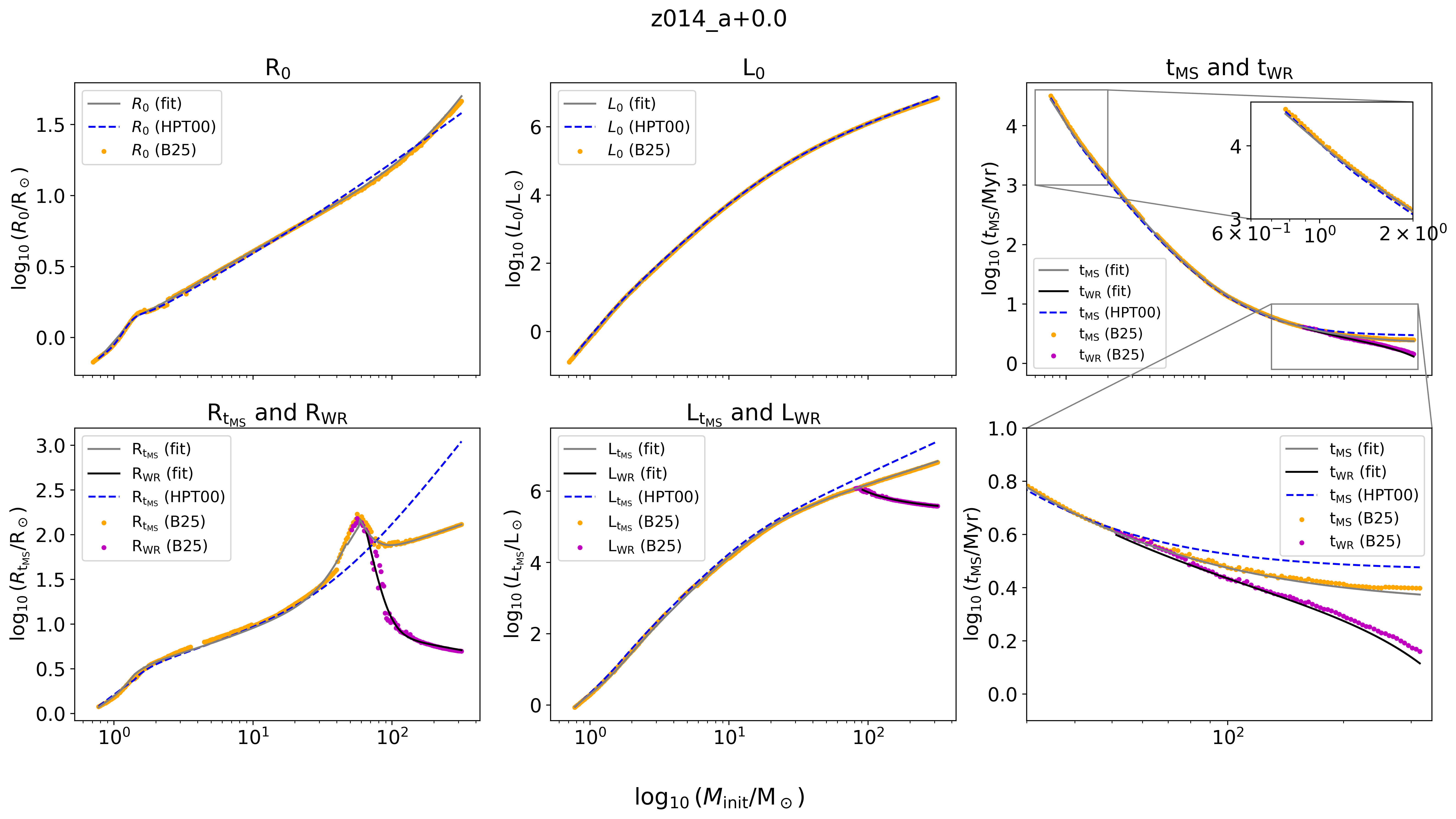}
    \caption{As Fig.~\ref{fig:fits_z003am2}, but for a composition of $Z=0.014$ and \afe=+0.0.}
    \label{fig:fits_z014a00}
\end{figure*}

\begin{figure*}
    \centering
    \includegraphics[width=0.95\linewidth]{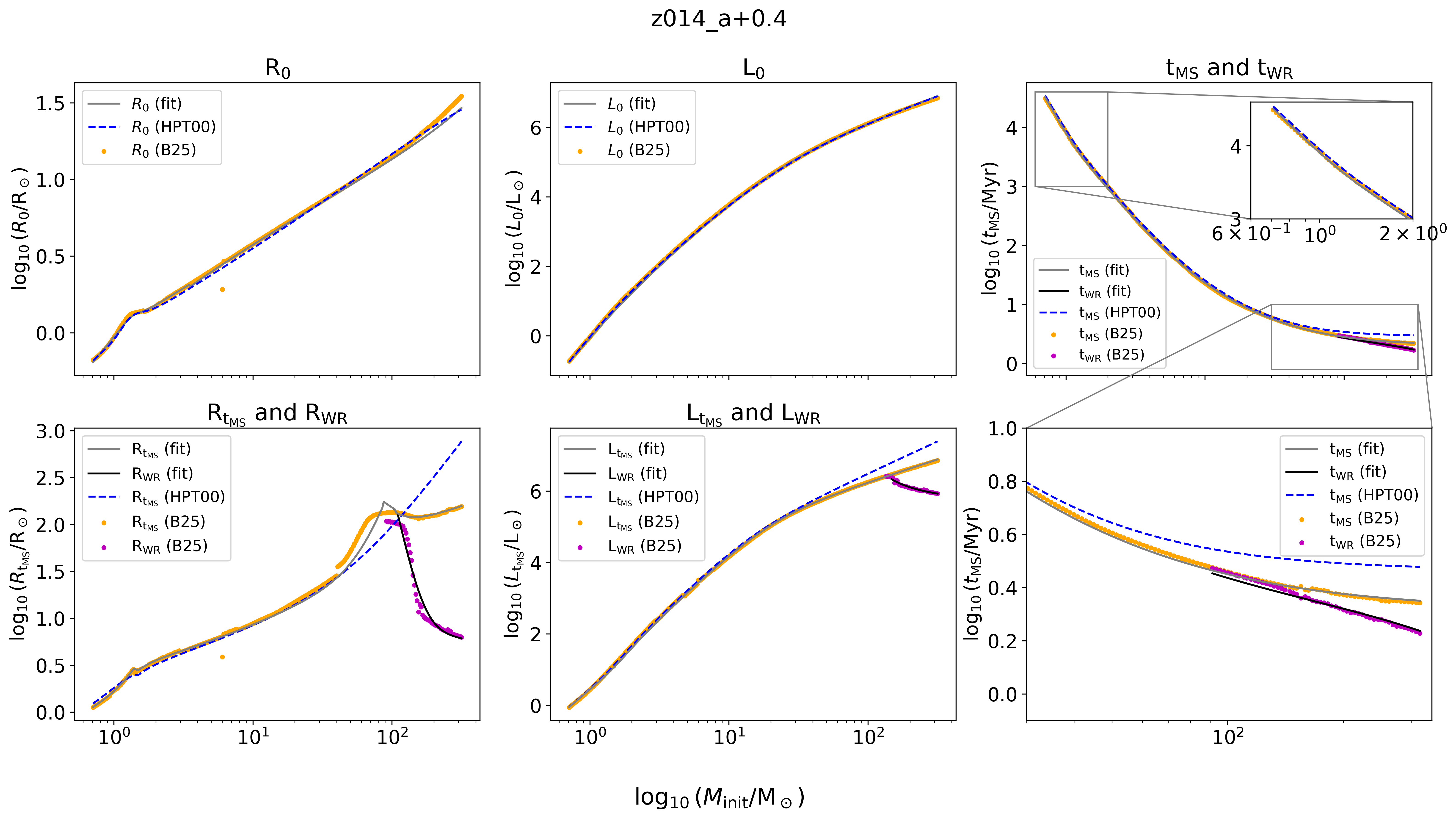}
    \caption{As Fig.~\ref{fig:fits_z003am2}, but for a composition of $Z=0.014$ and \afe=+0.4.}
    \label{fig:fits_z014a04}
\end{figure*}

\begin{figure*}
    \centering
    \includegraphics[width=0.95\linewidth]{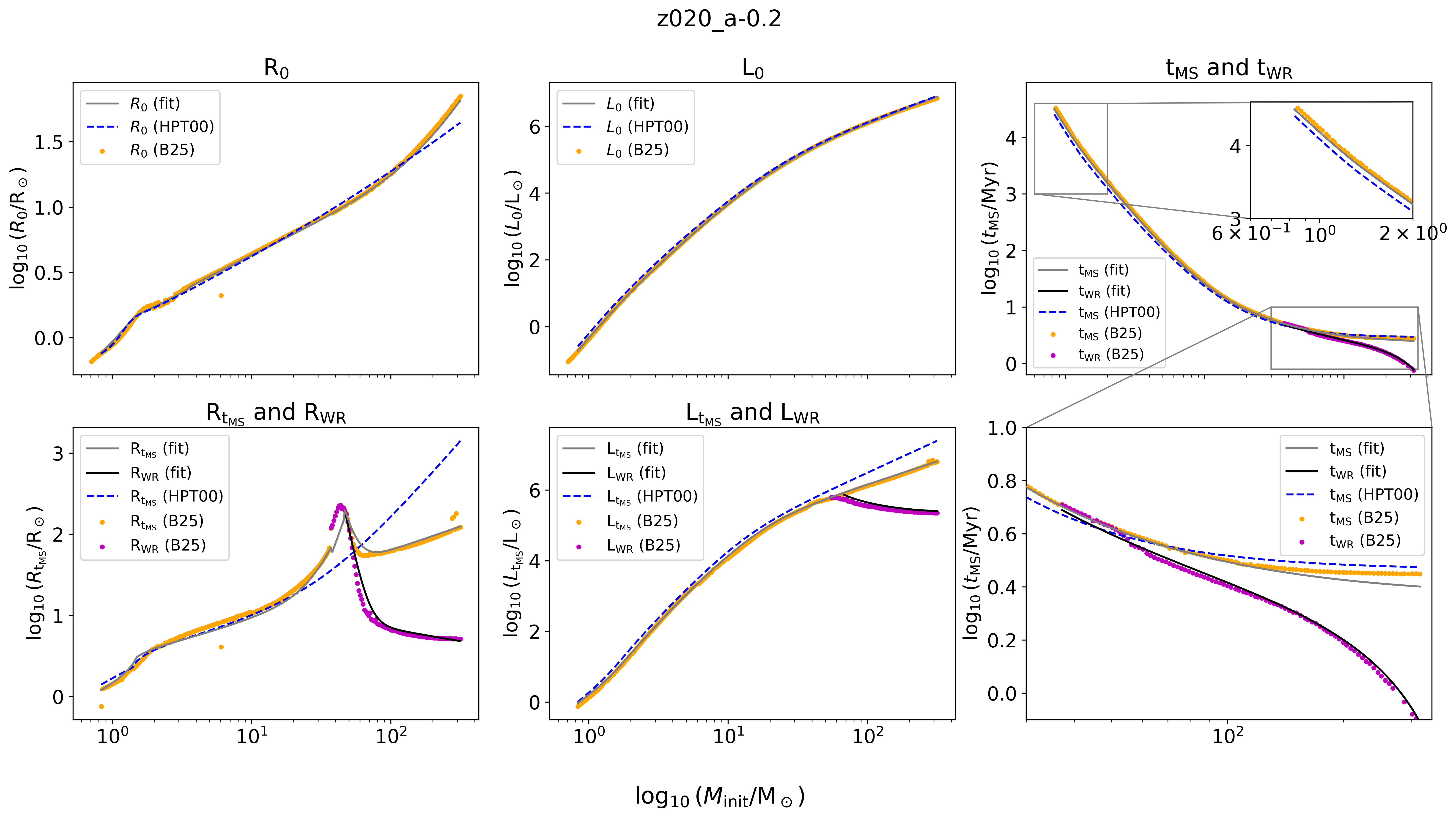}
    \caption{As Fig.~\ref{fig:fits_z003am2}, but for a composition of $Z=0.020$ and \afe=-0.2.}
    \label{fig:fits_z020am2}
\end{figure*}

\begin{figure*}
    \centering
    \includegraphics[width=0.95\linewidth]{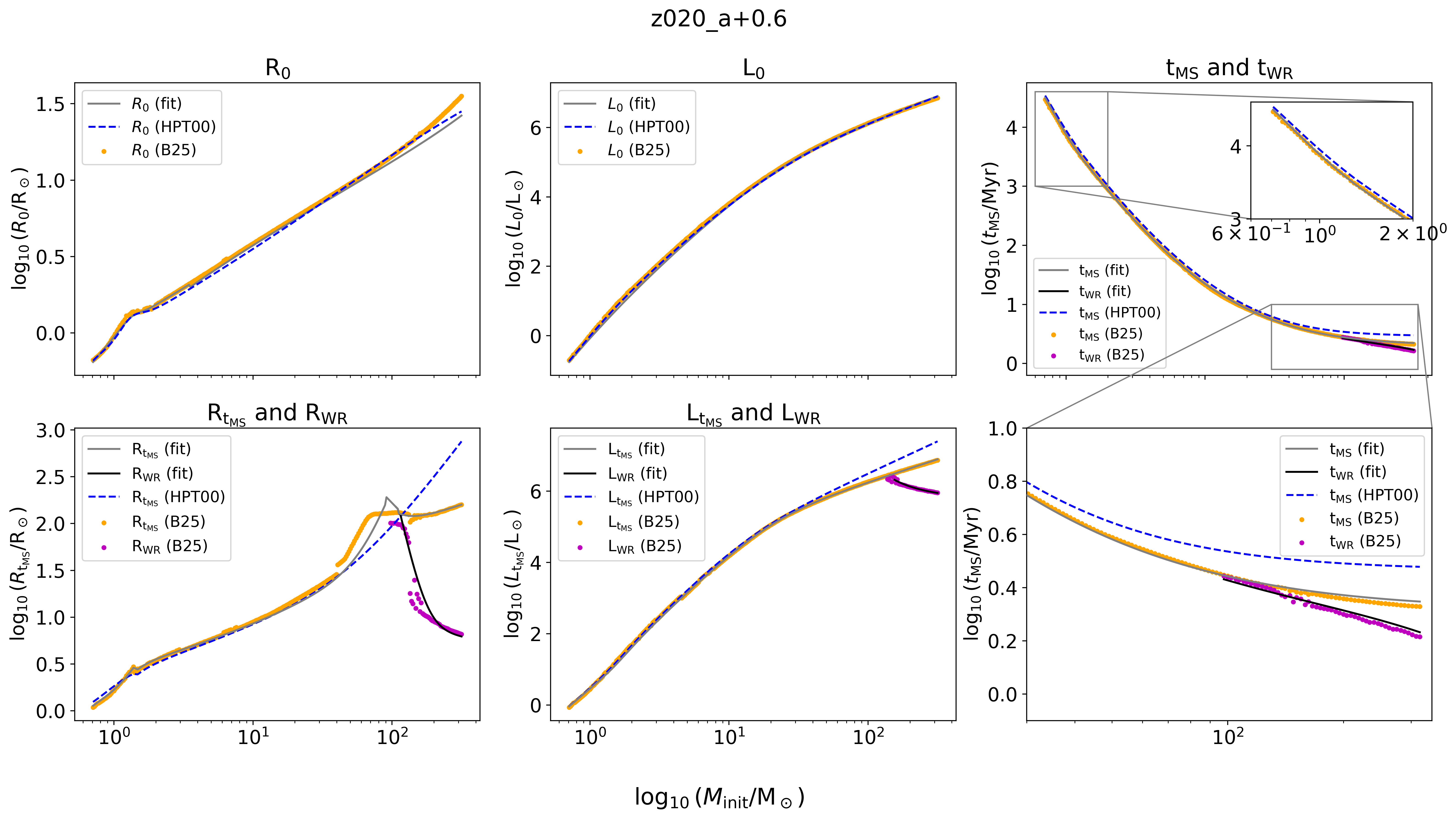}
    \caption{As Fig.~\ref{fig:fits_z003am2}, but for a composition of $Z=0.020$ and \afe=+0.6.}
    \label{fig:fits_z020a06}
\end{figure*}

\begin{figure*}
    \centering
    \includegraphics[width=0.95\linewidth]{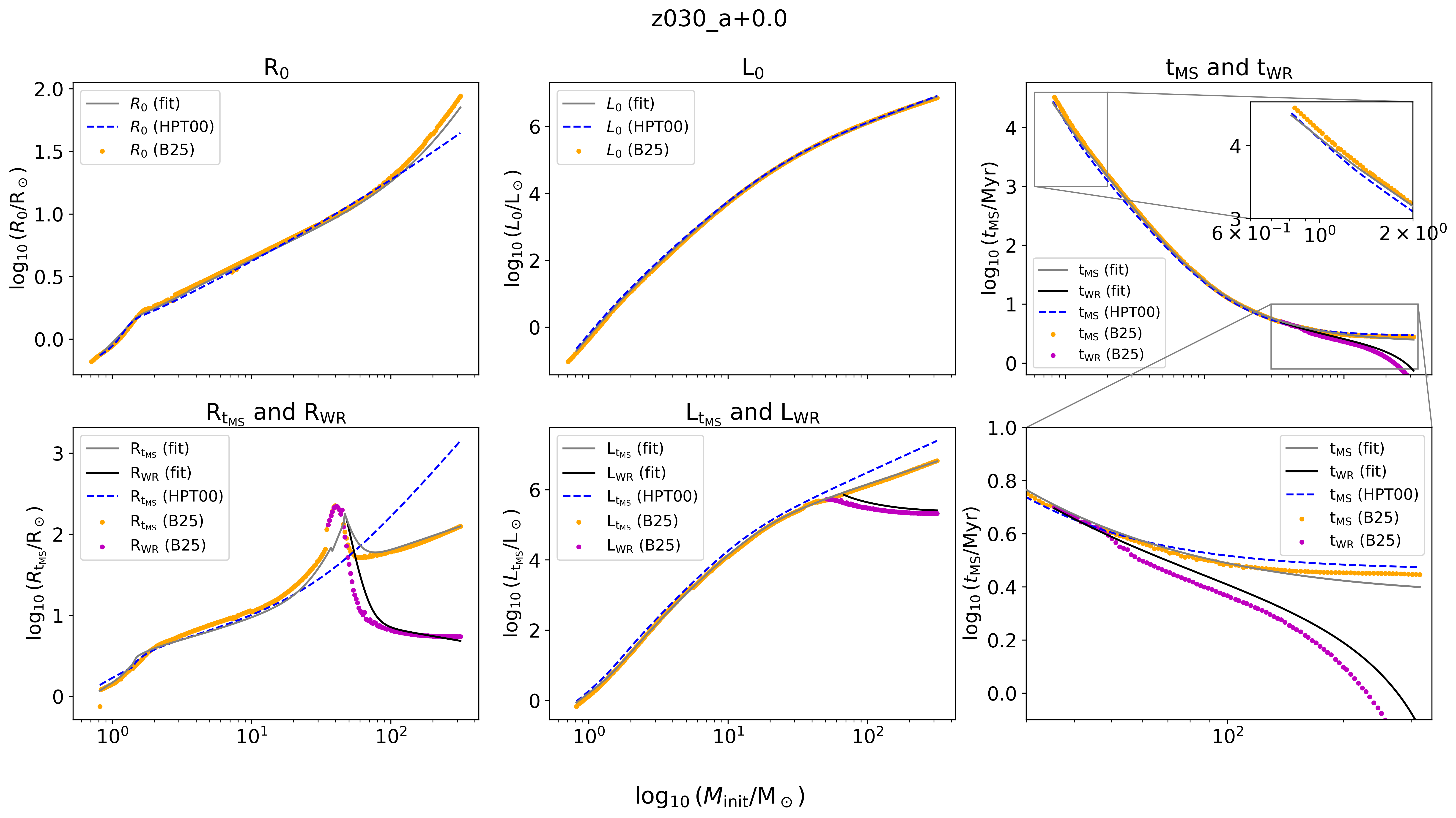}
    \caption{As Fig.~\ref{fig:fits_z003am2}, but for a composition of $Z=0.030$ and \afe=+0.0.}
    \label{fig:fits_z030a00}
\end{figure*}

\begin{figure*}
    \centering
    \includegraphics[width=0.95\linewidth]{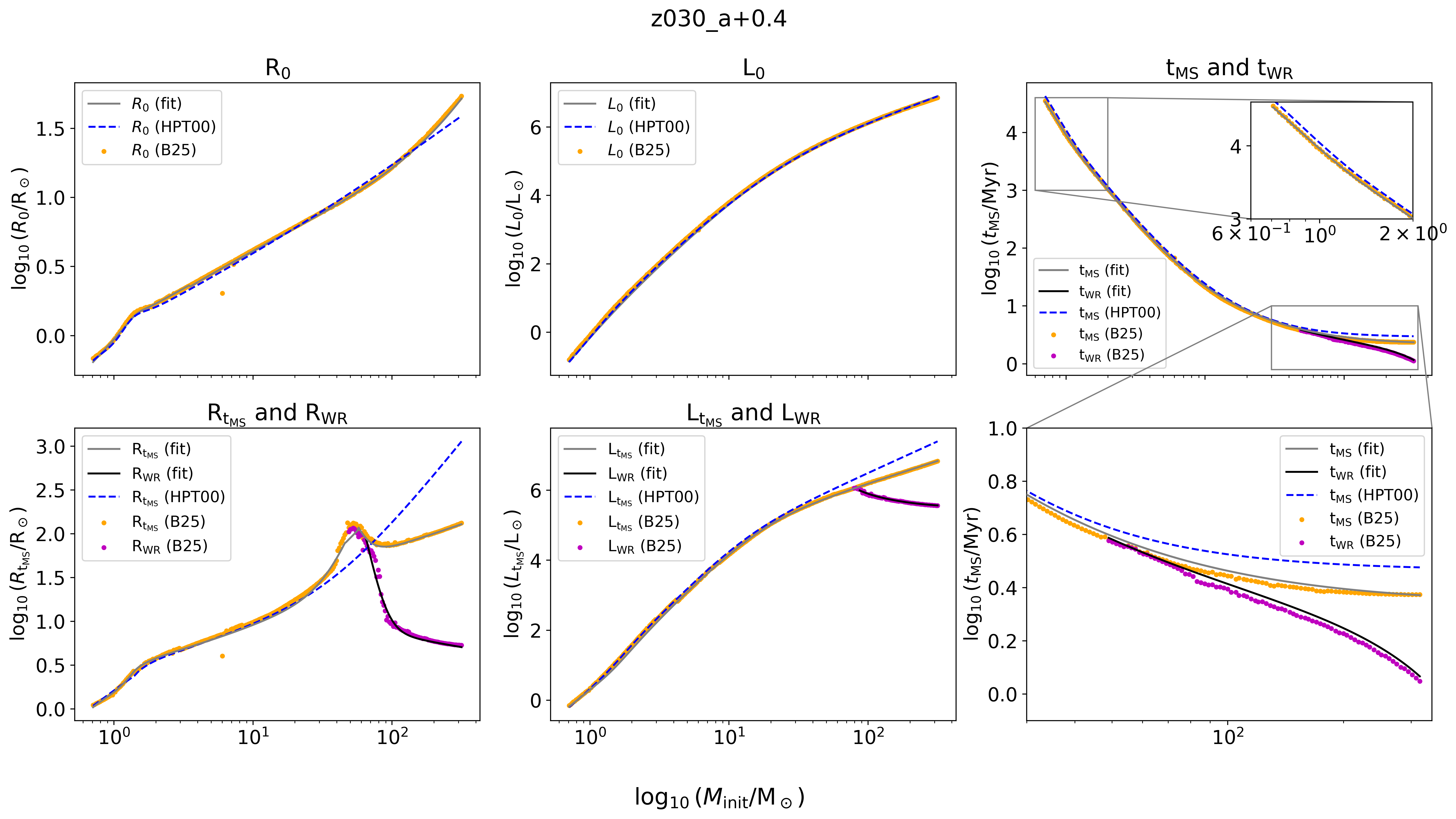}
    \caption{As Fig.~\ref{fig:fits_z003am2}, but for a composition of $Z=0.030$ and \afe=+0.4.}
    \label{fig:fits_z030a04}
\end{figure*}

Fig.~\ref{fig:fits_z003am2} to Fig.~\ref{fig:fits_z030a04} present the fitting results at a selection of metallicities, with the same layout as in Fig.~\ref{fig:fits_z020a02}. The compositions shown in each figure are listed in Table~\ref{tab:compplots}. These were chosen to show broad coverage of the composition parameter space including low metallicity ($Z=0.003$), moderate metallicity ($Z=0.006$), near-Solar ($Z=0.014$), Solar ($Z=0.020$), and super-Solar ($Z=0.030$), with multiple \afe\ values selected for each Z. 

\begin{table}
    \caption{Compositions used in the comparison figures found in main paper and this Appendix}
    \centering
    \begin{tabular}{l|c|c|c} \hline
    \label{tab:compplots}
    Figure     &  $Z$ & $Z'$ & \afe \\ \hline \hline
    Fig.~\ref{fig:fits_z003am2}     &  0.003 & 0.0054 & $-0.2$ \\
    Fig.~\ref{fig:fits_z003a02}     &  0.003 & 0.0030 & $+0.2$ \\
    Fig.~\ref{fig:fits_z003a06}     &  0.003 & 0.0014 & $+0.6$ \\
    Fig.~\ref{fig:fits_z006am2}     &  0.006 & 0.0110 & $-0.2$ \\
    Fig.~\ref{fig:fits_z006a02}     &  0.006 & 0.0060 & $+0.2$ \\
    Fig.~\ref{fig:fits_z006a06}     &  0.006 & 0.0027 & $+0.6$ \\
    Fig.~\ref{fig:fits_z014a00}     &  0.014 & 0.0190 & $+0.0$ \\
    Fig.~\ref{fig:fits_z014a04}     &  0.014 & 0.0095 & $+0.4$ \\
    Fig.~\ref{fig:fits_z020am2}     &  0.020 & 0.0360 & $-0.2$ \\
    Fig.~\ref{fig:fits_z020a02}     &  0.020 & 0.0200 & $+0.2$ \\
    Fig.~\ref{fig:fits_z020a06}     &  0.020 & 0.0091 & $+0.6$ \\
    Fig.~\ref{fig:fits_z030a00}     &  0.030 & 0.0410 & $+0.0$ \\
    Fig.~\ref{fig:fits_z030a04}     &  0.030 & 0.0200 & $+0.4$ \\ \hline
    \end{tabular}
\end{table}

\label{lastpage}
\end{document}